\documentclass[9pt,twocolumn,twoside]{article}

\usepackage{fullpage}
\usepackage{color}
\usepackage[export]{adjustbox}
\usepackage{authblk}
\usepackage{sidecap}
\usepackage{amsmath,mathtools}
\usepackage{amssymb}
\usepackage[normalem]{ulem}

\title{Covariations in ecological scaling laws fostered by \\ community \mbox{dynamics}}

\author[a,1]{Silvia Zaoli}
\author[a,b,1,2]{Andrea Giometto}
\author[c,2]{Amos Maritan}
\author[a,d,2]{Andrea Rinaldo}

\date{}

\affil[a]{\small Laboratory of Ecohydrology (ECHO/IIE/ENAC), EPFL, CH-1015 Lausanne, Switzerland}
\affil[b]{Department of Physics, Harvard University, MA-02138 Cambridge, United States}
\affil[c]{Dipartimento di Fisica ed Astronomia, Universit\`a di Padova, INFN and CNISM, I-35131 Padova, Italy}
\affil[d]{Dipartimento ICEA, Universit\`a di Padova, I-35131 Padova, Italy}
\affil[1]{S.Z. and A.G. contributed equally to this work.}
\affil[2]{To whom correspondence should be addressed. E-mail: andrea.rinaldo@epfl.ch, giometto@fas.harvard.edu or maritan@pd.infn.it.}

\begin{document}
\twocolumn[


\maketitle

\begin{@twocolumnfalse}
\begin{abstract}{Scaling laws in ecology, intended both as functional relationships among ecologically-relevant quantities and the probability distributions that characterize their occurrence, have long attracted the interest of empiricists and theoreticians. Empirical evidence exists of power laws associated with the number of species inhabiting an ecosystem, their abundances and traits. Although their functional form appears to be ubiquitous, empirical scaling exponents vary with ecosystem type and resource supply rate. The idea that ecological scaling laws are linked had been entertained before, but the full extent of macroecological pattern covariations, the role of the constraints imposed by finite resource supply and a comprehensive empirical verification are still unexplored. Here, we propose a theoretical scaling framework that predicts the linkages of several macroecological patterns related to species' abundances and body sizes. We show that such framework is consistent with the stationary state statistics of a broad class of resource-limited community dynamics models, regardless of parametrization and model assumptions. We verify predicted theoretical covariations by contrasting empirical data and provide testable hypotheses for yet unexplored patterns. We thus place the observed variability of ecological scaling exponents into a coherent statistical framework where patterns in ecology embed constrained fluctuations.}
\end{abstract}
\end{@twocolumnfalse}\twocolumn]

\begin{figure*}
\centering
\includegraphics[width=11.4cm]{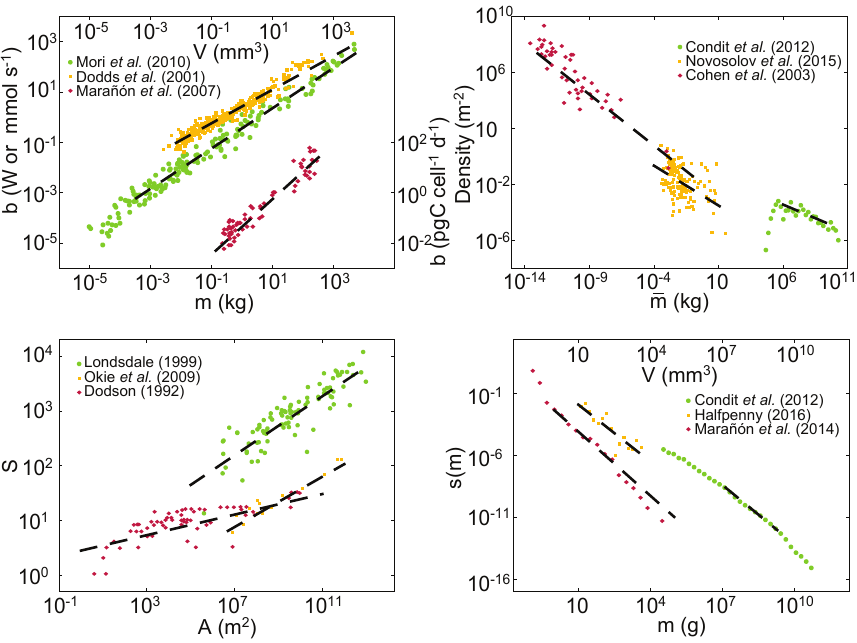}
\caption{Empirical evidence of scaling ecological patterns in different ecosystems: forests (green), terrestrial (yellow) and aquatic ecosystems (magenta). Regression lines are linear least square fits of log-transformed data. A) Kleiber's law: metabolic rates in $\mu$mol $s^{-1}$ (forests), W (terrestrial ecosystems),  pgC cell$^{-1}$ d$^{-1}$ (aquatic ecosystems). Size in kg (forests and terrestrial ecosystems) and in $\mu$m$^3$ (aquatic ecosystems); B) Damuth's Law ($\bar m$ is a species' mean mass); C) SAR; D) Community size-spectrum: size in g (forests and terrestrial ecosystems) and $\mu$m$^3$ (aquatic ecosystems). See SI (section 2) for scaling exponents estimates/errors. References to the datasets are provided in Table S2.} 
\label{fig:empiricalevidence}
\end{figure*}

A prototypical example of ecological scaling law is the species-area relationship (SAR) on which island biogeography is based \cite{Macarthur1963}. It states that the number of species $S$ inhabiting disjoint ecosystems increases as a power of their area, i.e. $S\propto A^z$, where $z$ is the SAR scaling exponent. The widespread interest in scaling laws \cite{Kleiber1932,Damuth1981,Brown2004,Marquet2005,Sole2006,White2007,Okie2009} lies in their intrinsic predictive power, e.g. the use of SAR to forecast how many species might go extinct if the available habitat shrinks or is fragmented into smaller unconnected parts. Precise estimates of the scaling exponents' values are thus crucial. Empirical evidence, however, shows that they vary considerably across ecosystems \cite{Cavender-Bares2001,Finkel2004,Maranon2015}, suggesting that exponents of scaling ecological laws are far from universal, although the power-law form proves remarkably robust (Fig. \ref{fig:empiricalevidence}).

Scaling patterns in ecology have mostly been studied within independent ecosystems, leading to canonical estimates of scaling exponents which may not be simultaneously achievable in a single ecosystem due to extant and consistency constraints. Although ecological scaling laws have historically been treated as disconnected, it is instructive to show by a simple example that they are functionally related. Consider a community hosted within a resource-limited ecosystem of area $A$ whose $i$-th species is characterized by abundance $n_i$ and typical body mass $m_i$. Empirical evidence suggests that the following patterns can be described at least approximately by power laws, disregarding possible cutoffs at large sizes: i) the community size spectrum \cite{Sheldon1972,Cavender-Bares2001,Rinaldo2002,White2007}, $s(m)\propto m^{-\eta}$, i.e., the fraction of individuals of body mass $m$ regardless of species; ii) the distribution of species' typical body masses \cite{Marquet1998,Marquet2005} $P(m)\propto m^{-\delta}$ and iii) the average abundance of a species with typical body mass $m$, $\langle n|m \rangle \propto m^{-\gamma}$ (Damuth's law \cite{Damuth1981,Marquet1990} or local size-density relationship \cite{White2007}). A back-of-the-envelope calculation suggests that the total number of individuals of mass $m$ (regardless of species) is the product of the number of species with typical mass $m$ and the average abundance of a species with typical mass $m$ (i.e., $s(m) \propto P(m) \langle n|m,A \rangle$). Thus, the scaling exponents must satisfy the consistency relationship: 
\begin{equation}
\label{eq:contoserva}
\eta=\delta+\gamma,
\end{equation}
which proves that exponents measured in the same ecosystem are not independent, unlike exponents measured in disparate ones. This example and a few others identified in earlier works  \cite{Rinaldo2002,Southwood2006,Banavar2007,Grilli2012} and in the context of MaxEnt \cite{Harte2011} highlight the need for a framework that comprehensively accounts for linking relationships among macroecological scaling laws.

\section*{\normalsize Results}
Here, we show that supply limitation imposes precise constraints on macroecological patterns, along with consistency relationships such as \eqref{eq:contoserva}. Assuming that individual resource consumption (metabolic) rates under field conditions, $b$, relate to body mass $m$ via Kleiber's law \cite{Kleiber1932,West1997,Banavar1999}, i.e. $b = c m^\alpha$ (with $\alpha \le 1$, $c$ constant), we argue that the constraint placed on the total community consumption rate $B$ by the finiteness of available resources translates into constraints on sustainable body sizes and abundances. 
To show this, we move from a scaling ansatz for the joint probability $P(n,m|A) dn dm$ of finding a species of abundance $n \in [n,n+dn]$ and typical mass $m \in [m,m+dm]$ within an ecosystem of area $A$, that postulates correlated fluctuations in mass and abundance for any species. Such joint distribution, which we term the `fundamental distribution', must be viable in the sense that its marginals must reproduce the empirical scaling observed in the field. Our conclusion (Methods, SI) is that a general, yet analytically tractable to some extent, form for $P(n,m|A)$ is: 
\begin{equation}
\label{eq:P(n,m|A)}
P(n,m|A)=(\delta-1) m^{-\delta} n^{-1} G\Bigl ( \frac{n}{\langle n|m,A \rangle} \Bigr )
\end{equation}
where:
\begin{equation}
\label{eq:P(m|A)}
P(m|A)=(\delta-1) m^{-\delta}
\end{equation}
is the probability density of finding a species of typical mass $m \in [m,m+dm]$,
\begin{equation}
P(n|m,A)=n^{-1} G\Bigl ( \frac{n}{\langle n|m,A \rangle} \Bigr )
\label{eq:P(n|m,A)}
\end{equation}
 is the probability density of finding a species of abundance $n \in [n,n+dn]$ among those of mass $m$ and
\begin{equation}
\label{eq:Damuth}
\langle n|m,A \rangle=m^{-\gamma} A^\Phi h\left(\frac{m}{A^{\lambda}}\right)
\end{equation}
is the average abundance of a species of typical mass $m$ within an ecosystem of area $A$. The properties of $G$ and $h$ are described in the Methods and in the SI (section 1.3).

Eqs. (\ref{eq:P(n,m|A)}--\ref{eq:Damuth}), through their marginals and moments, give rise to the empirically-observed set of macroecological scaling laws (SI section 1), namely: the SAR, $S \propto A^z$; Damuth's law, $\langle n|m,A \rangle \propto  A^\Phi m^{-\gamma}$, where the $A$ dependency is an addition to the original relationship proposed by Damuth; the community size-spectrum $s(m|A)\propto m^{-\eta}$; the species' mass distribution $P(m|A)\propto m^{-\delta}$; the scaling of the total biomass, $M\propto A^\mu$; the scaling of the total abundance \cite{Hubbell2001}, $N\propto A^\nu$; the scaling of the largest organism's mass \cite{Burness2001,Okie2009}, $m_{\max}\propto A^\xi$; the relative species' abundance (RSA) \cite{Preston1948}, defined as the probability of finding a species with abundance $n$; Taylor's law \cite{Cohen2012,Giometto2015}, linking mean and variance of a species' abundance as $\langle n^2 \rangle - \langle n \rangle^2 \propto \langle n \rangle ^\beta$. Note that the SAR and the scaling of $N$ and $M$ with $A$ are predictions (i.e., not assumptions) of our framework which follow from the imposed constraint on shared resources. 

In addition to Eq. \ref{eq:contoserva}, the scaling framework predicts the following exact relationships among scaling exponents: 
\begin{equation}
z=1-\Phi-\max\{0,\lambda(1+\alpha-\eta)\}
\label{zeta}
\end{equation}
\begin{equation}
\mu=1+\max \{ 0,\lambda(2-\eta) \}-\max \{ 0,\lambda(1+\alpha-\eta) \}
\label{mu}
\end{equation}
\begin{equation}
\nu=1-\max \{ 0,\lambda(1+\alpha-\eta) \}
\label{nu}
\end{equation}
\begin{equation}
\xi=\frac{z}{\delta-1}
\label{xi}
\end{equation}
where $\lambda$ accounts for a finite-size effect in Damuth's law: $\langle n|m,A \rangle = A^{\Phi} m^{-\gamma} h(m/A^\lambda)$, with $\lim_{x \to 0} h(x)=\mbox{const}$ and $\lim_{x\to\infty} h(x)=0$ (Methods). The exponent $\beta$ does not appear because its value is found to be independent from other exponents \cite{Giometto2015} (Methods).
Only $5$ of the $10$ observable exponents are thus independent.
\eqref{zeta} implies, in any ecosystem where $z>0$, as observed for forests \cite{Lonsdale1999}, mammals \cite{Okie2009} and lizards \cite{Novosolov2015}, that $\Phi<1$ and therefore species' densities decrease with increasing area. 
\eqref{xi} is compatible with the linking relationship derived in Southwood \textit{et al.} \cite{Southwood2006}, which is shown here to be one component of a broader set of linking relationships (SI section 1.9). Also, area-independent constraints to the maximum size of an organism may lead to a breakdown of Eq. \ref{xi} at large $A$ (SI, section 1.8.3).

To corroborate the validity of our framework, we investigated a broad class of stochastic models for the dynamics of a community limited by resource supply which is assumed to be proportional to the ecosystem area (Methods and SI section 3). Despite major changes in the speciation dynamics and regardless of  parametrization, all models are compatible with the finite-size scaling structure of $P(n,m|A)$ and therefore reproduce both the macroecological laws reported above and their covariations.

\begin{figure*}
\centering
\includegraphics[width=114mm]{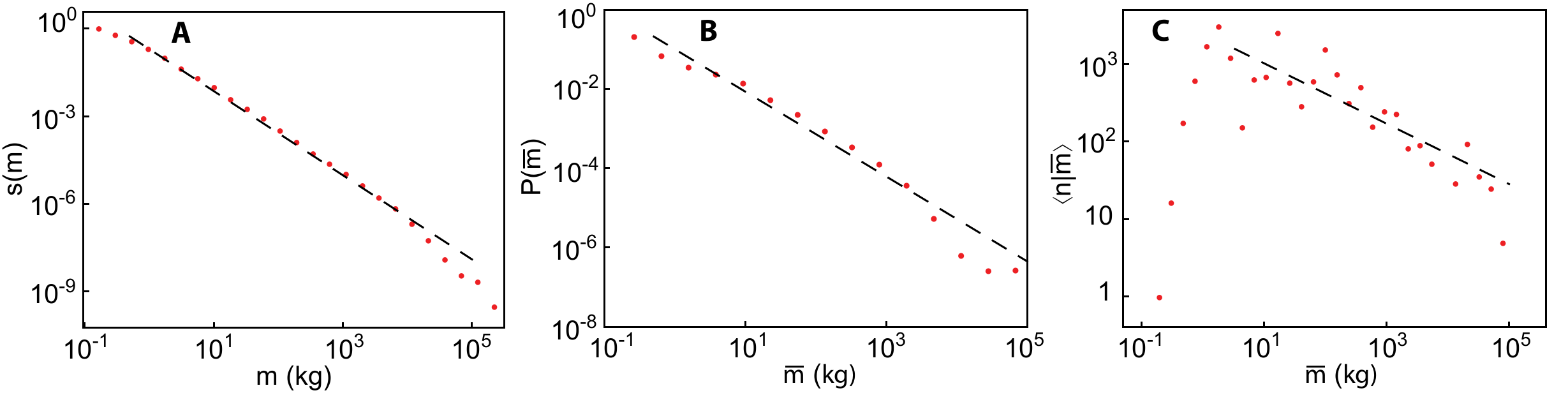}
\caption{Empirical evidence of scaling patterns in Barro Colorado Island \cite{Condit2012}, seventh census: A) Community size-spectrum, i.e. the probability distribution of individuals' mass regardless of species (red dots), B) Distribution of species mean masses $P(\bar{m})$ (red dots), C) Damuth's law, i.e. the average abundance $\langle n|\bar{m}, A \rangle$ of a species of mean mass $\bar{m}$ (red dots), where each point is the average abundance over bins of logarithmic size. Dashed black lines show power-functions with exponents as in Table S3. Details on exponents' estimates are reported in the Methods and SI (section 2.2.1).}
\label{fig:BCI}
\end{figure*}

The empirical verification of all the relationships (\ref{eq:contoserva}, \ref{zeta}--\ref{xi}) would require the simultaneous measurement within the same ecosystem of all scaling exponents. Unfortunately, such a comprehensive dataset does not seem to exist to date. Therefore, we searched for empirical data that would allow verifying, at least partially, Eqs. (\ref{eq:contoserva},\ref{zeta}--\ref{xi}). We found that \eqref{eq:contoserva} is verified within the errors in the tropical forests datasets of Barro Colorado Island (BCI, see Fig. \ref{fig:BCI}) \cite{Condit2012} and of the Luquillo forest \cite{Zimmerman2010} (Methods and SI section 2.2.1). \eqref{zeta} is verified within the errors in a dataset of lizard population densities on 64 islands worldwide (LIZ) \cite{Novosolov2015} (SI section 2.2.2). Finally, \eqref{xi} is verified within one standard error in a dataset of mammal body sizes in several islands in Sunda Shelf (SSI) \cite{Okie2009} (Methods and SI section 2.2.3). All the empirical tests performed are summarized in table S11.

 \section*{\normalsize Discussion}

The theoretical framework proposed here rationalizes the observed variability of ecological exponents across ecosystems. Jointly with empirical evidence, our framework supports the tenet that scaling exponents may vary across ecosystems but must satisfy consistency relationships that result in exact covariations of ecological patterns. When applying scaling laws, for example in conservation, care should be exerted not to combine exponents measured in different settings, which may not satisfy the relationships (\ref{eq:contoserva}, \ref{zeta}--\ref{xi}) leading to misled predictions for unmeasured patterns. 

Our framework adopts the minimum set of hypotheses allowing to reproduce widespread macroecological patterns found in empirical data, without compromising analytical tractability. Such analytical tractability is important in this context because it highlighs the relationships among macroecological patterns in simple terms, i.e. via algebraic relationships among their scaling exponents. However, there may be empirical examples where some of the patterns considered here deviate from pure power-laws. The framework presented here already comprises cut-offs in the community size-spectrum and in Damuth's law, allowing deviation from pure power-law behavior at large body sizes, and can be generalized to describe more complex ecological settings. For example, one can account for the fact that individuals' body sizes within the same species are characterized by intra-specific distributions \cite{Giometto2013}. Such generalization of the framework bears no modification to the linking relationships among macroecological laws, unless intra-specific size-distribution are heavy-tailed, in which case corrections apply (SI section 1.8.4 and 1.8.5). One can also account for curvatures in Kleiber's law \cite{Kolokotrones2010,Mori2010,Maranon2013,Banavar2014a}, which are found to induce curvatures in the species-area relationship (SI section 1.8.1). A cut-off or a non-power-law form for $P(m|A)$ can also be considered (SI section 1.8.6). Finally, the assumption that all individuals share the same resources would imply that our results apply to single trophic levels. However, we show in the SI (section 1.8.2) how our framework can be extended to describe multi-trophic systems. In the most general scenario in which the dependence of  $P(m|A)$ on $m$ and $\langle n|m,A \rangle$ on $m$ and $A$ cannot be expressed as in \eqref{eq:P(m|A)} and \eqref{eq:Damuth}  (which, however, are compatible with several empirical case studies) nor described by the generalizations treated in the SI section 1.8.6, one would have to rely on numerical methods to derive the covariations between macroecological patterns, following the same route adopted in our theoretical investigation. We anticipate that generalizations of Eqs. (\ref{eq:contoserva},\ref{zeta}--\ref{xi}) would hold in this scenario, although they would be expressed as integral equations in terms of the probability distributions introduced above.
The next step in the study of co-varying ecological patterns is the identification of the mechanisms that determine the values of the independent exponents. For example, theoretical evidence \cite{Bertuzzo2011} suggests that the value of $z$ is affected by topological constraints posed by the ecological substrate. 

\begin{figure*}
\centering
\includegraphics[width=114mm]{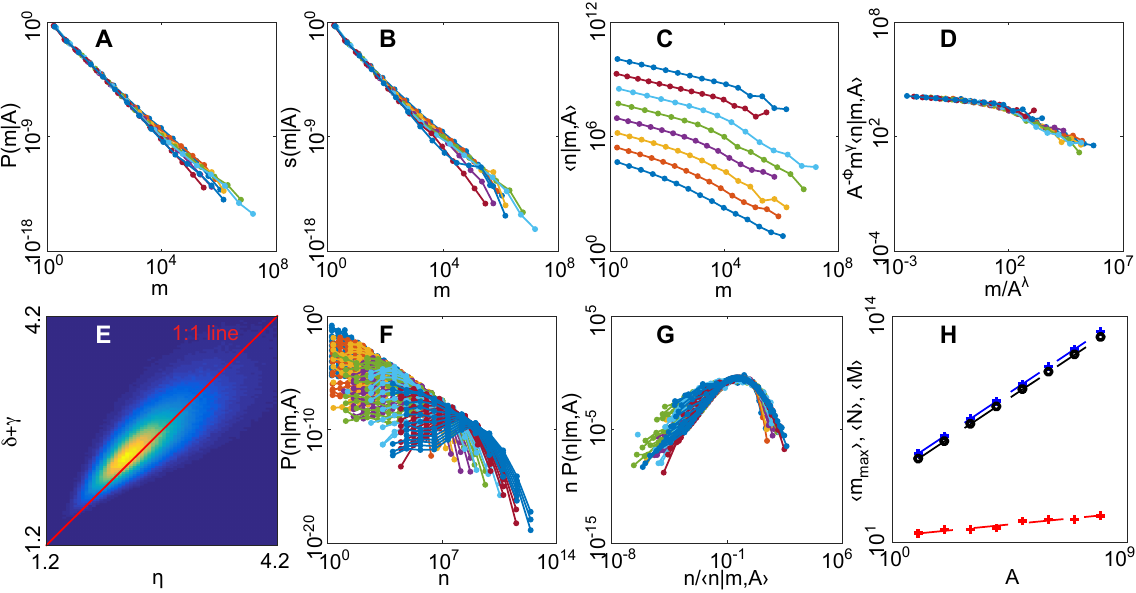}
\caption{Scaling patterns from the basic community dynamics model (Methods). Different colors refer to different values of $A= 10^i$, from $i=1$ (lower blue curve in panel C) to $i=8$ (upper blue curve in panel C). Panels A, B, C and f show respectively $P(m|A)$, $s(m|A)$, $\langle n|m,A \rangle$ and $P(n|m,A)$ at stationarity. Panels D and G show collapses of $\langle n|m,A \rangle$ and $P(n|m,A)$. Eqs.  (\ref{eq:P(n|m,A)}-\ref{eq:Damuth}) are verified because the curves $nP(n|m,A)$ versus $n/ \langle n|m,A\rangle$ collapse on the same curve for different $A$ (panel G), and so do the curves  $m^\gamma A^{-\Phi} \langle n|m,A\rangle$ versus $m/A^{\lambda}$  (panel D). Panel E: density histogram plot of $\eta$ vs $\delta+\gamma$ at different times. Panel H: scaling of $\langle m_{\max} \rangle$ (red crosses and dashed lines), $\langle N \rangle$ (black dots and dashed lines) and $\langle M \rangle$ vs $A$ (blue crosses and dashed lines).}
\label{fig:collapse}
\end{figure*}

\section*{\normalsize Methods}
\subsection*{\normalsize The fundamental distribution $P(n,m|A)$}

We consider an ecosystem of area $A$. We assume that the minimum viable mass for an organism is $m_0>0$ independent of $A$, so that $P(n,m|A)$ is zero for $m<m_0$. We measure mass and area in units of $m_0$ and of a reference unit area $a_0$, so that $m$ and $A$ are dimensionless. To comply with empirical evidence \cite{Marquet1998,Marquet2005}, we assume that $P(m|A)$ is a power function of $m$ (Eq. \ref{eq:P(m|A)}): $P(m|A)=(\delta-1) m^{-\delta}$,
where $\delta>1$ ensures integrability (see also the SI (section 1.8.6)).
For $P(n|m,A)$, in accordance with the community dynamics models and with the empirical observation of Damuth's law, we posit (Eq. \ref{eq:P(n|m,A)}): $P(n|m,A)=n^{-1} G \left( n/{\langle n|m,A \rangle} \right)$,
where $G(x)$ is such that $\int _0^\infty x^j G(x)dx<\infty$ for $j=-1,0,1$ and (Eq. \ref{eq:Damuth}) $\langle n|m,A \rangle=m^{-\gamma} A^\Phi h({m}/{A^{\lambda}})$
is the average abundance of a species of typical mass $m$ in an ecosystem of area $A$. The properties of $G$ ensure that $\int_0^\infty dn \ n \  P(n|m,A) = \langle n|m,A \rangle$, that is, Damuth's law is reproduced. 
The factor $n^{-1}$ in \eqref{eq:P(n|m,A)} is discussed in the SI. The function $h(x)$ describes an $A$-dependent cutoff on the abundances as observed in simulations of stochastic models of community dynamics (see, e.g., Fig. \ref{fig:collapse}C). $h(x)$ is such that $h(x) = o\left( x^{-2+\delta+\gamma} \right)$ as $x\to\infty$ to ensure  convergence of the moments we are interested in and $\lim_{x\rightarrow 0} h(x)=h_0$ constant to yield a power law regime before the cut-off.
Eqs. (\ref{eq:P(n,m|A)}--\ref{eq:Damuth}) constitute our  \emph{ansatz} on the scaling form of $P(n,m|A)$.

\subsection*{\normalsize Derivation of scaling ecological laws}
Eq. \eqref{eq:P(n,m|A)} can be used to compute the scaling of the ($j$,$k$)-th moment  with $A$ exactly (see SI section 1.6 for the detailed computation) as:  
\begin{equation}
\begin{split}
\label{moments}
I_{j,k}&=\int_1^\infty \int_0^\infty n^{j} m^{k} P(n,m|A) dn dm \\&\propto A^{j\Phi+\max \{ 0,\lambda(1+k-\delta- j \gamma) \}}.
\end{split}
\end{equation}
Scaling laws are derived from Eq. \ref{moments} as follows: 

\noindent i) \textit{Species-Area Relationship.} 
The total number of species $S$ is linked to the area $A$ via the constraint $B\propto A$ (SI sections 1.1 and 1.2). The total metabolic rate of the community is:
\begin{equation}
B\propto S \ I_{1,\alpha} =S \int_1^\infty \int_0^\infty n m^\alpha P(n,m|A) dn dm.
\label{speciesA1}
\end{equation}
where we have used Kleiber's law. 
The hypothesis $B \propto A$ (SI sections 1.1 and 1.2) leads to:
\begin{equation}
S \propto A^{z}, \mbox{ with } z=1-\Phi-\max \{0,\lambda(1+\alpha-\delta-\gamma)\},
\label{speciesA}
\end{equation}
which corresponds to Eq. \ref{zeta}. Note that, if $z>0$, this equation predicts that species' densities decrease with increasing $A$ (recall that $\langle n|m,A \rangle \propto A^\Phi$ with $\Phi<1$). This can be understood through a heuristic argument: if $N \propto A^\nu$ with $\nu \leq1$ and $S \propto A^z$, it follows that the average abundance per species scales sub-linearly as $\langle n|A \rangle=N/S\propto A^{\nu-z}$. Such scaling of $\langle n|A \rangle$ with $A$ is retained by the average abundance conditional on body size, $\langle n|m,A\rangle$, and thus back-of-the-envelope calculations suggest $\Phi=\nu-z$, which coincides with Eq. \ref{zeta} and \eqref{speciesA}, given Eq. \ref{nu}. This result is a novel prediction of our framework and implies that species' densities decrease with ecosystem area. Note also that we refer here to the so-called island SAR \cite{Preston1962}, obtained by counting species inhabiting disjoint patches of land (e.g. islands, lakes or, in general, areas separated by environmental barriers from the surroundings which we can think of as closed ecosystems) rather than to nested SARs where areas are sub-patches of a single larger domain \cite{Harte2009,Azaele2016}. The two SARs are quite different, as the nested SAR is related to the spatial distribution of individuals, while the island SAR stems from complex eco-evolutionary dynamics shaping the community;

\noindent ii) \textit{Damuth's law} is traditionally intended as the scaling of the average density of a species, ${\langle n|m,A \rangle}/{A}$, with its typical mass $m$. However, as discussed in \textit{i)}, the density of a species depends on the inhabited area, as found for example in our empirical analyses of the LIZ dataset \cite{Novosolov2015} (see SI section 2.2.4). Thus, we consider here a generalized version of Damuth's law, relating the average abundance $\langle n|m,A \rangle$ to the typical mass of the species and to the area of the ecosystem $A$.
Indeed, in our framework the average abundance of a species of characteristic mass $m$ in an ecosystem of area $A$ is:
\begin{equation}
\label{eq:metDamuth}
\begin{split}
\langle n|m,A \rangle&= \int_0^\infty n P(n|m,A) dn \\&= \int_0^\infty G\left[ \frac{n m^{\gamma}}{A^{\Phi}} \frac 1{h\left({m}/{A^{\lambda}} \right)} \right] dn \\
&= A^{\Phi}m^{-\gamma}h \left(\frac{m}{A^{\lambda}} \right) \int_0^\infty G\left(x \right) dx
 \\&\propto A^{\Phi}m^{-\gamma} h \left(\frac{m}{A^{\lambda}} \right),
\end{split}
\end{equation}
where the properties of $G$ ensure the convergence of the integral. The average abundance of a species of mass $m$, thus, has a power-law dependence on $m$ and $A$, as found in empirical data, and an $A$-dependent cutoff at large masses provided by the function $h$, as shown by our community dynamics models (Fig. \ref{fig:collapse}C).

\noindent iii) \textit{Scaling of total biomass.} The total biomass can be computed as:
\begin{equation}
\begin{split}
M&=S\langle n m \rangle =S \ I_{11} \\ &\propto A^{1+\max \{ 0,\lambda(2-\delta-\gamma) \}-\max \{ 0,\lambda(1+\alpha-\delta-\gamma) \}},
\label{totalbiomass}
\end{split}
\end{equation}
yielding Eq. \ref{mu} of the main text.

\noindent iv) \textit{Scaling of total number of individuals.} The total number of individuals $N$ in the ecosystem is given by:
\begin{equation}
N =S \langle n \rangle= S \ I_{10} \propto A^{1-\max \{ 0,\lambda(1+\alpha-\delta-\gamma) \}},
\label{totalabundance}
\end{equation}
yielding Eq. \ref{nu} of the main text.

\noindent v) \textit{Community size-spectrum.} 
The size spectrum $s(m|A)$ is the probability that a randomly sampled individual (regardless of its species) has mass in $[m,m+dm]$ and is therefore equal to:
\begin{equation}
\begin{split}
s(m|A) &= \frac SN \int_0^\infty n P(n,m|A) \ dn \\
& \propto A^{-\max \{ 0,\lambda(1-\delta-\gamma) \}} m^{-\delta-\gamma} h \left(\frac{m}{A^{\lambda}} \right)\\&=m^{-\delta-\gamma} h \left(\frac{m}{A^{\lambda}} \right),
\label{s(m)}
\end{split}
\end{equation}
where we have used \eqref{speciesA} and \eqref{totalabundance}, $\delta>1$ and the properties of $G$ ensure the convergence of the integral. 
The size spectrum has a power-law dependence on $m$ and we can identify $\eta=\gamma+\delta$, corresponding to Eq. \ref{eq:contoserva}. Furthermore, $s(m|A)$ displays a cutoff at $m\propto A^\lambda$.

\noindent vi) \textit{Scaling of the maximum body mass.}
The maximum body mass observed in an ecosystem is $m_{\max}$ such that $S\int_{m_{\max}}^\infty P(m|A)dm=1$, that is, the maximum mass extracted in $S$ samples drawn from $P(m|A)$ (see also the discussion in the SI section 1.9). Substituting $S\propto A^z$ we find $\int_{m_{\max}}^\infty x^{-\delta}  dx\propto A^{-z}$, leading to:
\begin{equation}
\label{eq:maxmass}
 m_{\max}\propto A^{\frac{z}{\delta-1}},
\end{equation}
which implies $z=\xi(\delta-1)$, i.e. Eq. \ref{xi}.

\noindent vii) \textit{Taylor's law.} Its exponent is given by:
\begin{equation}
\beta=\frac{\log \langle n^2 \rangle_m}{\log \langle n \rangle_m}=2+O\left(\frac{1}{\log (A)}\right).
\label{TLexp}
\end{equation}
In the large area limit $\beta=2$ which is the value typically found empirically \cite{Giometto2015}. Note that this computation of Taylor's law corresponds to the so-called `spatial Taylor's law' and not to its temporal counterpart \cite{Giometto2015}, in which case empirical estimates typically report values of $\beta \in [1,2]$. Deviations from $\beta=2$ may arise from the logarithmic correction in Eq. \ref{TLexp} and from the fact that the scaling of the variance (which is the second cumulant) and the second moment may differ \cite{Giometto2015}.

\noindent viii) \textit{Relative Species Abundance.} It is the distribution of species' abundances:
\begin{equation}
P(n|A)=\int_1^\infty  \ P(n,m|A) dm.
\end{equation}
There has been much interest in its analytical form. In our theoretical framework, $P_{RSA}$ cannot be computed in the general case where the exact form of $h$ and $G$ is unknown.  SI (section 1.7) reports an approximate analytical computation for a particular choice of the two functions satisfying the required properties, yielding a RSA with a tail well approximated by a lognormal.

\subsection*{\normalsize Data analysis}
\label{sec:data}

\textbf{Equation \ref{eq:contoserva}.} We verified Eq. \ref{eq:contoserva} on censuses of Barro Colorado Island (BCI) \cite{Condit2012,Condit1998,Hubbell1999} (Fig. 2) and of the Luquillo forest \cite{Zimmerman2010} (Fig. S4). Tree diameters were converted into mass using an established allometric relationship between mass and diameter \cite{Enquist2001,Simini2010}, $m \propto d ^{8/3}$. For each species, we used the mean mass of its individuals as our estimate of the typical species' mass $\bar{m}$. To account for possible deviations from the power-law behavior at small and large values of $\bar{m}$ we performed a maximum-likelihood estimation (SI section 2.2.1) of $\delta$ and $\eta$ by considering only the species with mass larger than a lower cutoff and by accounting for possible finite-size effects at large $\bar{m}$ in the form of a cut-off function (SI section 2.2.1). The estimation of the exponent $\gamma$ of Damuth's law in tropical forest datasets is affected by the sampling protocol and a correction is required to avoid sampling bias (SI section 2.2.1).  In our analysis, we used the fifth, sixth and seventh censuses of BCI and the five censuses of the Luquillo forest available online in the Center for Tropical Forest Science dataset collection. All censuses satisfy the relationship \eqref{eq:contoserva} within the errors. Whereas BCI censuses appear very similar to each other (and therefore also the exponent values estimated in different censuses, see Table S3), the Luquillo forest appears to be more dynamic (we note that the forest was hit by a major hurricane between the second and the third censuses), with values of $\gamma$ decreasing in time after $1998$ (second census, see Table S4). Because the estimate of $\delta$ remains constant, our framework would predict via Eq. \ref{eq:contoserva} that $\eta$ would also decrease in time, and this is found to be true. Finally, we note that both the BCI and the Luquillo datasets reject the linking relationship $\eta=\delta$ predicted earlier by a scaling framework \cite{Banavar2007} which is not capable of reproducing Damuth's law (Fig. S2).

\textbf{Equation \ref{zeta}.} Eq. \ref{zeta} is verified within one standard error in a dataset gathering population densities of several species of lizards on 64 islands worldwide (LIZ) \cite{Novosolov2015}, with areas ranging from 10$^{-1}$ to 10$^5$ km$^2$, where $\Phi=0.78 \pm 0.08$, $z=0.17 \pm 0.01$(mean$\pm$SE, $R^2=0.46$) and $\max\{0,\lambda(1+\alpha-\eta)\}=0$ because $\alpha \le 1$ and $\eta=\delta+\gamma=1.98 \pm 0.07$ (mean$\pm$SE), see Table S5 and Fig. S8 of the SI. Details of the fitting procedures and further discussion of the results can be found in section SI 2.2.2.
 
\textbf{Equation \ref{xi}.} To test the validity of Eq. \ref{xi} we used a dataset of mammals species presence/absence data on several islands in Sunda Shelf (SSI) \cite{Okie2009}, covering more than four orders of magnitude in island areas. 
The SAR and the scaling of the maximum body mass with the area were fitted by linear least-square regression on log-transformed data, while $P(m|A)$ was fitted by maximum-likelihood \cite{Clauset2009}. Scaling exponents in this dataset are reported in \textcolor{black}{Table S6}. Eq. \ref{xi} is verified in the SSI dataset within the errors, with $z=0.23\pm 0.02$ (mean$\pm$SE, $R^2=0.93$), $\delta= 1.6 \pm 0.2$ (mean$\pm$SE) and $\xi=0.49\pm0.09$ (mean$\pm$SE, $R^2=0.76$).

\subsection*{\normalsize Stochastic models of community dynamics}
We developed several community dynamics models accounting for the constraint on resource supply rate and incorporating empirically observed allometric relationships for the dependence of vital rates on individuals' body sizes \cite{Brown2004}. In all our models, the birth and death rates at which an individual of a species of mass $m_i$ and abundance $n_i$ is born or dies are, respectively:
\begin{equation}
\label{modello}
\begin{split}
u_i&=m_i^{-\theta} n_i ,\\
v_i&=\left[v_0+(1-v_0)c\frac{\sum_j n_j m_j^{\alpha}}{\mathcal{R}}\right]m_i^{-\theta} n_i ,
\end{split} 
\end{equation}
and thus the per-capita growth rate of species $i$ is \mbox{$\frac{u_i-v_i}{n_i}=(1-v_0)\left[ 1- \frac c{\mathcal R} {\sum_j n_j m_j^{\alpha}}\right] m_i^{-\theta}$},
which is equal to zero when \mbox{$c\sum_j n_j m_j^{\alpha}=\mathcal R\propto A$}, where $\mathcal R$ is the resource supply rate. 
At the stationary state, therefore, the total rate of resource consumption of the community fluctuates around $\mathcal R$ but the ecological dynamics continues and determines species' abundances through the constraints imposed by resources and by physiological rates.
Speciation was implemented in several ways (SI section 3), in order to test the robustness of our results to changes in the models' assumptions. We investigated models where we fixed the total number of species $S \propto A^z$ (SI section 3.1) and models where $\langle S \rangle \propto A^z$ is an emergent property of the community dynamics (SI section 3.2). By performing data-collapses (Fig. 3) of $P(m|A)$, $P(n|m,A)$ and $\langle n|m,A \rangle$  calculated using model data, we verified that they all comply with Eqs. (\ref{eq:P(n,m|A)}--\ref{eq:Damuth}). We note that the scaling exponents in Eqs. (\ref{eq:contoserva},\ref{zeta}--\ref{xi}) depend on model specifications, but the scaling properties of the fundamental distribution $P(n,m|A)$ specified in Eqs. (\ref{eq:P(n,m|A)}--\ref{eq:Damuth}) always hold.

\subsection*{\normalsize Basic community dynamics model}

In this section we describe the simplest model of community dynamics that reproduces the set of empirically-observed macroecological laws reported in the main text. We shall refer to such model as the basic model. Variations of the basic model assumptions, the exploration of parameters' space and other models are discussed in the SI (section 3).

In the basic model, each species speciates with probability $w$ per unit time (i.e., species-specific speciation events are Poisson-distributed with rate $w$). At each speciation event, a species is selected at random and a random fraction of individuals from such species is assigned to a new species $j$. The mass of the new species is obtained from the mass of the parent species as $m_j = \max\{ m_0; q m_i \}$ where $q$ is extracted from a lognormal distribution with mean and variance equal to unity so that the descendant has, on average, the same mass of the parent species. The maximum in the expression for $m_j$ ensures that the bound on the minimum mass $m_0$ that a species can attain is satisfied. The mass of the parent species is left unchanged. Species' masses thus undergo a process that is a combination of a multiplicative bounded process, known to produce power-laws \cite{Solomon1996, Sornette1997}, and of the birth/death dynamics.

The number of species $S$ is set to a constant value proportional to the area: $S=10 A^z$. Although the number of species in natural ecosystems may fluctuate in time, fixing it in the basic model allows us to vary the scaling exponent $z$ to effectively account for relevant ecological and evolutionary processes not included in the model which may affect the value of $z$ in natural ecosystems (SI, section 3.4). Note that fixing the number of entities in the model (here, $S$) is a common approximation in many related fields, such as population genetics (e.g., the Wright-Fisher model \cite{Wright1931} with fixed population size $N$) and neutral and metacommunity theory \cite{Azaele2016}. To maintain $S$ constant, we imposed that each extinction event causes a speciation event. Viceversa, at each speciation event, extinction is enforced on a species selected at random with probability inversely proportional to its abundance (i.e., more abundant species are less likely to go extinct) and proportional to the power $-\theta$ of its mass, which accounts for the fact that ecological rates are faster for smaller species. A variation on this extinction rule is discussed in the SI (section 3.1.2). Models where $\langle S \rangle \propto A^z$ is an emergent random variable are discussed in the SI (section 3.2).

The total number of individuals $N=\sum_{i=1}^S n_i$ and the total biomass $M=\sum_{i=1}^S n_i m_i$ are not fixed in the basic model (nor in the other models discussed in the SI section 3), but fluctuate in time around mean values that depend on the models' parameters and, most importantly, on the ecosystem area $A$. In other words, the mean biomass and the mean total abundance are given by a balance between birth, death and speciation events, with the constraint of resource supply limitation set by the ecosystem area $A$. The model thus allows to study the scaling of the total number of individuals and the total biomass as functions of $A$. 

The distribution $P(m|A)$ exhibits power-law behavior in $m$ (\eqref{eq:P(m|A)}) (Fig. 3A). 
The size spectrum is also a power-law across several orders of magnitude (Fig. 3B). 
The curves $\langle n |m,A \rangle$  exhibit power-law behavior in $m$ and $A$  with a cutoff at large $m$ (Fig. 3C). Data collapse (Fig. 3D) shows that its functional form is the one given by Eq. \ref{eq:Damuth}. In fact, the curves $m^\gamma A^{-\Phi} \langle n |m,A \rangle$ plotted versus $m/A^{\lambda}$ collapse onto the same curve for different values of $A$.
Moreover, Fig. 3F shows that the curves $nP(n|m,A)$ versus $n/\langle n|m,A \rangle$ collapse onto the same curve for different values of $m$ and $A$, implying that Eq. \ref{eq:P(n|m,A)} holds.
The mean total biomass $\langle M \rangle$, the mean total abundance $\langle N \rangle$ and the mean maximum mass $\langle m_{\max}\rangle$ were measured for each value of $A$ as the means across sampling times and are power functions of $A$. Parameter values used to generate the simulation data reported in Fig. 3 are reported in the SI, section 3.1.1. The stochastic model was simulated via a Gillespie tau-leap algorithm with estimated midpoint technique \cite{Gillespie2001}, with time step $\tau=1$. 

Because the \textit{ansatz} for the fundamental distribution $P(n,m|A)$ given by Eqs. (\ref{eq:P(n,m|A)}--\ref{eq:Damuth}) holds, the linking relationships among exponents (Eqs. \ref{eq:contoserva},\ref{zeta}--\ref{xi}) are satisfied at steady state by the basic model and by the other models studied in the SI (section 3). The linking relationship $\eta=\delta+\gamma$ is satisfied by the mean values of the exponents, and the density scatter-plot computed counting the occurrences of the pairs $(\eta,\delta+\gamma)$ during the temporal evolution of the community dynamics model (Fig. \ref{fig:collapse}E, shown are simulation data for the largest area value) is peaked along the 1:1 line. Thus, Eq. \ref{eq:contoserva} is satisfied, on average, during the temporal evolution of the community dynamics model.

A broad range of empirical evidence (see SI \textcolor{black}{section 2}) shows that ecological patterns are compatible with the predictions of our framework, which also agrees with heuristic calculations as shown in the main text and above. Thus, we hypothesize that our scaling framework describes not only the basic community dynamics model described here and the models described in the SI (\textcolor{black}{section 3}), but more generally any ecosystem subject to the constraint of finite resource supply rate. Further discussions on the specificity of our community dynamics models and the generality of our scaling framework are provided in the SI (\textcolor{black}{sections 3.3 and 3.4}). The basic model is thus arguably the simplest of a class of models that share the same scaling properties of the fundamental distribution, which in turn imply the same covariations of ecological patterns. This is akin to the concept of universality class \cite{Stanley1999,Stanley2000}, applied to the scaling form rather than to the exponents of the joint probability distribution and of ecological scaling laws.



{
\section*{\normalsize Acknowledgements}
\noindent Funds from the Swiss National Science Foundation Projects 200021\textunderscore 157174 and P2ELP2\_168498 are gratefully acknowledged. We thank Enrico Bertuzzo, Jayanth Banavar and Sandro Azaele for discussions. Part of the data used in section 2.2.1 of the SI were provided by the BCI forest dynamics research project, founded by S.P. Hubbell and R.B. Foster and now managed by R. Condit, S. Lao, and R. Perez under the Center for Tropical Forest Science and the Smithsonian Tropical Research in Panama. Numerous organizations have provided funding for the BCI forest dynamics research project, principally the U.S. National Science Foundation, and hundreds of field workers have contributed. The remaining part of the data in section 2.2.1 of the SI were provided by the Luquillo Long-Term Ecological Research Program, supported by grants BSR-8811902, DEB 9411973, DEB 0080538, DEB 0218039, DEB 0620910, DEB 0963447 AND DEB-129764 from NSF to the Department of Environmental Science, University of Puerto Rico, and to the International Institute of Tropical Forestry, USDA Forest Service. The U.S. Forest Service (Dept. of Agriculture) and the University of Puerto Rico gave additional support. The Luquillo Forest Dynamic Plot is part of the Smithsonian Institution Forest Global Earth Observatory, a worldwide network of large, long-term forest dynamics plots. Small mammals disturbance data shown in Fig. 1d were provided by the NSF supported Niwot Ridge Long-Term Ecological Research project and the University of Colorado Mountain Research Station.
}


\end{document}



\maketitle

The Supplementary Information is organized as follows.  Section 1 gives details concerning several theoretical results which underlie the derivation of the macroecological scaling linkages and generalizations of our scaling framework. Section 2 addresses the compatibility of our framework with empirical evidence of scaling macroecological laws. Section 3 provides details and results of community dynamics models which corroborate the validity and general applicability of our scaling framework.

\section{\normalsize Mathematical scaling framework}
The choice of variables suitable to describe an ecological system is not obvious. In our case, the choice of $n$, $m$ and $A$ is the one that better fits the scope of accounting for resource constraints. Individual metabolic rate is assumed to depend on body mass as $b=c m^\alpha$, with $\alpha \le 1$.  Thus, $m$ and $n$ determine the metabolic rate of a species, and by summing over all species one obtains the total ecosystem metabolic rate $B$, i.e. its resource consumption rate. By assuming that the resources supply rate $\mathcal{R}$ is proportional to the ecosystem area $A$ and that the total resource consumption rate (total metabolism $B$) is constrained by $\mathcal R$, we have $B\propto A$ (SI section 1.1 and 1.2).
The three variables $n$, $m$ and $A$ thus allow us to enforce the constraint of finite resources supply by imposing $B\propto \mathcal R \propto A$.

\subsection{\normalsize Proportionality of resources supply rate and ecosystem area}
\label{sec:supply2}
In our scaling framework and community dynamics models, we fix the resource-supply per unit area to unity, that is $\mathcal R/A=r=1$. The constant $r$ determines the type of ecosystem described by the model (endowed with abundant or scarce resources, which could be interpreted e.g. as a tropical forest or a desert). $\mathcal R$ will also affect the total number of species $S$ via the proportionality of the total consumption rate to the resources supply rate $B=c S I_{1,\alpha}=r A$, where $c$ is the proportionality constant that appears in Kleiber's law. Thus, the number of species $S$ varies linearly with $r$: $S=r/c \cdot A/I_{1,\alpha}$. Similarly, the total biomass and total abundance vary linearly with $r$ given that $M=S I_{1,1}$ and $N=S I_{1,0}$. The power-law scaling behavior for these quantities is valid at fixed values of $r$, that is, for ecosystems of similar type, for example a set of islands, lakes or forests sharing the same climate and environmental conditions.

We expect $\mathcal R \propto A$ to hold strictly for a community composed of a single trophic level, such as plants competing for light in a forest. In a generalized case, say describing multi-trophic levels, resources (and thus the total metabolism) might scale non-linearly with the area $\mathcal R \propto A^{\kappa}$ for certain trophic levels. For example, in a two-trophic levels community, the predator resource is the prey biomass, which might scale non-linearly with $A$ (this case is discussed in section 1.8.2). However, if the prey biomass scales linearly with the area, which is most often the case (see section 1.8.2 and Eq. 14 of the main text), then $\kappa=1$.

\subsection{\normalsize Proportionality of total metabolic rate and resources supply rate}
\label{sec:supply}
In the derivation of macrecological laws, we assume that the total metabolism $B$ of a community (i.e. the total resource consumption rate) is proportional to the resource supply rate $\mathcal R$. Such assumption is motivated by the following reasoning. Assume $B\propto \mathcal{R}^{\iota}$. The total metabolism per unit resources is thus $B/\mathcal R \propto \mathcal{R}^{\iota-1}$. If $\iota=1$, $B/\mathcal R$ is constant in the limit of large $A$. If $\iota>1$, the total metabolism per unit resources would diverge in the limit of large $A$, which would not be sustainable. If $\iota<1$, conversely, the total metabolism per unit resources would tend to zero in the limit of large $A$, so that resources would be completely unexploited. We expect that, at stationarity, the total metabolism of the community will be the maximum sustainable one, which corresponds to a linear dependence of the total metabolic rate on the resources supply rate $\mathcal R$. 

\subsection{\normalsize  Ansatz for the joint probability distribution of mass and abundance}

Consider an ecosystem of area $A$, and let $p(n,m|A) dm$ be the joint probability of finding a species of abundance $n \in  \mathbb{N}$ (including $n=0$) and typical mass $m \in [m,m+dm]$. The joint probability distribution $p(n,m|A)$ differs from $P(n,m|A)$ presented in the main text, as described in this section. We assume that the minimum viable mass for an organism is $m_0>0$ which is independent of $A$, so that $p(n,m|A)$ is null for $m<m_0$. We measure mass and area in units of $m_0$ and a reference area $a_0$, so that $m$ and $A$ are dimensionless.

For a fixed ecosystem area $A$ and in the limit of infinite mass, the probability that a species has abundance $n>0$ must go to zero, i.e. $\lim_{m \to \infty} p(n>0|m,A) =0$. Because $p(n>0|m,A) = 1- p(n=0|m,A)$, one has $\lim_{m\to\infty} p(n=0|m,A) = 1$, which implies that for large $m$ there is a finite probability that $n=0$, i.e. $p(n=0|m,A)>0$. Therefore, when approximating $p(n,m|A)$ with a continuous (in both $n$ and $m$) probability density\footnote{i.e., for $n>0$, $p(n,m|A) dn dm$ is the probability that a species has abundance in $(n,n+dn)$ and mass in $(m,m+dm)$.}, care must be taken to separate the value $n=0$ from $n>0$. Such procedure is analogous to the separation of the lowest energy state from the excited ones in the computation of the energy distribution of a Bose-Einstein condensate \cite{Huang1987}. For example, the normalization condition for $p(n,m|A)$ reads:
\begin{equation}
1 = \int_1^\infty p(0,m|A) dm + \int_1^\infty \int_0^\infty p(n,m|A) dn dm,
\end{equation}
where we evaluated the contribution of $n=0$ separately. We indicate with $p(m|A)$ and $p(n|m,A)$ the distributions derived from $p(n,m|A)$ via the identity $p(n,m|A)=p(m|A)p(n|m,A)$.
When estimating the joint probability distribution of mass and abundance in real datasets or in simulations of community dynamics models, one only observes species that exist, i.e. one evaluates probability distributions conditional on $n>0$. Thus, we indicate with $P(n,m|A) = p(n,m|A,n>0)$ the joint probability distribution that one can estimate in real datasets or in community dynamics models simulations. Correspondingly, we indicate with $P(m|A)=p(m|A,n>0)$ and $P(n|m,A)=p(n|m,A,n>0)$ the distributions derived from $P(n,m|A)$.
The relationship between $P(m|A)$ and $p(m|A)$ can be derived using Bayes' theorem as follows:
\begin{equation}
\begin{split}
P(m|A)&=p(m|A,n>0)=\frac{p(m|A)p(n>0|m,A)}{p(n>0|A)}=p(m|A)\frac{1-p(n=0|m,A)}{1-p(n=0|A)},
\label{Bayes1}
\end{split}
\end{equation}
which can be recast as:
\begin{equation}
p(m|A)=p(n=0,m|A)+\left[ 1-p(n=0|A) \right] P(m|A),
\end{equation}
where we have used the fact that $p(n=0|m,A)=p(n=0,m|A)/p(m|A)$ to express $p(n=0|m,A)$ in terms of $p(n=0,m|A)$. $P(n|m,A)$, in turn, is $p(n|m,A)$ renormalized for $n>0$, that is:
\begin{equation}
P(n|m,A)=\frac{p(n|m,A)}{\sum_{n>0} p(n|m,A)}=\frac{p(n|m,A)}{1-p(n=0|m,A)}.
\label{Bayes2}
\end{equation}
Eqs. \ref{Bayes1} and \ref{Bayes2} allow us to express $p(n,m|A)$ in terms of the experimentally or numerically observable distributions $P(m|A)$ and $P(n|m,A)$ as:
\begin{equation}
\begin{split}
p(n,m|A)=& p(m|A)p(n|m,A)\\
=&  \left[1-p(n=0|A)\right] P(m|A) P(n|m,A)\\
 =& \left[1-p(n=0|A)\right] P(n,m|A),
\label{jointBayes}
\end{split}
\end{equation}
which holds for all $n>0$. Note that $P(n,m|A)$ is simply $p(n,m|A)$ renormalized for $n>0$. Moments of $P(n,m|A)$ are computed as:
\begin{equation}
I_{j,k} = \int_1^\infty \int_0^\infty n^j  m^k P(n,m|A) dn dm,
\end{equation}
whereas moments of $p(n,m|A)$ are computed as:
\begin{equation}
\begin{split}
i_{j,k} = & 0^j \int_1^\infty m^k p(0,m|A) dm + \int_1^\infty \int_0^\infty n^j  m^k p(n,m|A) dn dm \\
 =&  0^j \int_1^\infty m^k p(0,m|A) dm + \left[ 1-p(n=0|A) \right] \\
& \times \int_1^\infty \int_0^\infty n^j  m^k P(n,m|A) dn dm \\
 = &  0^j \int_1^\infty m^k p(0,m|A) dm + \left[ 1-p(n=0|A) \right] I_{j,k},
\end{split}
\end{equation}
and therefore for $j>0$ one has: $i_{j,k} / I_{j,k}=\left[ 1-p(n=0|A) \right]$. Because $1-p(n=0|A)$ is limited, $i_{j,k}$ and $I_{j,k}$ have the same scaling with $A$.

Supported by the characterization of the stationary state of our stochastic community dynamics models (Section 3) and by the empirical scaling behavior of ecological patterns (Section 2), we put forward  an \textit{ansatz} for the analytical form of the joint probability distribution of species  abundances and masses $P(n,m|A)$.
Thus, for $P(m|A)$, $P(n|m,A)$ and $\langle n|m,A \rangle$ we assume:
\begin{align}
\label{eq:ansatz}
P(m|A)&=(\delta-1) m^{-\delta} \\
P(n|m,A)&= \hat{G} \left( \frac{n}{\langle n|m,A \rangle} \right) g(m,A),
\end{align}
where:
\begin{equation}
\label{eq:meannma}
\langle n|m,A \rangle = m^{-\gamma} A^{\Phi} h \left( \frac{m}{A^{\lambda}} \right).
\end{equation}
The term $g(m,A)$ allows further dependencies on $A$ and $m$ required for normalization and is characterized in the next section. The functions $\hat G(x)$ and $h(x)$ have the properties:
\begin{subequations} 
\begin{align}
\int_0^\infty&  \ x^j \hat{G}(x) dx<\infty & j =0,1,2 \label{propG}\\
h(x) &= o\left( x^{-2+\delta+\gamma} \right) & \mbox{as } x \to \infty \label{proph1}\\
\lim_{x\rightarrow 0} & h(x)= h_0, \label{proph2}
\end{align}
\end{subequations}
where $h_0$ is a positive constant. 
The rate of the decay of  $h$ and $\hat{G}$ for large arguments is such as to allow convergence of the $(j,k)$-th moment of $P(n,m|A)$ for $j =0,1,2$ and $k\in [0,1]$. These are all the moments needed to derive ecological scaling laws (see Methods). An example of functions satisfying the above requests is $h(x)=\hat G(x)=e^{-x}$. 

\subsection{\normalsize Normalization of $P(n,m|A)$}
\label{normalization}
Here we derive the normalization condition for $P(n,m|A)$. 
The marginal distribution $P(m|A)$ given in \eqref{eq:ansatz} is already normalized, as $\int_1^{\infty} dm ({\delta-1}){m^{-\delta}}=1$. It remains to impose normalization on $P(n|m,A)$:
\begin{equation}
\begin{split}
\label{normG}
1&=\int_0^\infty P(n|m,A)dn =g(m,A) \int_0^{\infty}  \hat{G}\left( \frac{n}{\langle n|m,A \rangle}  \right) dn= g(m,A) \langle n|m,A \rangle \int_0^\infty \hat{G}(x) dx.
\end{split}
\end{equation}
Thus, Eq. \ref{normG} reads:
\begin{equation}
1=g(m,A) \langle n|m,A \rangle,
\end{equation}
where we have imposed $\int_0^\infty \hat G(x) dx = 1$ without loss of generality. Therefore:
\begin{equation}
P(n|m,A)=\frac{1}{\langle n|m,A \rangle} \hat{G}\left( \frac{n}{\langle n|m,A \rangle} \right)=n^{-1}G\left(\frac{n}{\langle n|m,A \rangle}\right),
\label{pn}
\end{equation}
where we defined $G(x) = x\hat{G}(x)$.
In conclusion, the normalized joint probability density distribution $P(n,m|A)$ reads:
\begin{equation}
P(n,m|A)= (\delta-1) n^{-1} m^{-\delta} G\left[ \frac{n m^{\gamma}}{A^{\Phi}} \frac{1}{h(m/A^{\lambda}) } \right].
\label{pnm}
\end{equation}
This result motivates the choice of the factor $n^{-1}$ in Eq. 2 of the main text. 
Note that \eqref{pnm} can be recast as:
\begin{equation}
P(n,m|A)= (\delta-1)n^{-1} m^{-\delta} G \left[ \frac n{A^\Psi} \frac{(m/A^\lambda)^\gamma}{h(m/A^\lambda)} \right],
\label{pnm2}
\end{equation}
with $\Psi=\phi-\gamma \lambda$. For scaling to hold, \eqref{pnm2} must be valid for $x=n/A^\Psi$, $y=m/A^\lambda$ fixed and $n,m,A \to \infty$. Therefore, one has $\Psi=\phi-\gamma \lambda>0$.

\subsection{\normalsize Average abundance conditional on mass and area $\langle n|m,A \rangle$}
\label{averageNm}
We show here that the distributions:
\begin{equation}
\begin{aligned}
P(m|A)&=(\delta-1)m^{-\delta}\\
P(n|m,A)&=n^{-1}G\left[ \frac{n m^\gamma}{ A^\Phi} \frac{1}{h(m/A^\lambda)} \right]
\end{aligned} 
\end{equation}
lead to the average abundance conditional on mass and area $\langle n|m,A \rangle = A^\Phi m^{-\gamma }h(m/A^\lambda)$. In fact:
\begin{equation}
\begin{aligned}
\langle n | m,A \rangle &=\int_0^\infty  n P(n|m,A) dn = \int_0^\infty  G\left[ \frac{n m^\gamma}{ A^\Phi} \frac{1}{h(m/A^\lambda)} \right] dn \\
&=A^\Phi m^{-\gamma} h\left( \frac{m}{A^\lambda}\right) \int_0^\infty  G(x) dx = A^\Phi m^{-\gamma} h\left( \frac{m}{A^\lambda}\right),
\end{aligned} 
\end{equation}
where we have redefined without loss of generality $G(x)$ as $G'(x)=G(C x)$, where $C=\int_0^\infty G(x) dx$, so that $\int_0^\infty G'(x) dx = \int_0^\infty x^{-1} G'(x) dx=1$. We will drop the prime symbol in the remainder of the text to simplify the notation.

\subsection{\normalsize  Moments of $P(n,m|A)$}
\label{sec:derivation}
One can compute the scaling of the $(j,k)$th-moment ($j=1,2$, $k\in[0,1]$) with $A$, for large $A$, as:
\begin{equation}
\begin{split}
I_{j,k}&=\int_1^\infty \int_0^\infty n^{j} m^{k} P(n,m|A)\  dn\  dm \\
&= (\delta-1)\int_1^\infty  m^{k-\delta}\int_0^\infty  n^{j-1}  G\left[ \frac{n m^{\gamma}}{A^{\Phi}} \frac1{h\left( {m}/{A^{\lambda}} \right)}  \right] dn\  dm\\
&= (\delta-1)A^{j\Phi}\int_1^\infty  m^{k-\delta-j\gamma}h^j \left( \frac{m}{A^{\lambda}} \right) dm \int_0^\infty  x^{j-1}  G\left(x \right)dx \\
&\propto A^{j\Phi}\int_1^\infty  m^{k-\delta-j\gamma} h^j \left( \frac{m}{A^{\lambda}} \right) dm\\
&\propto A^{j\Phi +\lambda(1+k-\delta- j \gamma) }\int_{1/A^\lambda}^\infty y^{k-\delta-j\gamma} h^j(y) dy \\
&  \begin{multlined} \propto A^{j\Phi +\lambda(1+k-\delta- j \gamma) } \left[ h_0^j \int_{1/A^\lambda}^\epsilon y^{k-\delta-j\gamma} dy  \right. \\
                      \left. + \int_{\epsilon}^\infty y^{k-\delta-j\gamma} h^j(y) dy\right] \end{multlined}\\
&\propto A^{j\Phi +\lambda(1+k-\delta- j \gamma) } \left[ c_1 A^{-\lambda(1+k-\delta- j \gamma)}+ c_2 \right]\\
&\propto A^{j\Phi+\max \{ 0,\lambda(1+k-\delta- j \gamma) \}},
\end{split}
\label{moments}
\end{equation}
with $\epsilon \ll 1$, $c_1$ and $c_2$ constants. We used the property \eqref{propG} to ensure that the integral $\int_0^\infty  x^{j-1}  G\left(x \right) dx$ converges for $j=1,2$, and properties (\ref{proph1}-\ref{proph2}) to evaluate the integral in $m$ (in particular, $h(x)\simeq h_0$ constant for $x\in(0,\epsilon]$). \eqref{moments} can be used to derive several macroecological scaling laws, as outlined in the Methods section.

\subsection{\normalsize Relative species abundance (RSA)}
\begin{figure*}
\begin{center}
\includegraphics[width=114mm]{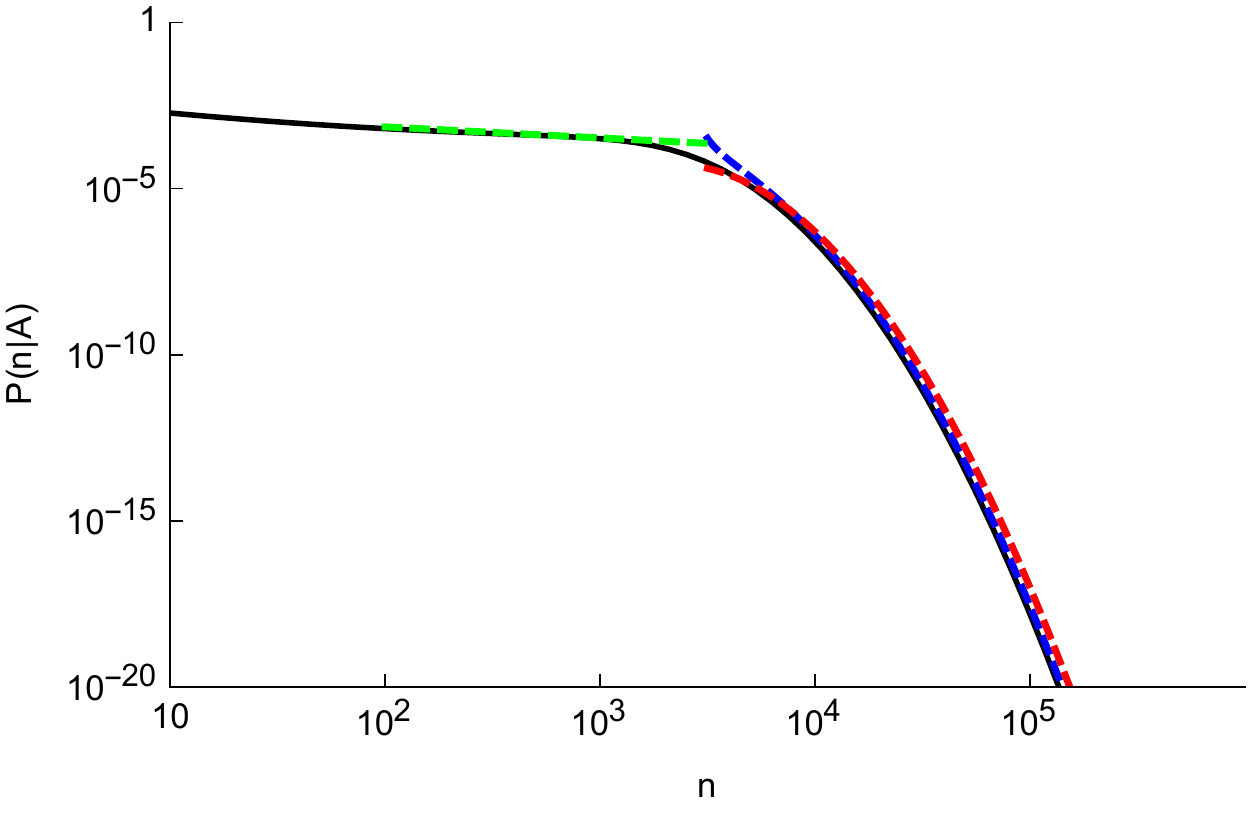}
\caption{Relative species abundance $P(n|A)$ computed numerically via \eqref{RSA} (solid black curve), with
$A = 100$, $h_0 = 100$, $\sigma = 1/10$, $\delta=3/2$, $\gamma=3/4$, $\lambda=3/4$ and $\Phi=3/4$. Shown are the approximations to the RSA computed via \eqref{tail} (dashed blue curve) and \eqref{body} (dashed green curve). The blue and red curves are plotted for $n>A^\Phi h_0 e^{-\sigma/4}$. The green curve is plotted for $100 \leq n \leq A^\Phi h_0 e^{-\sigma/4}$.
A lognormal tail is plotted for comparison (dashed red curve).}
\label{fig:RSA}
\end{center}
\end{figure*}

The RSA is the distribution of species abundances $P(n|A)$. In our framework it can be obtained by marginalizing $P(n,m|A)$ over $m$:
\begin{equation}
\begin{aligned}
P(n|A)&=\int_1^\infty \ P(n,m|A)  dm =(\delta-1) n^{-1}\int_1^\infty  \ m^{-\delta} G\left(\frac{n}{\langle n|m,A \rangle} \right)dm.
\end{aligned}
\end{equation}
This integral cannot be computed in the general case, that is, without specifying $G$ and $h$. We compute it here for the particular choice $G(x)=\frac{1}{\sqrt{\pi \sigma}} \ e^{-\frac{1}{\sigma}(\log x + \sigma/4)^2}$, with $\sigma>0$ constant. Note that $\int_0^\infty G(x) dx = \int_0^\infty G(x)/x dx= 1$ as prescribed in sections 1.4 and 1.5. For $h(x)$, it is sufficient to know that $h(x)$ is monotonically decreasing to carry out the calculations, but to simplify the expressions we take $h(x)=h_0 e^{-x}$ with $h_0>0$ constant.
With these assumptions,  $P(n|A)$ reads:
\begin{equation}
\label{RSA}
P(n|A)=\frac1n \frac{\delta-1}{\sqrt{\pi \sigma}} \int_1^\infty  \ m^{-\delta} e^{f(m)} dm,
\end{equation}
where we defined
\begin{equation}
f(m)=-\frac{1}{\sigma}\left[\log \left( \frac{nm^\gamma}{h_0 A^{\Phi}} \right) +\frac{m}{A^\lambda}+\frac \sigma4 \right]^2.
\end{equation}
The integral in \eqref{RSA} cannot be computed analytically. However, noticing that the contribution to the integral is maximum when $m=m^*$ where $m^*$ maximizes $f(m)$, we can approximate the integral for certain values of $n$. The approximation is akin to the Laplace method, but it is not possible to give an upper bound on the error made by the approximation. Nonetheless, the approximation can be compared to the numerical computation of $P(n|A)$ (see Fig. \ref{fig:RSA}).
The derivative of $f(m)$ reads:
\begin{equation}
\label{f'm=0}
f'(m)=-\frac{2}{\sigma }\left[\log \left( \frac{nm^\gamma}{h_0A^\Phi} \right)+\frac{m}{A^\lambda} +\frac \sigma4 \right] \left(\frac{\gamma}{m}+\frac{1}{A^\lambda} \right).
\end{equation}
Note that the derivative $f'(m)$ is negative for any $m\in [1, \infty]$ if $n > A^\Phi h_0 e^{-1/A^{\lambda}} e^{-\sigma/4} \simeq A^\Phi h_0 e^{-\sigma/4}$. Thus, for $n\gg A^\Phi h_0 e^{-\sigma/4}$ (i.e. in the tail of the distribution) $f(m)$ is maximum at $m^*_{\small \mbox{tail}}= 1$ and the approximation gives:
\begin{equation}
\begin{aligned}
P(n|A)_{\small \mbox{tail}}&=\frac1n \frac{\delta-1}{\sqrt{\pi \sigma}}  \frac{e^{-\frac{1}{\sigma} \left[ \log \left( \frac{n}{h_0A^\Phi} \right)+\frac{1}{A^\lambda} +\frac \sigma4  \right]^2}}{\frac{2}{\sigma}\left[\log \left( \frac{n}{h_0A^\Phi} \right)+\frac{1}{A^\lambda} +\frac \sigma4 \right] \left( \gamma+\frac{1}{A^{\lambda}}\right)} \simeq \frac1n \frac{\delta-1}{\sqrt{\pi \sigma}}  \frac{e^{-\frac{1}{\sigma} \left[ \log \left( \frac{n}{h_0 A^\Phi} \right) +\frac \sigma4 \right]^2}}{\frac{2\gamma}{\sigma  } \left[ \log \left( \frac{n}{h_0 A^\Phi} \right) + \frac \sigma 4 \right]}.
\label{tail}
\end{aligned}
\end{equation}
Note that the tail of the RSA resembles that of a lognormal distribution, which is typically found empirically \cite{Azaele2016}, plus a correction of the form $C_1+C_2 \log n$ at the denominator, where $C_1$ depends on $A$. For comparison, we plotted in Fig. \ref{fig:RSA} the (rescaled) lognormal tail $\frac1n \frac{\delta-1}{\sqrt{\pi \sigma}} \frac{\sigma}{2 \gamma} e^{-\frac{1}{\sigma}\left[ \log \left( \frac{n}{h_0 A^\Phi} \right) + \frac \sigma4 \right]^2}$.

If $n < A^{\Phi} h_0 e^{-1/A^{\lambda}} e^{-\sigma/4} \simeq A^{\Phi} h_0 e^{-\sigma/4}$, the maximum of $f(m)$ occurs at a value $\hat m>1$. However, one cannot solve
$f'(m) = 0$ analytically to determine $\hat m$. We can approximate the RSA at small and intermediate values of $n$ as follows. The behavior of $P(n|A)$ can be characterized for $n\simeq A^{\Phi} h_0 e^{-\sigma/4}$ by recognizing that, at such values of $n$, the value $m^*_{\small \mbox{body}}$ maximizing $f(m)$ is close to 1. Therefore, we approximate  $\log \left( \frac{n m^\gamma}{h_0 A^\Phi} e^{m/A^\lambda}\right)\simeq \log \left( \frac{n m^\gamma}{h_0 A^\Phi} e^{1/A^\lambda}\right) $ in $f'(m) = 0$ (see Eq. \ref{f'm=0}) and solve for $m$, yielding $m^*_{\small \mbox{body}}\simeq\left[ \frac{A^\Phi h_0}{n} e^{-1/{A^{\lambda}}} e^{-\sigma/4}\right] ^{\frac{1}{\gamma}}$.
By applying the approximation method, one finds that the approximation for the RSA for $n \simeq A^{\Phi} h_0 e^{-\sigma/4}$ is:
\begin{equation}
\label{body}
P(n|A)_{\small \mbox{body}} = \frac{\delta-1}{\gamma} \left[ A^{\Phi}h_0 e^{-\frac{\sigma}{4}} \right]^{\frac{1-\delta}{\gamma}}  n^{-1-\frac{1-\delta}{\gamma}},
\end{equation}
which is a power law with exponent $-1-\frac{1-\delta}{\gamma}$. The extents of the tail and body of the RSA distribution depend on the values of $h_0$ and $\sigma$. Fig. \ref{fig:RSA} shows the RSA computed numerically via \eqref{RSA} (black curve) and its approximations computed via equations \eqref{tail} (blue curve) and  \eqref{body} (green curve). 

\subsection{\normalsize  Generalizations of the scaling framework}
\noindent \textbf{\\1.8.1 Consequences of curvature in Kleiber's law}\\
Various studies have claimed that Kleiber's law, the relationship linking metabolic rates to body size, displays curvature in a log-log plot \cite{Kolokotrones2010,Mori2010,Maranon2013,Banavar2014a}, implying departures from a power-law behavior. For example, respiration rates of mammals have been claimed to increase with body size as $m^{2/3}$ until $m_1 \simeq 400$ g and as $m^{3/4}$ above \cite{Banavar2014a}. Conversely, respiration rates of trees have been claimed to increase linearly with tree biomass $m$ until $m_1 \simeq 40$ g and as $m^{3/4}$ above \cite{Mori2010,Banavar2014a}. Our framework can be used to infer the implications of such curvature on other macroecological patterns, as we show here.

In order to perform analytical calculations, we will make the simplifying assumption that Kleiber's law is described by the piecewise power-law:
\begin{equation}
b(m) = \begin{cases} c_0 m^{\alpha_1}	\mbox{ if $m<m_1$}\\
c m^{\alpha}	\mbox{ if $m\geq m_1$}
\end{cases}
\label{Kleiber_piecewise}
\end{equation}
where $c_0=c \ m_1^{\alpha-\alpha_1}$ is such that $b(m)$ is continuous in $m=m_1$. Using \eqref{Kleiber_piecewise} we can compute the total community consumption rate per species:
\begin{equation}
\begin{aligned}
B/S = & \int_1^\infty \int_0^\infty n b(m) p(n,m|A) dn dm \\
    = & (\delta-1)\int_1^\infty  \int_0^\infty  b(m) m^{-\delta} G \left[ \frac{n m^\gamma}{A^\Phi} \frac 1{h\left(                         m/{A^\lambda} \right)} \right] dn dm \\
    \propto &  A^\Phi \int_1^\infty  b(m) m^{-\gamma-\delta} h\left( \frac m{A^\lambda} \right) dm\int_0^\infty  G(x)  dx\\
     \propto & A^\Phi \int_1^{m_1}  c_0 m^{\alpha_1-\gamma-\delta} h\left( \frac m{A^\lambda} \right) dm\\
     &+ A^\Phi \int_{m_1}^\infty  c m^{\alpha-\gamma-\delta} h\left( \frac m{A^\lambda} \right)dm\\
     \propto & A^{\Phi + \lambda \left(1+ \alpha_1 - \gamma - \delta \right)} \int_{A^{-\lambda}}^{m_1 A^{-\lambda}} c_0 x^{\alpha_1-\gamma-\delta} h(x) dx \\
     &+  A^{\Phi + \lambda \left(1+ \alpha - \gamma - \delta \right)} \int_{m_1 A^{-\lambda}}^{\infty} c x^{\alpha-\gamma-\delta} h(x) dx.
\end{aligned}
\end{equation}
We now consider separately two possible scenarios: 
\begin{itemize}
\item[a)] $A \ll m_1^{1/\lambda}$:
\begin{equation}
\begin{split}
B/S \propto & \begin{multlined}[t] \ c_0 A^{\Phi+ \lambda \left(1+ \alpha_1 - \gamma - \delta \right)}  \left[ \int_{A^{-\lambda}}^{\epsilon} x^{\alpha_1-\gamma-\delta} h(x) dx \right. \\
\left. + \int_\epsilon^{m_1 A^{-\lambda}}  x^{\alpha_1-\gamma-\delta} h(x) dx \right] \end{multlined} \\ 
+& c A^{\Phi + \lambda \left( 1+\alpha-\gamma-\delta \right]} \int_{m_1 A^{-\lambda}}^\infty  x^{\alpha-\gamma -\delta} h(x) dx  \\
 \propto & \begin{multlined}[t] \ c_0 A^{\Phi+\lambda  \left( 1+\alpha_1-\gamma-\delta \right)} \left[c_3 + c_4 A^{-\lambda \left(1+ \alpha_1 - \gamma - \delta \right)} \right.\\
 \left. +  \int_\epsilon^{m_1 A^{-\lambda}}  x^{\alpha_1-\gamma-\delta} h(x) dx \right] \end{multlined} \\ 
 +& c A^{\Phi+\lambda \left( 1+\alpha-\gamma-\delta \right)} \int_{m_1 A^{-\lambda}}^\infty  x^{\alpha-\gamma-\delta} h(x) dx\\
\propto & \ A^{\Phi+\max \{0,\lambda \left( 1+\alpha_1 -\delta - \gamma \right) \}},
\end{split}
\end{equation}
where $\epsilon \ll 1$, $c_3$ and $c_4$ are constants. In the first line we have used the limiting behavior $\lim_{x\to0} h(x)=h_0$ constant and in the last line we have taken the limit $m_1 \to \infty$ before evaluating the integral.
\item[b)] $A \gg m_1^{1/\lambda}$:
\begin{equation}
\begin{split}
B/S  \propto & c_0 A^{\Phi+ \lambda \left(1+ \alpha_1 - \gamma - \delta \right)}  \int_{A^{-\lambda}}^{m_1 A^{-\lambda}} x^{\alpha_1-\gamma-\delta} h(x) dx  \\ 
&+ c A^{\Phi+\lambda \left( 1+\alpha-\gamma-\delta \right)} \Bigg( \int_{m_1 A^{-\lambda}}^\epsilon x^{\alpha-\gamma-\delta} h(x) dx \\& \ +  \int_{\epsilon}^\infty x^{\alpha-\gamma-\delta} h(x) dx \Bigg)\\
\propto & c_0 c_4 A^\Phi \left( 1-m_1^{1+\alpha_1-\gamma-\delta} \right) \\
&+ c A^{\Phi+\lambda \left( 1+\alpha-\gamma-\delta \right)} \left[ c_5 + c_6 m_1^{1+\alpha-\gamma-\delta} A^{-\lambda \left( \gamma+\delta-1-\alpha \right)}  \right]\\
\propto & A^{\Phi+\max\{ 0,\lambda(1+\alpha-\gamma-\delta) \}},
\end{split}
\end{equation}
where $\epsilon \ll 1$, $c_5$ and $c_6$ are constants and we have used the properties of $h$.
\end{itemize}

Thus, in the limit of large area (i.e. above the crossover value $m_1^{1/\lambda}$) the scaling of $B/S$ with $A$ is independent on $\alpha_1$, whereas below such crossover value $B/S$ scales as $A^{\Phi+\max\{ 0,\lambda(1+\alpha_1-\gamma-\delta) \}}$. 
Thus, the curvature in the relationship between individual metabolic rates and body mass translates into a curvature in the scaling of the specific community consumption rate $B/S$ with $A$. The scaling of the total number of species $S$ with $A$ is in turn determined by the proportionality of the total consumption rate to the ecosystem area, $B \propto A$. Imposing such proportionality we find the following scaling of $S$ with $A$:
\begin{itemize}
\item[a)] $A \ll m_1^{1/\lambda}$:
$
S \propto A^{1-\Phi-\max \{0,\lambda \left( 1+\alpha_1 -\delta - \gamma \right) \}}.
$
\item[b)] $A \gg m_1^{1/\lambda}$:
$
S \propto A^{1-\Phi-\max\{ 0,\lambda(1+\alpha-\delta-\gamma) \}}.
$
\end{itemize}
These equations are generalizations of the linking relationship Eq. 6 of the main text and show that the curvature in the relationship between individual metabolic rates and body mass causes a curvature in the scaling of $S$ with $A$. The difference in scaling exponents between the two regimes at small and large areas (i.e. $A \ll m_1^{1/\lambda}$ and $A \gg m_1^{1/\lambda}$) is $\Delta z=\max \{0,\lambda \left( 1+\alpha_1 -\delta - \gamma \right) \} - \max\{ 0,\lambda(1+\alpha-\gamma-\delta) \}$. Recalling the values of $\alpha_1$ and $\alpha$ for mammals reported at the beginning of this section one finds $\Delta z=\max \{0,\lambda \left( 5/3  -\delta - \gamma \right) \} - \max\{ 0,\lambda(7/4-\gamma-\delta) \}$, which if $\delta+\gamma<5/3$ and $\lambda >0$ is equal to: $\Delta z = -\lambda/12$, i.e. a concave curvature. In general, if $\lambda(1+\alpha_1-\delta-\gamma )> 0$ and $\lambda(1+\alpha-\delta-\gamma )> 0$, a convex curvature in Kleiber's law implies a concave curvature in the species-area relationship, and viceversa. Note that for tropical forests, where we find $\lambda=0$ (see Methods), a curvature in Kleiber's law does not imply any curvature in the species area relationship. 

Of course, the curvature in Kleiber's law is likely to be described by a smoother functional relationship than the piecewise-power law assumed in \eqref{Kleiber_piecewise}. However, such assumption should not affect the asymptotic estimates for small and large areas, that is, the scaling of $S$ with $A$ for $A \ll m_1^{1/\lambda}$ and $A \gg m_1^{1/\lambda}$.

\textcolor{black}{We note here that a recent investigation by Chisholm \textit{et al.} \cite{Chisholm2016} has highlighted curvatures in species-area relationships that are qualitatively similar to the one we described here, arguing that the origin of such curvature relies in a transition from a niche-structured regime on small islands to a colonization-extinction balance regime on large ones \cite{Chisholm2016}. Despite the qualitative similarities between the predicted curvatures in the two studies, we believe that further research is required to understand whether the empirical curvatures reported in Chisholm \textit{et al.} \cite{Chisholm2016} can be ascribed to a curvature in Kleiber's law as reported in this section. Because our results hold strictly in the limit of large areas, however, we believe that the `small-island effect' described in Chisholm \textit{et al.} \cite{Chisholm2016} is a fine detail that cannot be reconciled with our framework in its present form. We do not believe that this observation should call the general framework into question, but rather it highlights how our results should be applied to large ecosystems where scaling arguments can be informative. May we note that our framework deals with several macroecological patterns at once and not exclusively with the species-area relationship, and one must trade the level of detail with which each pattern is reproduced with the capability to derive the covariations of several macroecological patterns analytically. Finer details of some of these patterns may be addressed by including additional ecological processes in our approach, although we deem that such path is beyond the scopes of this investigation, which aims at establishing a general null model framework.}\\

\noindent \textbf{1.8.2 Multiple trophic levels}
\label{multitrophic}

So far we have assumed that all individuals in the ecosystem share the same resources. Thus, our results hold strictly for organisms within the same trophic level. Power law distributions and relationships, however, are empirically observed even on multi-trophic level contexts. For example, power-law size spectra across trophic levels are routinely observed in marine microbial ecosystems over several orders of magnitude \cite{Sheldon1972}. Here, we show that the conceptual design of our framework can be applied to multi-trophic level systems by means of a simple example. We assume that species' interactions are described by a simple foodweb made of two trophic levels, with one level feeding on abiotic factors like light (i.e. the producers) and the latter feeding on the lower trophic level (i.e. the consumers). In such an ecosystem, producers would be described by our framework with resources limited by ecosystem area ${\mathcal R} \propto A$, while consumers would be described by the same framework where the limiting resource is now the total producers' biomass, which scales as $A^{\mu_p}$ (the subscript $p$ identifies the producers). Thus, one can describe the two trophic levels separately, with the producers' level being described by our current framework and the consumers level being described by a similar scaling framework where resources scale as ${\mathcal R}\propto A^{\mu_p}$. The effect of such modification on scaling relationship is easy to compute and affects the species-area relationship via a modification of \eqref{second}:
\begin{equation}
z_c=\mu_p-\Phi_c-\max\{0,\lambda_c(1+\alpha_c-\eta_c)\},
\end{equation}
which differs from \eqref{second} only for the exchange of $1$ with $\mu_p$ (the subscript $c$ identifies the consumers). Such modification propagates to equations \eqref{third} and \eqref{fourth} (recall that $N=S I_{1,0}$ and $M=S I_{1,1}$) which become:
\begin{equation}
\begin{aligned}
&\mu_c=\mu_p+\max \{ 0,\lambda_c(2-\eta_c) \}-\max \{ 0,\lambda_c(1+\alpha_c-\eta_c) \},\\
&\nu_c=\mu_p-\max \{ 0,\lambda_c(1+\alpha_c-\eta_c) \}.
\end{aligned}
\end{equation}

We note, however, that $\mu_p$ is most likely equal to one in most ecosystems, as found for example in our empirical analysis of tropical forests datasets. In such a case, the consumers level is described exactly by the framework that we presented in the main text. Obviously, more complicated schemes of multi-level foodwebs may be envisioned in the proposed framework. On this, research is forthcoming.\\

\noindent \textbf{1.8.3 Area-independent limitation on maximum size}
\label{physiological}
Eq. 5 of the main text predicts that, if $z>0$, the maximum species' mass increases with $A$ as $m_{\max}=A^\xi$. 
When $A$ is very large, area-independent constraints could settle in to limit the maximum body size, either due to physiological limits or due to ecological dynamics making larger body sizes unfavorable.
Because $\xi=z/(\delta-1)$, the critical value of $A$ above which the maximum body size is independent of $A$ is equal to $A_c=M_0^{z/(\delta-1)}$. This observation can be reconciled with our framework by generalizing Eq. 7 as:
\begin{equation}
\label{eq:pmcutoff}
P(m|A)=m^{-\delta}H\left( \frac m{M_0} \right),
\end{equation}
where the cutoff function $H$ is such that $\lim_{x\to0} H(x)=\mbox{const}$, $\lim_{x\to\infty} {H(x)}/{x^{{\max \{ 2-\delta-\gamma,1-\delta \}}}}=0$ and is such that $\int_1^\infty P(m|A) dm = 1$. This generalization of Eq. 7 does not affect our results for $A<A_c$. In fact, the joint probability distribution in such a generalized setting reads:
\begin{equation}
\begin{split}
\label{eq:jointgen}
P(n,m|A)&=n^{-1}m^{-\delta}H\left( \frac{m}{M_0} \right) G \left( \frac n{\langle n|m,A \rangle} \right) \\
&= n^{-1}m^{-\delta}H\left( \frac{m}{M_0} \right) G \left[ \frac {n m^\gamma}{A^\Phi h(m/A^\lambda)} \right]
\end{split}
\end{equation}
and integrals of this distribution (e.g. marginals and moments) depend on which of the two finite-size cutoffs ($h$ and $H$) sets in at the lowest value of $m$. We show this by calculating the moment $I_{j,k}$ with $j,k>0$:
\begin{equation}
\label{eq:momentgen}
\begin{aligned}
I_{j,k}&=\int_1^\infty \int_0^\infty n^{j-1} m^{k-\delta} G\left[ \frac{n m^\gamma}{A^\Phi h(m/A^\lambda)} \right] H\left( \frac{m}{M_0}\right) dn dm\\
&= A^{j \Phi} \int_1^\infty m^{k-\delta -j \gamma} h^j\left( \frac{m}{A^\lambda} \right) H\left( \frac{m}{M_0} \right) dm \int_0^\infty x^{j-1} G(x) dx,
\end{aligned}
\end{equation}
 where the properties of $G$ ensure the convergence of the integral. The scaling of $I_{j,k}$ with $A$ is thus determined by which of the two functions $h^j(m/A^\lambda)$ and $H(m/M_0)$ decays earlier in $m$. For intermediate values of $A$ (i.e. $A^\lambda<M_0$) one can replace $H(m/M_0)$ with $(\delta-1)$ in equations \eqref{eq:jointgen} and \eqref{eq:momentgen}, and is left with the framework described in the main text. For larger values of $A$ (i.e. $A > M_0^{1/\lambda}$), instead, one replaces the term $h(m/A^\lambda)$ with the constant $h_0^j$ in equations \eqref{eq:jointgen} and \eqref{eq:momentgen}. In the limit of very large area, therefore, the maximum mass $M_0$ ceases to increase with $A$, due to attained physiological or ecological constraints. Furthermore, Damuth's law displays no cutoff in $m$, i.e. $\langle n|m,A \rangle \simeq h_0 m^{-\gamma} A^\Phi$ (more precisely, the cutoff in Damuth's law would be unobservable due to the extremely low probability to observe a species with $m>M_0$), and the ecosystem is effectively described by the joint probability distribution:
\begin{equation}
\begin{split}
P(n,m|A)&=n^{-1}m^{-\delta}H\left( \frac{m}{M_0} \right) G \left( \frac n{\langle n|m,A \rangle} \right) \\
&= n^{-1}m^{-\delta}H\left( \frac{m}{M_0} \right) G \left( \frac {n m^\gamma}{h_0 A^\Phi } \right),
\label{jointphysio}
\end{split}
\end{equation}
and thus by the modified $P(n|m,A)= n^{-1} G[n m^\gamma / (h_0 A^\Phi)]$. One can show that in this limit of very large area the linking relationship (1) goes unchanged, whereas equations (6--9) are replaced by: $z=1-\Phi$, $\mu=1$, $\nu=1$ (i.e. the maxima in equations 6--8 disappear) and $\xi=0$. 
This could be the situation for the two forest datasets analyzed in section 2.2.2, where one should have $\lambda=0$ to verify the reasonable assumption $\mu=\nu=1$ (see section 2.2.2 for details).\\

\noindent \textbf{1.8.4 Intra-specific size distributions}

\textcolor{black}{One assumption of our scaling framework is that all individuals of the same species have the same mass, although in reality individuals masses within the same species are distributed according to intra-specific size distributions. The scaling properties of intra-specific size distributions have been studied by Giometto \textit{et al.} \cite{Giometto2013}, where it was shown that protist species belonging to four different phyla and covering five orders of magnitude in mass are characterized by a universal size distribution:
\begin{equation}
p(m | \bar m)=\frac 1{m} F\left(\frac {m}{\bar m} \right),
\label{intraspecific_eq_protists}
\end{equation}
where $m$ is the mass of an individual, $\bar m$ is the characteristic mass of the species and $F(x)\to0$ suitably fast for $x\to0$ and $x\to\infty$, as detailed in Giometto \textit{et al.} \cite{Giometto2013}. When intra-specific size distributions are described by \eqref{intraspecific_eq_protists}, the results of our scaling frameworks hold exactly. The moment ($j$,$k$) is computed as: 
\begin{equation}
\begin{aligned}
\label{eq:momento_intraspec_general}
I_{j,k}&=\int_0^\infty  \int_1^\infty \int_1^\infty  \ n^j m^k P(n,\bar{m}|A) p(m|\bar{m})\ dm \ d\bar{m}\ dn=\int_0^\infty  \int_1^\infty \ n^j  P(n,\bar{m}|A) \langle m^k | \bar m  \rangle\   d\bar{m}\ dn
\end{aligned}
\end{equation}
where :
\begin{equation}
\label{eq:<mk>}
\langle m^k | \bar m \rangle:=\int_1^\infty  \  m^k p(m|\bar{m}) dm.
\end{equation}
Usinq \eqref{intraspecific_eq_protists}, we obtain:
\begin{equation}
\begin{aligned}
\langle m^k | \bar m  \rangle &= \int_1^\infty  m^{k-1}F \left( \frac{m}{\bar{m}}\right) dm=\bar{m}^{k}\int_{1/\bar{m}}^\infty x^{k-1} F(x) dx \simeq \bar{m}^{k}\int_0^\infty x^{k-1} F(x)dx \propto \bar{m}^{k}
\end{aligned}
\end{equation}
where $x=m/ \bar{m}$ and $\int_0^\infty x^{k-1} F(x) dx$ is a constant. Substituting this result in \eqref{eq:momento_intraspec_general}, we have $I_{j,k}=\int_0^\infty  \int_1^\infty \ n^j \bar{m}^k P(n,\bar{m}|A)  d\bar{m}\ dn$, which corresponds to \eqref{moments}.
Thus, the moments $I_{j,k}$ computed in this generalized framework have the same scaling with the area as the ones computed by assuming that all individuals within a species have the same mass. Therefore, the linking relationships (2--5), whose derivation relies on the scaling of $I_{j,k}$ with $A$, are unchanged. Furthermore, we show that the linking relationship in Eq. 1 is also unchanged. In fact, assuming that intra-specific size distributions are given by \eqref{intraspecific_eq_protists}, the size spectrum is given by:
\begin{equation}
\begin{split}
s( m|A) &= \frac SN \int_0^\infty n \int_1^\infty P(n,\bar m|A)  m^{-1} F\left( \frac {m}{\bar m} \right) d\bar m dn\\
&= \frac SN \int_0^\infty \int_1^\infty \bar m^{-\delta}  G\left[ \frac{n \bar m^\gamma}{A^\Phi h(\bar m/A^\lambda)} \right]  m^{-1} F\left( \frac {m}{\bar m} \right) d\bar m dn\\
&=  m^{-\delta} \frac SN \int_0^\infty \int_{1/m}^\infty  x^{-\delta} G\left[ \frac{n (x m)^\gamma }{A^\Phi h( x m/A^\lambda)} \right]  F\left( \frac 1x \right) dx dn\\
&= m^{-\delta-\gamma} A^\Phi \frac SN \int_{1/m}^\infty  x^{-\delta-\gamma} h\left( \frac{xm}{A^\lambda} \right) F\left( \frac 1x \right) dx \int_0^\infty G(y) dy,
\end{split}
\label{sm_intraspecific_protists}
\end{equation}
where $x=\bar m/ m$, $y=\left[{n (x m)^\gamma }\right]/\left[{A^\Phi h( x m/A^\lambda)}\right]$ and $\int_0^\infty G(y)dy$ is a constant. We note that we cannot compute analytically the scaling of $s(m|A)$ with $m$ from \eqref{sm_intraspecific_protists}, because $m$ appears both at the lower limit of the integral in $x$ and in the argument of $h$. However, for large $m$, the lower limit of the integral in $x$ tends to $0$ and thus $s(m|A) \propto m^{-\delta-\gamma} \tilde h(m/A^\lambda)$ in the limit of large $m$, where $\tilde h (y)=\int_0^\infty x^{-\delta-\gamma} F(1/x) h(x y) dx$ has the same limiting behavior of $h(y)$ at $y\to0$ and $y\to\infty$\footnote{\textcolor{black}{
Because $\lim_{x\to\infty } h(x)=0$, $\lim_{x\to0} h(x)=h_0$, $h(x) \leq h_0$ and $\int_0^\infty x^{-\delta-\gamma}F(1/x)dx<\infty$ if $\delta+\gamma>1$, then $\lim_{y\to\infty} \tilde h(y) = 0$ and $\lim_{y \to 0} \tilde h(y)=\mbox{const}$ follows from the Lebesgue's dominated convergence theorem.}}, and thus the linking relationship $\eta=\delta+\gamma$ (Eq. 1) still holds. Thus, introducing intra-specific variability in mass according to \eqref{intraspecific_eq_protists} does not alter the linking relationships (1, 6--9).}\\

\noindent \textbf{1.8.5 Tree intraspecific size distributions}

As we discuss in section 2.2.1, trees are an exception to \eqref{intraspecific_eq_protists}, given that a single species can cover several orders of magnitude in mass. We show in section 2.2.1 that the intra-specific size distributions of the most abundant tree species in tropical forests are characterized by the finite-size scaling form:
\begin{equation}
p(m|\bar m)= m^{-\Delta} \mathcal F\left( \frac{m}{\bar{m}^\Omega} \right),
\label{intraspecific_eq_trees}
\end{equation}
where $\Omega=1/(2-\Delta)$ ensures that $\int m p(m|\bar{m}) dm = \bar{m}$, $\Delta = 1.12 \pm 0.06$ and $\mathcal F(x)$ is a scaling function with limiting behaviors $\mathcal F(x)\to \mbox{const}$ for $x\to0$ and $\mathcal F(x)\to0$ more rapidly than any power of $x$ for $x\to\infty$. The ($j$,$k$)th moment $I_{j,k}$ can still be computed exactly and some small corrections to the scaling exponents arise, compared to \eqref{moments}. In fact, using \eqref{intraspecific_eq_trees} we obtain:
\begin{equation}
\begin{aligned}
\langle m^k | \bar m \rangle &= \int_1^\infty  m^{k-\Delta} \mathcal F \left( \frac{m}{\bar{m}^\Omega}\right) dm\\& =\bar{m}^{\Omega(1+k-\Delta)}\int_{1/{\bar{m}^\Omega}}^\infty x^{k-\Delta} \mathcal F(x) dx\\ 
                    & \propto \bar{m}^{\max\{0,\Omega(1+k-\Delta)\}}.
\end{aligned}
\label{eq:intraspecific_moment}
\end{equation}
Note that, among the consequences of this result, we obtain that the variance of the intra-specific size distribution increases with the average body size: $\langle m^2 | \bar m \rangle \propto \bar{m}^{\Omega(3-\Delta)}$.
Substituting \eqref{eq:intraspecific_moment} in \eqref{eq:momento_intraspec_general} gives:
\begin{equation}
\begin{aligned}
I_{j,k}&=\int_0^\infty  \int_1^\infty \ n^j \bar{m}^{\max\{0,\Omega(1+k-\Delta)\}} P(n,\bar{m}|A)    d\bar{m}\ dn\\
       &\propto A^{j\Phi+\max \{0,\lambda(1-\delta-j\gamma+\max\{0,\Omega(1+k-\Delta)\})\}},
\end{aligned}
\label{moments_trees}
\end{equation}
where the calculations have been performed as in section 1.6. \eqref{moments_trees} differs only slightly from \eqref{moments}. As a result, a small correction applies to the linking relationship (2): the term $\alpha$ in Eq. 2 is substituted by $\frac{1+\alpha -\Delta}{2-\Delta}$, which for the empirical value $\Delta=1.12\pm0.06$ results in a very small correction $C=\alpha-\frac{1+\alpha -\Delta}{2-\Delta}=(\alpha-1)\left( \frac{1-\Delta}{2-\Delta} \right)\simeq0$ which is compatible with zero, given that $\Delta$ is compatible with one. Equations (3--5) are unchanged.   
The size spectrum is given by:
\begin{equation}
\begin{split}
s(m|A) =& \frac SN \int_0^\infty n \int_1^\infty P(n,\bar m|A) m^{-\Delta} \mathcal F\left( \frac {m}{\bar m^\Omega} \right) d\bar m dn\\
=& \frac SN \frac{1}{m^{\Delta}} \int_1^\infty  \frac{1}{\bar m^{\delta}} \mathcal F\left( \frac{m}{\bar m^\Omega} \right) \int_0^\infty  G\left[ \frac{n \bar m^\gamma}{A^\Phi h(\bar m/A^\lambda)} \right] dn \  d\bar m\\
=&\frac SN m^{-\Delta+(1-\delta)/\Omega}\int_0^m x^{(\delta-1)/\Omega-1} \mathcal F(x) \\
& \times \int_0^\infty G\left[ \frac{n (m/x)^{\gamma/\Omega}}{A^\Phi h \left[ (m/x)^{1/\Omega}/A^\lambda \right]} \right] dn \  dx\\
=&\frac SN A^\Phi m^{-\Delta+(1-\delta-\gamma)(2-\Delta)} \int_0^\infty G(y) dy  \\
& \times \int_0^m x^{(2-\Delta)(\delta+\gamma-1)-1} \mathcal F(x) h\left[ \frac{(m/x)^{1/\Omega}}{A^\lambda} \right] dx
\end{split}
\label{sm_trees}
\end{equation}
where $x=m/\bar m^\Omega$, $\Omega=1/(2-\Delta)$, \\
$y = \left[ n (m/x)^{1/\Omega} \right] / \left\{ A^\Phi h\left[ (m/x)^{1/\Omega} / A^\lambda \right] \right\}$ and $\int_0^\infty G(y)dy$ is a constant. Note that we cannot compute the scaling exponent of $s(m|A)$ with $m$ exactly, because $x$ and $m$ are found in the arguments of both $\mathcal F$ and $h$. We can, however, derive an approximation that holds for large $m$. This is most easily seen if we assume that no finite-size effect is found in Damuth's law, i.e. $h=\mbox{const}$, for which we have:
\begin{equation}
\begin{split}
s(m|A) &\propto \frac SN A^\Phi m^{(2-\Delta)(1-\delta-\gamma)-\Delta} \int_0^{m} x^{-(2-\Delta)(1-\delta-\gamma)-1}\mathcal F(x) dx,
\end{split}
\end{equation}
which, for large $m$, is a power-law ($s(m|A) \propto m^{-\eta}$) with:
\begin{equation}
\eta=\Delta - (2-\Delta)(1-\delta-\gamma),
\label{generalized_eq1}
\end{equation}
which generalizes Eq. 1 to the case of intra-specific size distributions as in \eqref{intraspecific_eq_trees}. We note, in analogy to the calculations performed in the previous section, that for large $m$ the upper limit of the integral in $x$ in \eqref{sm_trees} tends to $+\infty$ and thus $s(m)\propto m^{(2-\Delta)(1-\delta-\gamma)-\Delta}\tilde h\left(m/A^{\lambda\Omega}\right)$ in the limit of large $m$, where the function $\tilde h(y)=\int_0^\infty x^{-(-2-\Delta)(1-\delta-\gamma)}\mathcal F(x)h\left[ (y/x)^{1/\Omega}\right]dx$ has the same limiting behavior of $h(y)$ for $y\to0$ and $y\to\infty$\footnote{
Because $\lim_{x\to\infty } h(x)=0$, $\lim_{x\to0} h(x)=h_0$, $h(x) \leq h_0$ and $\int_0^\infty x^{-\delta-\gamma}F(1/x)dx<\infty$ if $\delta+\gamma>1$, then $\lim_{y\to\infty} \tilde h(y) = 0$ and $\lim_{y \to 0} \tilde h(y)=\mbox{const}$ follows from the Lebesgue's dominated convergence theorem.}.
 Therefore, the linking relationship in \eqref{generalized_eq1} holds also when $h$ is not identically constant. The correction to the linking relationship (1) is small because the difference between $\eta_1=\delta+\gamma$ computed via Eq. 1 and $\eta_2=\Delta - (2-\Delta)(1-\delta-\gamma)$ computed via \eqref{generalized_eq1} is $\eta_1-\eta_2 = (\Delta-1)(\delta+\gamma-2)$ and $\Delta = 1.12 \pm 0.06$ is compatible with one, in which case equations (1) and \eqref{intraspecific_eq_trees} coincide. Using the values reported in Table \ref{tab:bci}, we find $\eta_1-\eta_2 \simeq-0.07 \pm 0.03$, which is compatible with zero. Because the differences between the linking relationships (1) and (6) reported in the main text and the generalizations reported in this section are negligible and compatible with zero, we will neglect such generalizations in the rest of this study.\\

\noindent \textbf{1.8.6 Alternative forms of $P(m|A)$}

Our \textit{ansatz} for $P(m|A)$ is a pure power function, \eqref{eq:ansatz}. There are two possible relaxation of this hypothesis that do not compromise analytical tractability, in addition to the one already explored in section \ref{physiological}. The first is the addition of a cut-off at large masses, $P(m|A)=m^{-\delta} F_1 \Bigl( \frac{m}{A^\lambda}\Bigr)$  with $\delta>1$ (for normalization purposes), where the exponent $\lambda$ is the same as the one of the cut-off in Damuth's law (\eqref{eq:meannma}) and where $F_1(x)$ is such that $\lim_{x \to 0} F_1(x)=\mbox{const}$ and $lim_{x \to \infty} F_1(x)=0$. In this case, the computation of the moments $I_{j,k}$ is performed similarly to \eqref{eq:momentgen}, where the role of $h^j\Bigl( \frac{m}{A^\lambda}\Bigr)$ is now played by the product $F_1\Bigl( \frac{m}{A^\lambda}\Bigr) h^j\Bigl( \frac{m}{A^\lambda}\Bigr)$, behaving similarly. The final result is the same as in \eqref{eq:momentgen}. In the computation of the community size-spectrum, Eq. 16 of the main text, the cut-off is now given by the product of the two functions $F_1$ and $h$. In conclusion, this generalization does not change the linking relationships in Eq. 1, 6--8 of the main text. The generalization of Eq. 9 of main text to this case cannot be computed analytically. 
Some empirical studies \cite{Siemann1996,Labra2015} found a $P(m|A)$ similar to a log-normal. We can describe this case as $P(m|A)=m^{-\delta} F_2\Bigl( \frac{m}{A^\lambda}\Bigr)$ where $F_2(x)$ is such that $\lim_{x \to 0,\infty} x^j F_1(x)=0$  $\forall j$. Normalization requires $\delta=1$. Moments can be computed, again, as in \eqref{eq:momentgen} but noticing that in this case $\int_{1/A^\lambda}^\infty x^{k-\delta-\gamma j} h^j(x) F(x) \approx const$ in the limit of large $A$, as the expression inside the integral tends to 0 faster than any power of $x$ when $x\to 0$. Therefore, we obtain $I_{j,k}\propto A^{\Phi j +\lambda(k- \lambda j)}$. Eq. 6--8 of the main text are then replaced by: $z=1-\Phi-\lambda(\alpha-\gamma)$, $\nu=1-\lambda\alpha$ and $\mu=1+\lambda(1-\alpha)$. The community size-spectrum is computed as:
\begin{equation}
\begin{split}
s(m|A)&=\frac{S}{N}\int_0^\infty n P(n,m|A) dn \\
      &=\frac{S}{N} m^{-1} F_2\Bigl( \frac{m}{A^\lambda}\Bigr) \int_0^\infty dn G\Bigl( \frac{n m^\gamma}{A^\Phi h(m/A^\lambda)}\Bigr)\\
      &\propto m^{-1} \frac{A^\lambda}{m}^\gamma F_2\Bigl( \frac{m}{A^\lambda}\Bigr) h\Bigl( \frac{m}{A^\lambda}\Bigr)\\
      &=m^{-1} H\Bigl( \frac{m}{A^\lambda}\Bigr)
\end{split}
\end{equation}
where $H(x):=x^{-\gamma}F_2(x)h(x)$ is such that $\lim_{x \to 0,\infty} x^j H(x)=0$  $\forall j$. Therefore, $s(m|A)$ is not a power-law, but has an internal mode, similarly to $P(m|A)$. The generalization of Eq. 9 of main text to this case cannot be computed analytically.

\subsection{\normalsize  Compatibility with previous works}

Southwood \textit{et al.} \cite{Southwood2006} derived a linking relationship that is equivalent to our Eq. 5. Here, we briefly review their result and illustrate how it must be corrected due to a few miscalculations.

The authors start from the observation \cite{May1988} that the total number of species of length class $L$ scale as $S_L(L)\propto S L^{-\Delta}$, where $\Delta=3/2$ and $S$ is the total number of species in the ecosystem. The estimate $\Delta=3/2$ made in that study \cite{Southwood2006} is incorrect, because the correction needed to account for logarithmic binning \cite{Stegen2008,White2008} is missing. The correct estimate that accounts for logarithmic binning \cite{Stegen2008} is $\Delta=5/2$. First, the authors of reference 9 say that the maximum length is obtained by imposing $S_L(L_{\max})=S L_{\max}^{-\Delta}=1$, i.e. $S\propto L_{\max}^\Delta$. This calculation is also incorrect, because it makes an improper use of probability distributions. In fact, $S_L(L)$ is the fraction of species with length in $[L,L+dL]$ and the maximum species' length is found by imposing that the probability of finding a species with length larger than $L_{\max}$ be equal to $1/S$ (this is a widely employed argument to estimate the largest random number drawn from a specified distribution \cite{Redner1990,Giometto2015}) which results in $S\propto L_{\max}^{\Delta-1}$. Incidentally, these two miscalculations compensate each other and together lead to the estimate $S \propto L_{\max}^{3/2}$. However, and this is crucial to derive the correct linking relationship, the correct equation is $S \propto L_{\max}^{\Delta-1}$ and not the one suggested in reference 9, i.e. $S \propto L_{\max}^{\Delta}$. The equation relating $S$ and $L$ converts into a species-mass relationship via the scaling $L \propto m^{1/3}$, where we have assumed that body density does not scale with body mass. Specifically, one finds $S \propto m_{\max}^{\delta-1}$ with $\delta=(\Delta+2)/3$. Then, one can use the finding by Burness \textit{et al.} \cite{Burness2001}, $m_{\max}\propto A^\xi$, to derive the linking relationship between $z$, $\xi$ and $\delta$. In fact, by comparing $S \propto m_{\max}^{\delta-1} \propto A^{\xi(\delta-1)}$ with the SAR $S \propto A^z$ one obtains the linking relationship $z=\xi(\delta-1)$, which coincides with our Eq. 5. However, we note that, in reference 9, $z$ was meant as the exponent of the nested species-area relationship, differently from here. Nonetheless, the linking relationship is the same.\looseness=-1

Our results thus agree with the earlier result by Southwood \textit{et al.} \cite{Southwood2006}, where one of the linking relationship was discovered. Our investigation reveals that such linking relationship is only one component of the broader set of linking relationships, Eqs. 1,6--9 of the main text.

\begin{figure*}
\begin{center}
\includegraphics{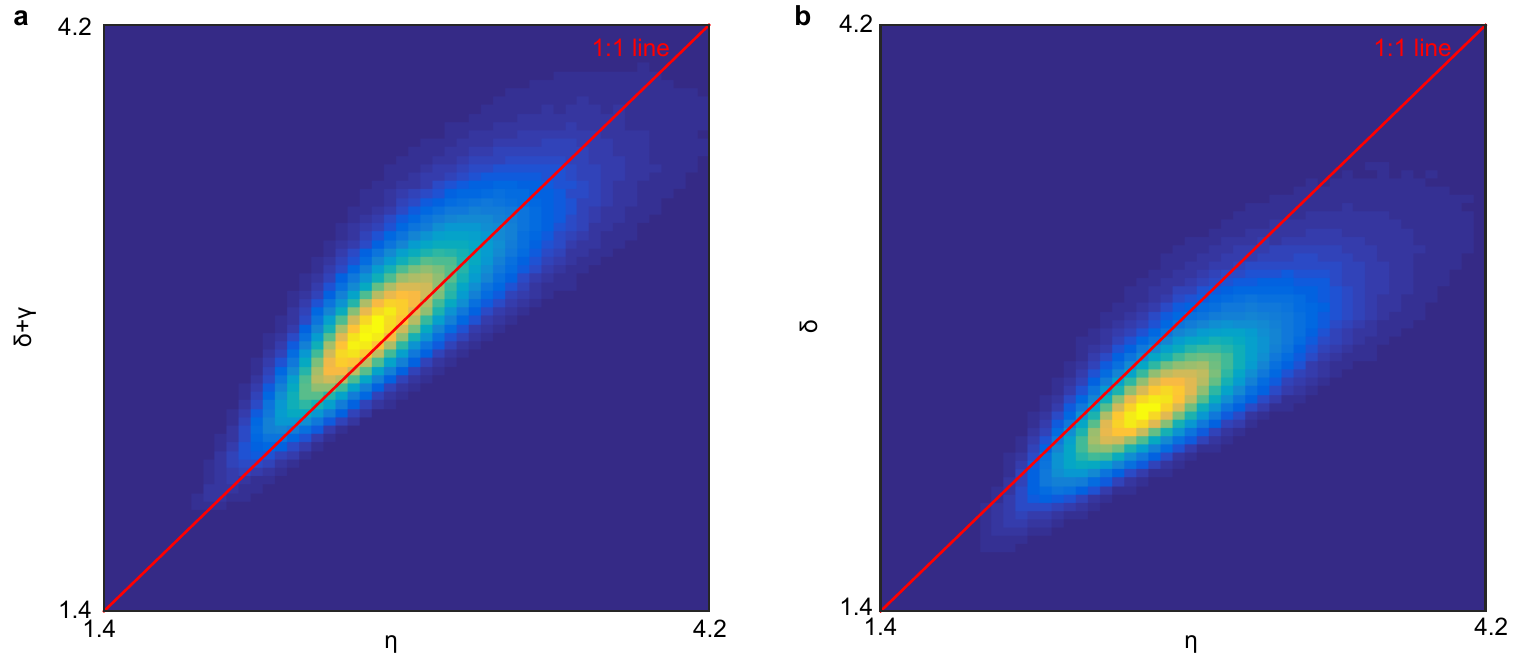}
\caption{Density-scatter plot of $\delta+\gamma$ (panel a) and $\delta$ (panel b) versus $\eta$ in simulations of the stochastic community dynamics model (model \textit{a}, see section 3.2), with the exponents estimated at each sampling time-point. The parameters of the stochastic community dynamics model are reported in Fig. \ref{fig:nonscaling_conserved_fission_no_n}; shown are simulation data for the largest simulated area $A=10^3$. Density histograms are normalized to one, with blue representing the value zero and yellow the value one. At each time-step, the exponent $\gamma$ was extimated by linear least-squares fit of ($\log (m_i)$,$\log(n_i)$) where the index $i$ runs on all the species present in the ecosystem at that time-step. Note that the small deviations from the 1:1 line are also due to statistical errors in the estimation of the exponents at each time step. Panel b) shows that the prediction \cite{Banavar2007} that $\eta = \delta$  is not supported by simulation data.}
\label{fig:scatter}
\end{center}
\end{figure*}

We finally note the differences of the current work with a previous scaling framework \cite{Banavar2007}, namely: 1) the enforcement and the implications of resource limitation, 2) the validation based on a broad class of community dynamics models, 3) the extensive empirical verification, and 4) the richer set of macroecological laws that the scaling framework accounts for, most importantly Damuth's law, which allows us to reconcile the predicted linkages with empirical data and community dynamics models (see, e.g. Fig. \ref{fig:scatter}). Such framework \cite{Banavar2007} took as starting point an equation compatible with our assumption on $P(n,m|A)$ (Eq. 2 of the main text). Specifically, Eq. 1 in Banavar \textit{et al.} \cite{Banavar2007} is more general than our Eq. 6, but one needs to specify further properties of $P(n,m|A)$ in order to recover Damuth's law, as we did here. Furthermore, the scaling properties hypothesized for the function $F$ that appears in Eq. 1 of Banavar \textit{et al.} \cite{Banavar2007} (hypothesis $\mathcal H_1$ therein) is incompatible with Damuth's law and leads to different predictions for the pattern covariations (e.g. the linking relationship $\eta=\delta$) that are falsified by empirical data and by our model simulations (Fig. \ref{fig:scatter}). Note that the introduction of a constraint on the total community consumption rate in the framework of Banavar \textit{et al.} would not affect the relationship $\eta=\delta$, which is instead a byproduct of the assumptions on $P(n,m|A)$.


\section{\normalsize Compatibility with macroecological data}
\label{data}
\subsection{\normalsize  Empirical evidence of scaling ecological laws}
Abundant empirical evidence exists of scaling ecological patterns in diverse types of ecosystems: forests, terrestrial (including mammals in particular) and aquatic ecosystems. A comparison of these empirical results highlights the non-universality of the values of scaling exponents. Fig. 1 in the main text shows evidences of Kleiber's law (panel a), Damuth's law (panel b), the Species-Area Relationship (SAR) (panel c) and the community size-spectrum $s(m)$ (panel d) for the three types of ecosystems. Regression lines in Fig. 1 are fits provided in the original papers (see legends), except for the patterns for forests and terrestrial ecosystems in panel b, which were fitted by linear least-squares fits on log-transformed data, and for the community size-spectra from BCI and Niwot Ridge datasets in panel d, which were fitted with maximum likelihood \cite{Clauset2009}. Table \ref{exponents} reports the estimates for the scaling exponents. Table \ref{citations} contains a compilation of references to empirical measures of the ecological patterns referred to in the main text.

\subsection{\normalsize  Compatibility of linking relationships and data}
\label{sec:verification}

Our framework predicts five relationships linking the scaling exponents of ecological laws (Eqs. 1, 6--9 of the main text): \looseness=-1
\begin{subequations}\label{linkingrel}
\begin{align} 
&\eta=\gamma+\delta,\label{first}\\
&z=1-\Phi-\max\{0,\lambda(1+\alpha-\eta)\},\label{second}\\
&\mu=1+\max \{ 0,\lambda(2-\eta) \}-\max \{ 0,\lambda(1+\alpha-\eta) \},\label{third}\\
&\nu=1-\max \{ 0,\lambda(1+\alpha-\eta) \},\label{fourth}\\
&\xi=\frac{z}{\delta-1}.\label{fifth}
\end{align}
\end{subequations}
Eqs. \ref{second} and \ref{fourth} also imply
\begin{equation}
\label{eq:zphinu}
z+\Phi=\nu.
\end{equation}

\begin{figure*}
\centering
\includegraphics[width=178mm]{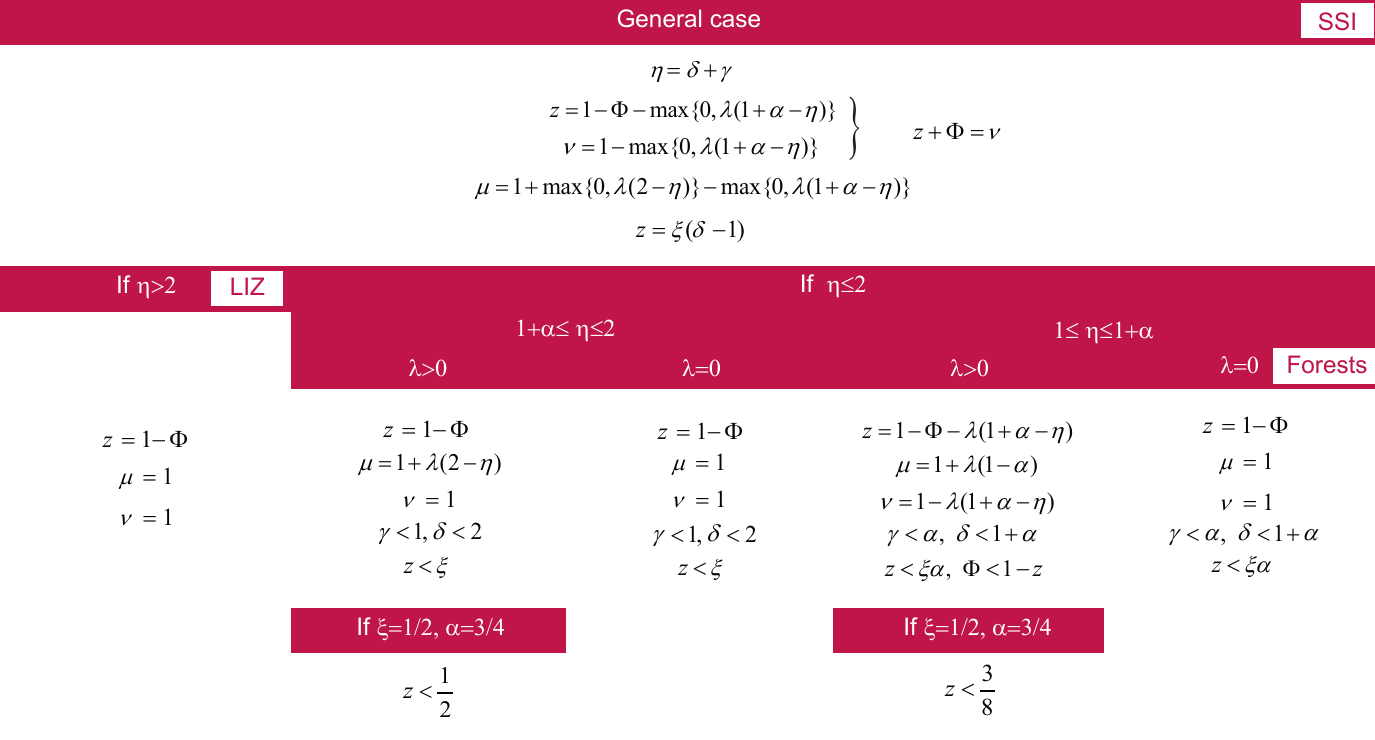}
\caption{Scheme of predictions on the values or bounds of scaling exponents based on the linking
relationships (\ref{linkingrel}a-e). Datasets names are located in different columns according to the available information
on their exponents’ values. Forests include Barro Colorado Island \cite{Condit2012}, Luquillo \cite{Zimmerman2010}, SSI stands for Sunda Shelf Islands \cite{Okie2009}, LIZ for
the dataset of lizard densities on islands worldwide \cite{Novosolov2015}. Note that the relationship $z<\xi\alpha$ is valid for forests only before the physiological constraint has been attained (see section 1.8.3). }
\label{fig:table}
\end{figure*}

Notice that there are only 5 independent exponents (e.g. $\gamma$, $\delta$, $\Phi$, $\alpha$ and $\lambda$), whereas the observable laws amount to $10$: Kleiber's law and Eqs. 3, 12-18 of the main text. Note that Eq. 13 contains three laws because it describes the scaling of the average abundance of a species with body mass and with the area of the ecosystem and the scaling of its cut-off with the area of the ecosystem. Figure \ref{fig:table} summarizes the predictions on the values or bounds  of scaling exponents based on the linking relationships (\ref{linkingrel}a-e), for different possible values of the independent exponents. As each of the exponents appearing in Eqs. (\ref{linkingrel}a-e) could have different values in different ecosystems and in different environmental conditions, in order to verify the validity of Eqs. (\ref{linkingrel}a-e) each relationship must be verified on a single dataset, where all the exponents  have been measured simultaneously.\\

\begin{figure*}
\begin{center}
\includegraphics[width=178mm]{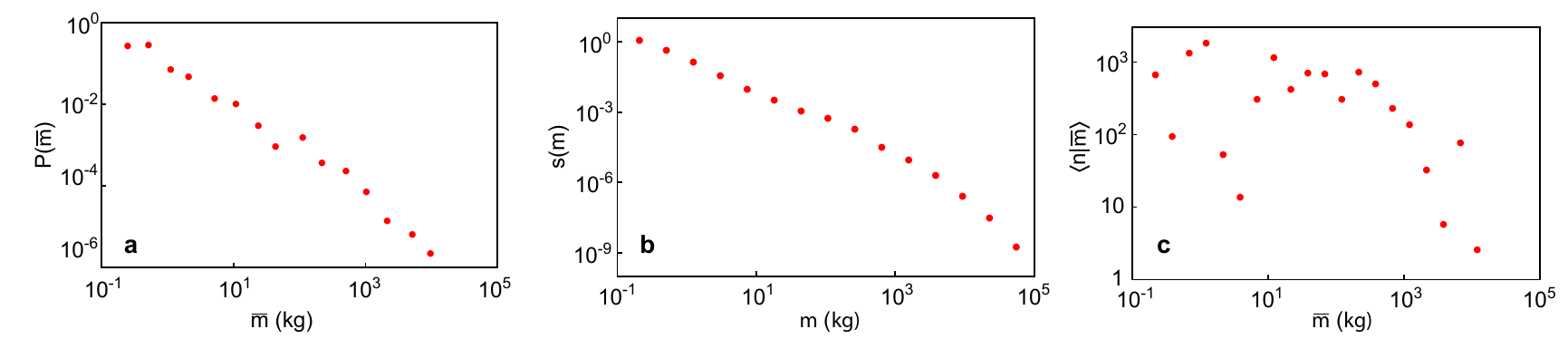}
\caption{Scaling patterns in the Luquillo forest, first census. a) $P(\bar{m})$ vs $\bar{m}$, b) $s(m)$ vs $m$, c) $\langle n | \bar{m} \rangle$ vs $\bar{m}$. Scaling exponents estimates are reported in the text. Note that finite-size effects may be present both at small and large values of $m$ and $\bar{m}$, for example due to the sampling protocol (see text).}
\label{fig:luquillo}
\end{center}
\end{figure*}

\begin{figure*}
\begin{center}
\includegraphics[width=178mm]{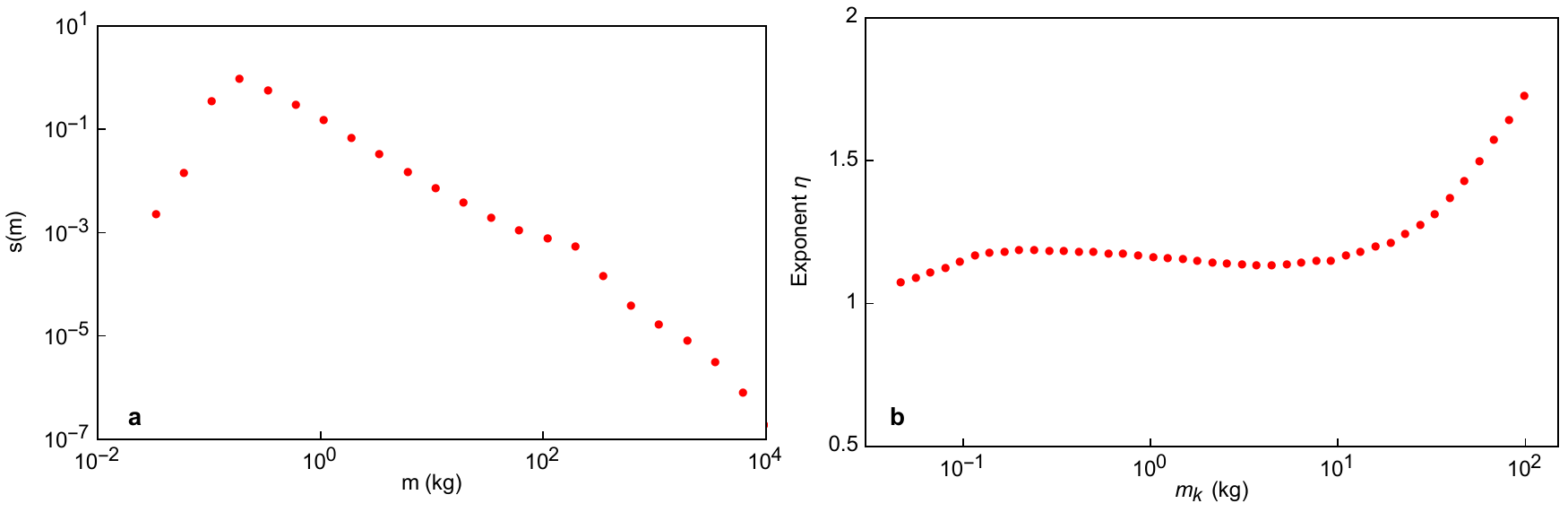}
\caption{a) Size spectrum in the Luquillo forest, second census. Finite-size effect are present both at small (i.e. $m<0.2$ kg) and large (i.e. $m>10^2$ kg) values of $m$. b) Size-spectrum exponent $\eta$ estimated using only data with mass $m>m_k$. The estimated exponent initially increases (until $m_k\simeq 0.2$ kg) due to a finite-size effect, then is rather stable until the statistics is not sufficient to properly estimate it ($m_k>20$ kg).}
\label{fit_distributions}
\end{center}
\end{figure*}

\noindent \textbf{2.2.1 Equation \ref{first} (Eq. 1 of the main text)}
\label{datafirst}
We verified \eqref{first} on censuses of Barro Colorado Island (BCI) \cite{Condit2012} (Fig. 2) and the Luquillo forest \cite{Zimmerman2010} (Fig. \ref{fig:luquillo}). These datasets report the trunk diameter and the species' identity of every tree having a diameter at breast height (dbh) $>$10 mm contained in a plot of 50 ha within the BCI forest (Panama), and a plot of 16 ha within the Luquillo forest (Puerto Rico). Diameters were converted into mass using an established allometric relationship between mass and diameter \cite{Enquist2001,Simini2010}, $m = 0.124 d ^{8/3}$ kg, with $d$ expressed in cm. To compute $P(\bar{m}|A)$, where $\bar{m}$ is the typical species' mass, we used the mean mass of the species' individuals as our estimate of $\bar{m}$. To obtain estimates of $\delta$ and $\eta$ the probability distributions $P(\bar{m}|A)$ and $s(m|A)$ were fitted via maximum likelihood to the functional forms $P(\bar{m}|A)=a_1 \bar{m}^{-\delta} e^{-b_1 \bar{m}}$ and $s(m|A)=a_2 m^{-\eta} e^{-b_2 m}$, where $a_1$ and $a_2$ are the normalization constants (which can be expressed in terms of $b_1$ and $b_2$), $b_1$ and $b_2$ are constants that accounts for possible finite-size effects, and $\delta$ and $\eta$ are the power-law exponents\footnote{\textcolor{black}{Specifically, one has: $a_1=b_1^{1-\delta}/\Gamma(1-\delta, \bar m_0\delta)$, where $\Gamma$ is the incomplete Gamma function. We maximized the log-likelihood: $\log L(\delta,b_1)=S \log \left[ b_1^{1-\delta} / \Gamma(1-\delta, \bar m_0\delta)\right] -\delta \sum_{i=1}^S \log \bar m_i - b_1 \sum_{i=1}^S \bar m_i$ with respect to $\delta$ and $b_1$, where $\bar m_0$ is the minimum species mass, $S$ is the total number of species and $i$ is an index that identifies the species. The maximum likelihood fitting of $s(m|A)$ was performed analogously.}}. Section 1.8.3 justifies the possible presence of an upper cutoff in $P(\bar{m}|A)$.  To account for deviations from the power-law behavior at low values of $m$ or $\bar{m}$ (these may arise for various reasons, like e.g. sampling protocols affecting the estimates of mean masses and mean abundances at small masses, as described in the next paragraph) we performed the maximum-likelihood estimation of $\delta$ and $\eta$ by considering only the data with $m>m_k$ at various values of $m_k=a^k \ 0.124$ kg in the range $(0.124-10^2)$ kg (with $1<a<2$ and $k$ integer). Note that $0.124$ kg is the mass of a tree with dbh=$10$ mm, i.e. the lower limit of the sampling protocol. If the data were distributed according to a pure power-law with no finite-size effects, such procedure would return approximately the same value of the exponent for any $m_k$. If the data were distributed according to a power-law with finite-size effects at small and large values of $m$, instead, one would observe an approximately constant estimate of the exponent at intermediate $m_k$ and deviations from such estimate at small and large values of $m_k$ (see e.g. Fig. \ref{fit_distributions}). For each fit, we identified the extent of the power-law regime and our estimate of the exponent and the associated error are, respectively, the mean and standard deviation of the maximum-likelihood exponent at different values of $\bar{m}_k$ in the power-law regime. 

\begin{figure*}
\begin{center}
\includegraphics[width=178mm]{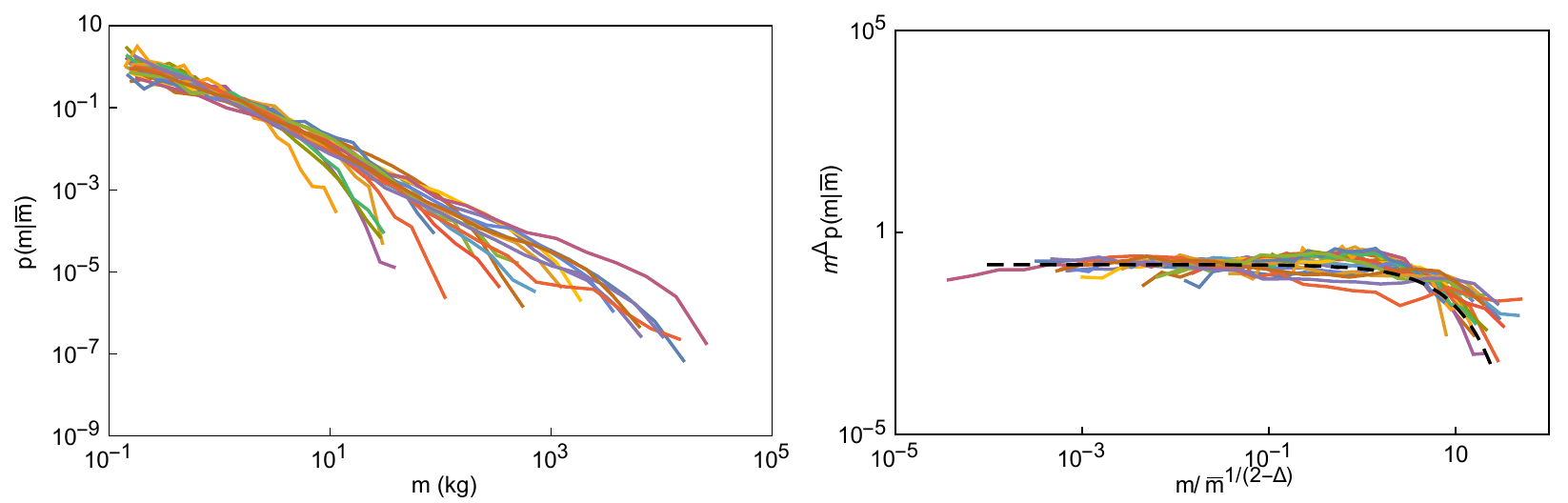}
\caption{a) Intraspecific tree size distributions in BCI (most abundant species). Each color corresponds to a different species. b) Intraspecific tree size distributions collapse according to \eqref{eq:intraspecific} onto the same universal curve. The dashed black line is the best fit of $\mathcal F(x)=q_0 e^{-q_1 x}$ (see text).}
\label{fig:intraspecific}
\end{center}
\end{figure*}
\begin{figure*}
\begin{center}
\includegraphics[width=178mm]{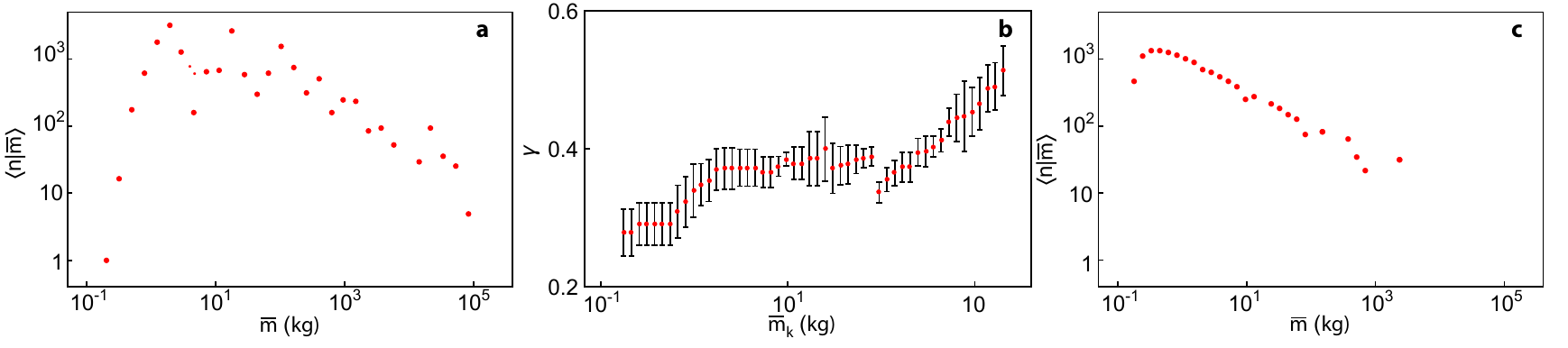}
\caption{a) Damuth's law ($\langle n| \bar{m} \rangle$ vs $\bar{m}$) in BCI, seventh census. Finite-size effect are present both at small (i.e. $\bar{m}<10$ kg) and large (i.e. $\bar{m}>10^{3}$ kg) values of $\bar{m}$. b) Damuth's law exponent $\gamma$ estimated using only the species with typical mass $\bar{m}>\bar{m}_k$. The estimated exponent initially increases, then is stable until the statistics is not sufficient to properly estimate it (or finite-size effects become relevant). c) Damuth's law in an artificial forest after mimicking the sampling protocol (see text for details). The effect of the sampling bias is visible at small masses.}
\label{fig:fit_damuth}
\end{center}
\end{figure*}

The estimate of Damuth's law exponent $\gamma$, describing the decay of mean abundances with species' typical masses $\bar{m}$, requires a correction for a bias introduced by the sampling protocol on the estimates of mean abundances and mean masses at small values of $\bar{m}$. In fact, the sampling protocol in tropical forests censuses instructs to sample only the trees with dbh larger than $10$ mm. The measured abundances of small species (i.e. those with typical diameter close to $10$ mm) are therefore lower than the true ones because individuals with diameter $d \leq 10$ mm were not censused. As a result, the average abundance as a function of a typical species' mass initially increases with $\bar{m}$ and is followed by the decreasing power-law regime where the effect of the sampling protocol becomes unimportant (fig. 2A). The initial increase is a sampling artifact. In fact, we verified that this is the case by creating an artificial forest dataset where species' mean abundances follow Damuth's law exactly. Within such artificial forest, we distributed species mean masses $\bar{m}$ according to the power-law $p(\bar{m})=(\delta-1)\bar{m}^{-\delta}$.  We drew the abundance of each species from a Poisson distribution with mean $\bar{m}^{-\gamma}$. Finally, we needed to assign a mass to each individual of each species. To do so, we characterized the intra-specific mass distributions of tropical trees $p(m|\bar{m})$, i.e. the probability that an individual has mass in $(m,m+dm)$ given that it belongs to a species with mean mass $\bar{m}$. We computed intra-specific mass distributions in the BCI and Luquillo forests (Fig. \ref{fig:intraspecific}a) and found, looking at the species with more than $1500$ (BCI) and $400$ (Luquillo) individuals, that most species have intraspecific size distributions characterized by the finite-size scaling form:
\begin{equation}
\label{eq:intraspecific}
p(m|\bar{m})= m^{-\Delta} \mathcal F\left( \frac{m}{\bar{m}^\Omega} \right),
\end{equation}
where $\Omega=1/(2-\Delta)$ ensures that $\int m p(m|\bar{m}) dm = \bar{m}$, $\Delta = 1.12\pm0.06$\footnote{\textcolor{black}{The exponent estimate is computed using the method described in Bhattacharjee and Seno \cite{Bhattacharjee2001}. The error is computed as the value of the exponent at which the error functional $P_b$ defined in Bhattacharjee and Seno \cite{Bhattacharjee2001} is $1\%$ larger than its value at the minimum.}} and $\mathcal F(x)$ is a scaling function with limiting behaviors $\mathcal F(x)\to \mbox{const}$ for $x\to0$ and $\mathcal F(x)\to0$ more rapidly than any power of $x$ for $x\to\infty$. A similar result was found for unicellular protists \cite{Giometto2013}. Furthermore, via data collapse (i.e. plotting $m^{\Delta}p(m|\bar{m})$ vs $ {m}/{\bar{m}^\psi} $) we found that $\mathcal F(x)=q_0 e^{-q_1 x}$ provides a good fit to the data, where $q_0=0.17$ and $q_1=0.21$ are constants. Note that $q_0$ is not a parameter of the fit, as it is fixed by normalization. Having characterized the scaling form of intraspecific distributions, we could then randomly sample from such distributions the masses of individuals belonging to each species in our artificial forest. We then mimicked the sampling protocol by eliminating all individuals with mass lower than $0.124$ kg (corresponding to a dbh of $10$ mm) and computed Damuth's law in such a filtered dataset. Fig. \ref{fig:fit_damuth}c shows that, despite the fact that mean species abundances in the artificial forest follow Damuth's law exactly, the sampling protocol causes the emergence of a new regime at small $\bar{m}$ where the relationship between $\langle n|\bar{m} \rangle$ and $\bar{m}$ is monotonically increasing. This demonstrates that the sampling protocol introduces an artificial deviation from the power-law regime which has to be considered with care while interpreting empirical data.
The sampling artifact can be corrected as follows. To derive our estimate for $\gamma$ and the associated error, we binned the typical species masses logarithmically and computed the mean abundance of all species within each bin. Then, we varied the number of bins $n_{bin}$ and computed the Damuth's law exponent $\gamma_{n_{bin}}$ via least-squares fitting of log-transformed data, weighted by the standard deviation of abundances within each bin. Our estimate for $\gamma$ is the mean $\gamma=\langle \gamma_{n_{bin}} \rangle$ across several values of $n_{bin}$. To correct for the bias caused by the sampling protocol, we repeated such computation by considering only the species with mean mass $\bar{m}>\bar{m}_k$, with $\bar{m}_k=a^k \ 0.124$ kg in the range $(0.124 - 10^2)$ kg (with $1<a<2$ and $k$ integer). If mean abundances followed Damuth's law exactly (in the absence of sampling bias and with sufficient statistics), such procedure would return the same value of $\gamma$ for any $\bar{m}_k$. If finite-size effects were present at small and large values of $\bar{m}$, instead, one would observe an approximately constant estimate of the exponent at intermediate $\bar{m}_k$ and deviations from such estimate at small and large values of $\bar{m}_k$ (see e.g. Fig. \ref{fig:fit_damuth}b). At large values of $\bar{m}_k$ a finite-size effect may also be induced by low statistics. For each fit, we identified the extent of the power-law regime and our estimate of $\gamma$ and the associated error are, respectively, the mean and standard deviation of the exponents estimated at the different values of $\bar{m}_k$ in the power-law regime.

In our analysis, we used the fifth, sixth and seventh censuses of BCI and the five censuses of the Luquillo forest available online in the Center for Tropical Forest Science (CTFS) dataset collection. All these censuses satisfy the linking relationship \eqref{first} within the errors. Whereas BCI censuses appear very similar to each other (and therefore also the exponent values estimated in different censuses, see Table \ref{tab:bci}), the Luquillo forest appears to be more dynamic (we note that the forest was hit by a major hurricane between censuses 2 and 3), with the values of $\gamma$ decreasing in time after $1998$ (census 2, see Table \ref{tab:luquillo}). Because the estimate of $\delta$ remains constant suggesting that climatic, ecological or anthropogenic dynamics affected only species' abundances in this forest, our framework would predict via eq. \eqref{first} that $\eta$ would also decrease in time, and in fact this is also found in the data, with eq. \eqref{first} being verified in all censuses. We note that both the BCI and the Luquillo datasets reject the linking relationship $\eta=\delta$ predicted in a previous theoretical work \cite{Banavar2007}.

Although we do not have an estimate of the exponents $\mu$ and $\nu$ for tropical forests, a reasonable assumption is $\mu=1=\nu$. Given that $\eta<2$ for the BCI and the Luquillo datasets, such assumption is only verified if $\lambda=0$, i.e. if the maximum body size does not scale with the ecosystem area. An analysis of this situation is provided in section 1.8.3.\\

\noindent \textbf{2.2.2 Equations \eqref{second} and \eqref{fourth} in other datasets (Eqs. 6, 8 of the main text)}
\label{sec:LIZ}

\begin{figure*}
\begin{center}
\includegraphics[width=178mm]{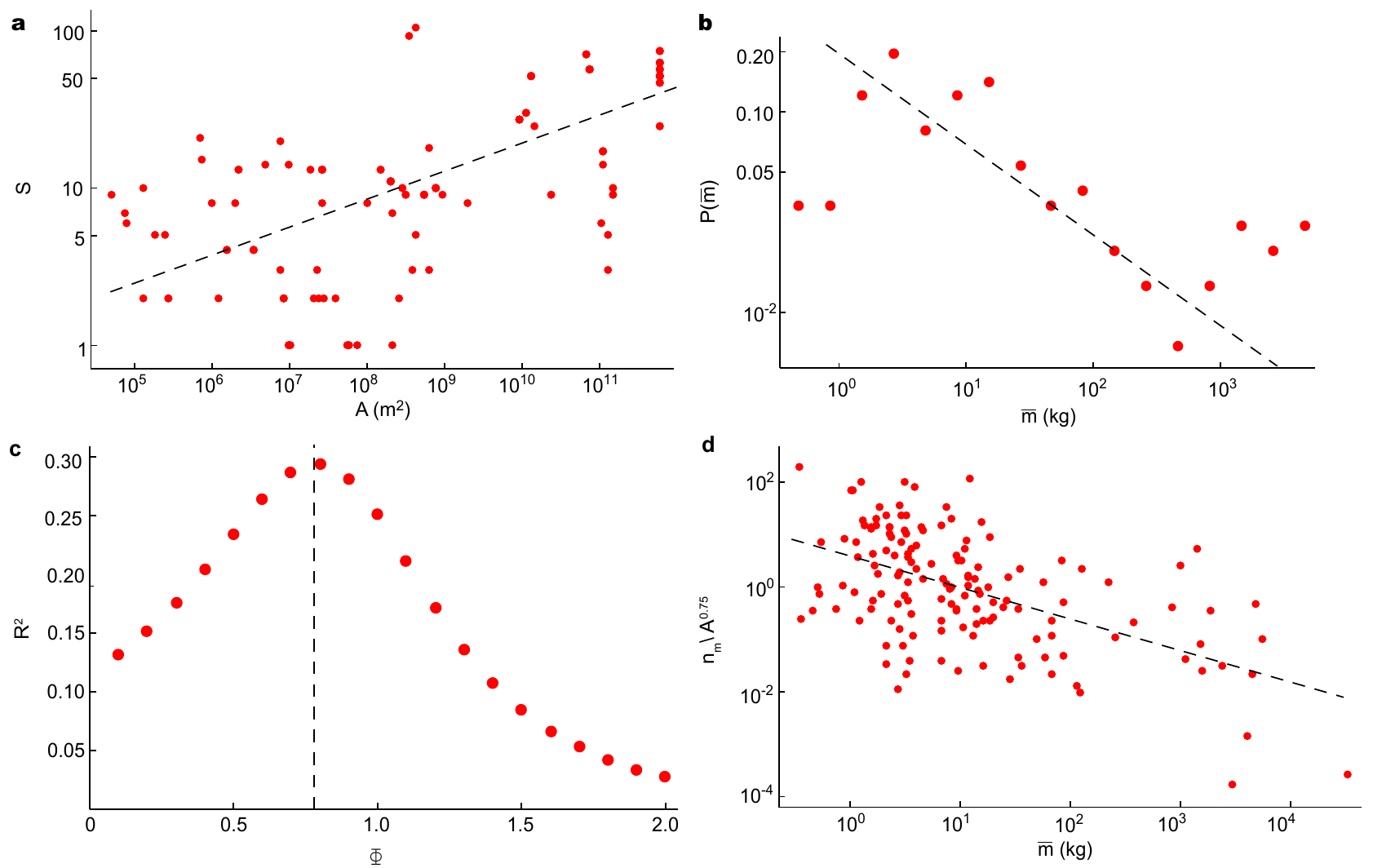}
\caption{Empirical evidence of scaling patterns in LIZ dataset \cite{Novosolov2015}: a) SAR; b) P($\bar{m}$); c) Coefficient of determination $R^2$ of the linear least-squares fit of $\left( \log \bar{m}, \log \frac{n(\bar{m},A)}{A^\Phi} \right)$ for $\Phi \in [0,2]$. Dashed line in correspondence of the value $\Phi=0.75$ giving the best fit; d) Non-binned Damuth's law plotted using the estimated value of $\Phi$ ($\bar m$ are species' mean masses). Best fit parameters are reported in Table \ref{tab:LIZ}, details on the fit in the SI text.  }
\label{fig:liz}
\end{center}
\end{figure*}

To test the validity of \eqref{second} we used a dataset gathering population densities of several species of lizards on 64 islands worldwide (LIZ) \cite{Novosolov2015}, with areas ranging from 10$^{-1}$ to 10$^5$ km$^2$. In this dataset, we fitted the SAR with linear least-squares regression on log-transformed data and $P(\bar m|A)$ via maximum-likelihood \cite{Clauset2009} (Fig. \ref{fig:liz}a and b), where $\bar m$ are species' mean masses. The exponent $\Phi=0.78$ (describing the dependence of $\langle n|\bar{m},A \rangle$ on $A$) was obtained by maximizing the coefficient of determination $R^2$  of the linear least-squares regression of the pairs $\left( \log \bar{m}, \log \frac{n}{A^\Phi} \right)$ obtained by varying $\Phi$ in the interval [0,2] (Fig. \ref{fig:liz}c and d). The estimate of $\gamma$ is obtained from the pairs $\left( \log \bar{m}, \log \frac{n}{A^\Phi} \right)$ computed with the optimal value of $\Phi$ with the same methods used for forests, with $\bar{m}_k=2 c^k 10^{-1}$ kg in the range $(10^{-1}-10^2)$ kg ( with $1<c<2$ and $k$ integer). Note that this estimate of $\gamma$ is different from the one given in Table \ref{exponents}, which was obtained by plotting densities versus typical masses (equivalent to taking $\Phi=1$) in order to allow comparison with other data from the literature. The estimates for the scaling exponents in this dataset are reported in Table \ref{tab:LIZ}. Because in this dataset $\eta=\delta+\gamma$ is compatible with 2, Eqs. \ref{third} and \ref{fourth} imply $\mu=\nu$. Furthermore, because in general $\alpha \leq 1$, one has $\max\{0,\lambda(1+\alpha-\eta)\}=0$ and therefore our framework predicts:
\begin{equation}
\label{eq:numu}
\nu=\mu=1.
\end{equation} 
\eqref{second} (or, equivalently, \eqref{eq:zphinu}) thus implies that in order to have $z>0$, as found in the dataset analyzed here, $\Phi$ needs to be smaller than one. Since $\Phi$ describes the scaling of $\langle n|\bar{m},A \rangle$ with $A$, our framework predicts that species' densities should decrease with increasing ecosystem area. This is indeed found in LIZ and the values of $z$ and $1-\Phi$ are compatible within the errors (see Table \ref{tab:LIZ}). Equivalently, using \eqref{eq:zphinu} and our estimate of $z$ we find $\nu=0.95 \pm 0.08$, which is compatible with the prediction (\eqref{eq:numu}) $\nu=1$.\\

\noindent \textbf{2.2.3 Equation \ref{fifth}  (Eq. 9 of the main text)}

To test the validity of \eqref{fifth} we used a dataset of mammals species presence/absence data on several islands in Sunda Shelf (SSI) \cite{Okie2009}, covering a wide range of island areas (10$^1$ to 10$^6$ km$^2$). The SAR and the scaling of the maximum body mass with the area were fitted with linear least-squares regression on log-transformed data, while $P(\bar{m}|A)$ was fitted with maximum-likelihood \cite{Clauset2009}. Scaling exponents in this dataset are reported in Table \ref{tab:SSI}. We find that eq. \eqref{fifth} is verified in the SSI dataset within one standard error, as $z=0.23\pm 0.02$ and $\xi(\delta-1)=0.29\pm 0.1$.
 
\section{\normalsize Mathematical community dynamics models}
\label{model}
\subsection{\normalsize Fixed number of species}
\begin{figure*}
\begin{center}
\includegraphics[width=17.8cm,center]{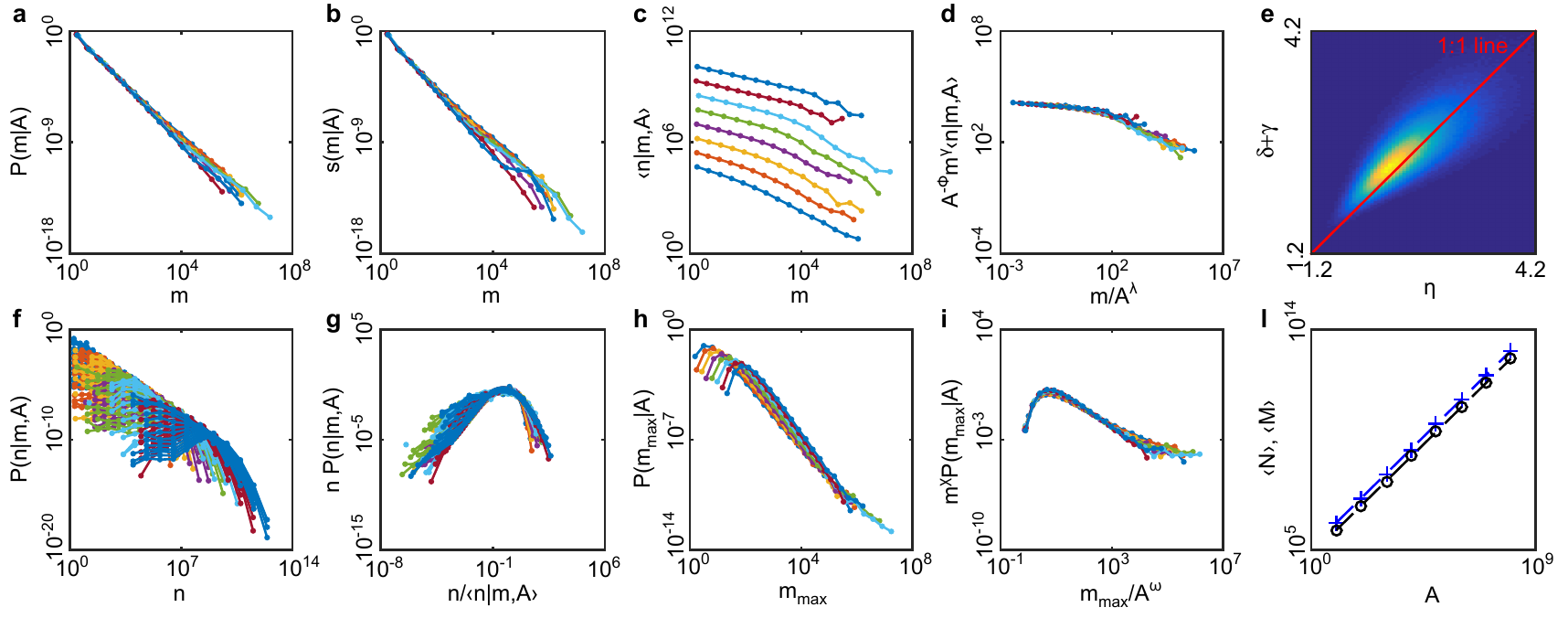}
\caption{Basic model statistics with parameters $z=1/4$, $w=10^{-3}$, $\alpha=3/4$, $\theta=1/4$ (simulation results shown in Fig. 3 of the main text). Different colors refer to different values of $A= 10^i$, from $i=1$ (lower blue curve in panel c) to $i=8$ (upper blue curve in panel c). Panels a-c, f and h show respectively $P(m|A)$, $s(m|A)$, $\langle n|m,A \rangle$,  $P(n|m,A)$ and $P(m_{\max}|A)$ estimated at stationarity. Panels d, g and i show collapses of simulation data for $\langle n|m,A \rangle$,  $P(n|m,A)$ and $P(m_{\max}|A)$, respectively. Panel e shows the density scatter-plot of $\delta+\gamma$ versus $\eta$. The density histogram is normalized to one, with blue representing the value zero and yellow the value one. Shown are simulation data for the largest area $A=10^8$. Panel j shows the scaling of the average total biomass $\langle M \rangle$ (blue crosses and dashed lines) and average total abundance $\langle N \rangle$ (black dots and dashed lines) with $A$. See Table \ref{tab:exploration} for estimates of exponents' values.}
\label{fig:z14standard}
\end{center}
\end{figure*}

\begin{figure*}
\begin{center}
\includegraphics[width=17.8cm,center]{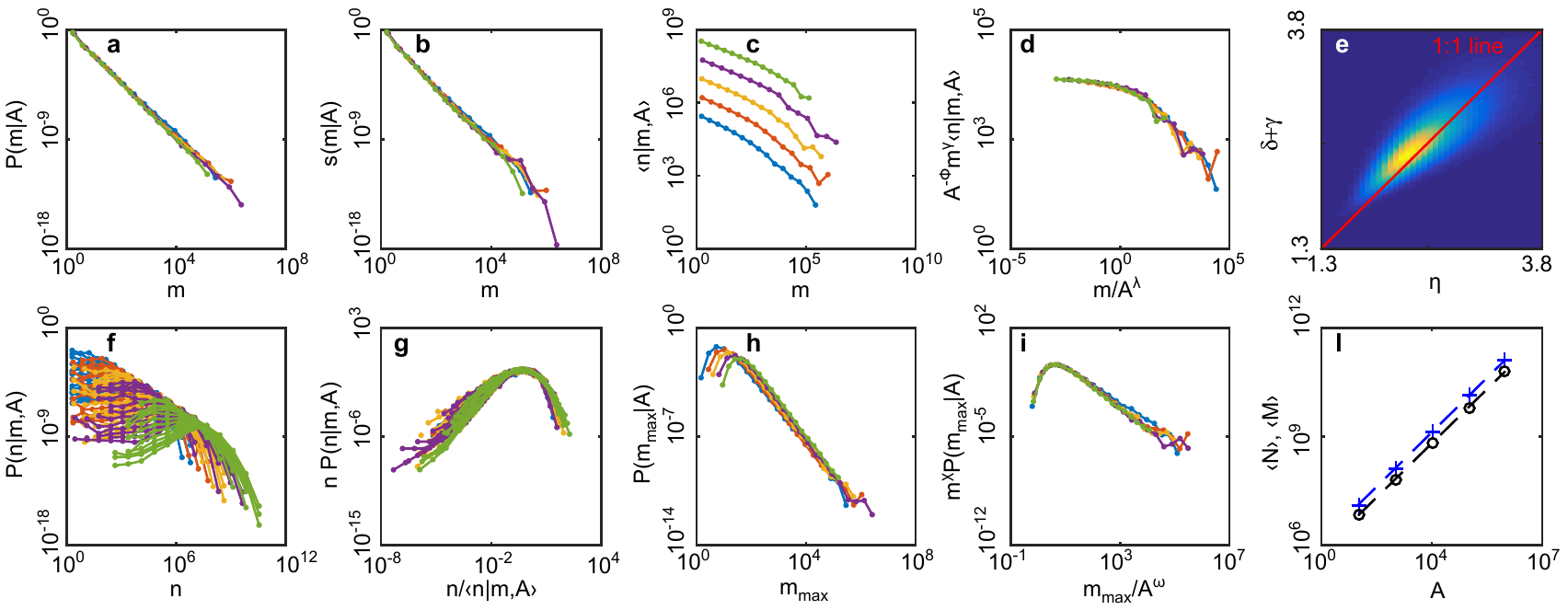}
\caption{Basic model statistics with parameters $z=1/4$, $w=10^{-4}$, $\alpha=3/4$, $\theta=1/4$. Different colors refer to different values of ecosystem area $A= 10^i$, with $i=2$ (cyan)$; 3$ (orange)$; 4$ (yellow)$; 5$ (purple)$; 6$ (green). Panels a-c, f and h show respectively $P(m|A)$, $s(m|A)$, $\langle n|m,A \rangle$,  $P(n|m,A)$ and $P(m_{\max}|A)$ estimated at stationarity. Panels d, g and i show collapses of simulation data for $\langle n|m,A \rangle$,  $P(n|m,A)$ and $P(m_{\max}|A)$, respectively. Panel e shows the density scatter-plot of $\delta+\gamma$ versus $\eta$. The density histogram is normalized to one, with blue representing the value zero and yellow the value one. Shown are simulation data for the largest area $A=10^6$. Panel j shows the scaling of the average total biomass $\langle M \rangle$ (blue crosses and dashed lines) and average total abundance $\langle N \rangle$ (black dots and dashed lines) with $A$. See Table \ref{tab:exploration} for estimates of exponents' values.}
\label{fig:z14mu4}
\end{center}
\end{figure*}

\begin{figure*}
\begin{center}
\includegraphics[width=17.8cm,center]{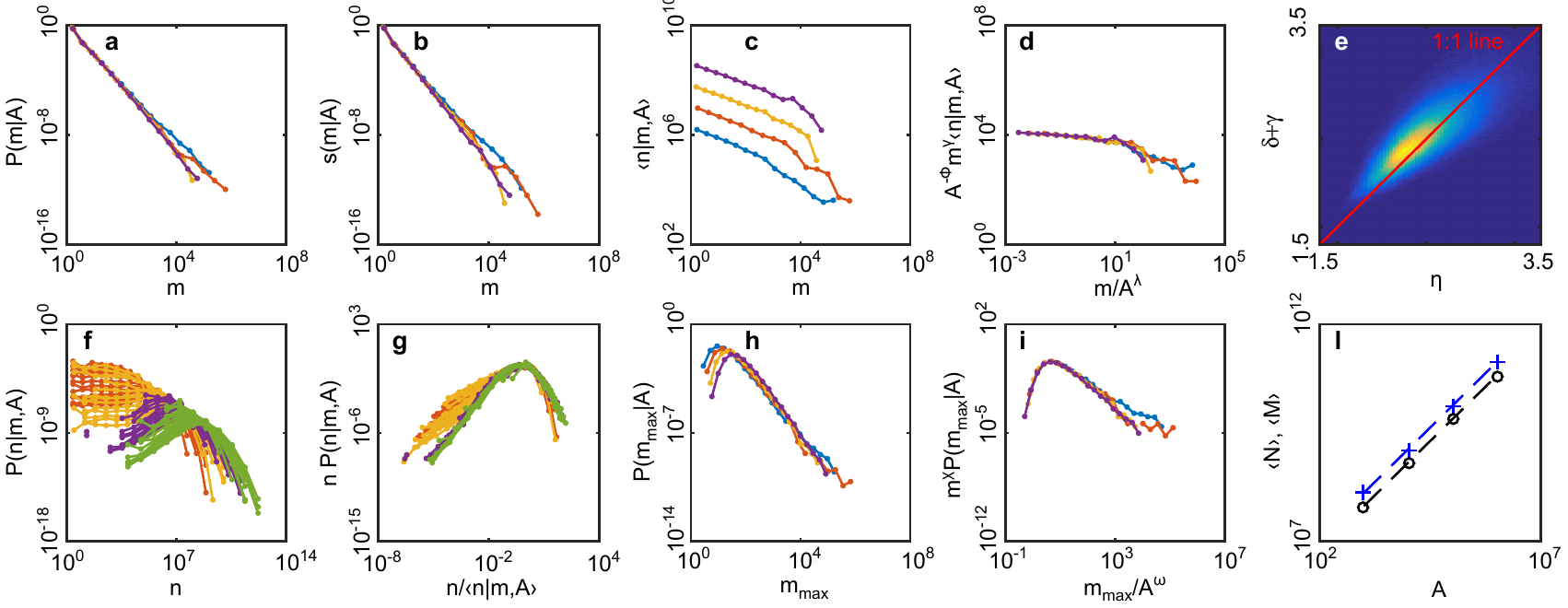}
\caption{Basic model statistics with parameters $z=1/4$, $w=10^{-5}$, $\alpha=3/4$, $\theta=1/4$. Different colors refer to different values of ecosystem area $A= 10^i$, with $i=3$ (cyan)$; 4$ (orange)$; 5$ (yellow)$; 6$ (purple). Panels a-c, f and h show respectively $P(m|A)$, $s(m|A)$, $\langle n|m,A \rangle$,  $P(n|m,A)$ and $P(m_{\max}|A)$ estimated at stationarity. Panels d, g and i show collapses of simulation data for $\langle n|m,A \rangle$,  $P(n|m,A)$ and $P(m_{\max}|A)$, respectively. Panel e shows the density scatter-plot of $\delta+\gamma$ versus $\eta$. The density histogram is normalized to one, with blue representing the value zero and yellow the value one. Shown are simulation data for the largest area $A=10^6$. Panel j shows the scaling of the average total biomass $\langle M \rangle$ (blue crosses and dashed lines) and average total abundance $\langle N \rangle$ (black dots and dashed lines) with $A$. See Table \ref{tab:exploration} for estimates of exponents' values.}
\label{fig:z14mu5}
\end{center}
\end{figure*}

\begin{figure*}
\begin{center}
\includegraphics[width=17.8cm,center]{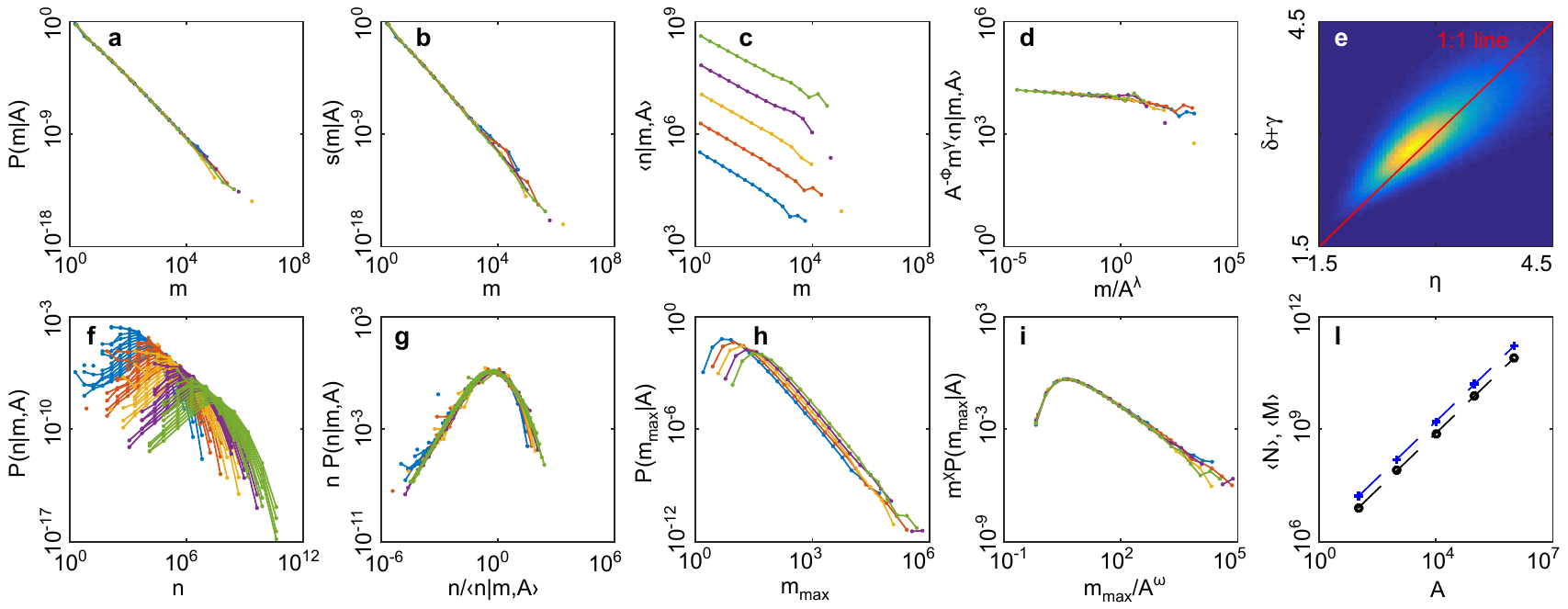}
\caption{Basic model statistics with parameters $z=1/4$, $w=10^{-3}$, $\alpha=1/2$, $\theta=1/4$. Different colors refer to different values of ecosystem area $A= 10^i$, with $i=2$ (cyan)$; 3$ (orange)$; 4$ (yellow)$; 5$ (purple)$; 6$ (green). Panels a-c, f and h show respectively $P(m|A)$, $s(m|A)$, $\langle n|m,A \rangle$,  $P(n|m,A)$ and $P(m_{\max}|A)$ estimated at stationarity. Panels d, g and i show collapses of simulation data for $\langle n|m,A \rangle$,  $P(n|m,A)$ and $P(m_{\max}|A)$, respectively. Panel e shows the density scatter-plot of $\delta+\gamma$ versus $\eta$. The density histogram is normalized to one, with blue representing the value zero and yellow the value one. Shown are simulation data for the largest area $A=10^6$. Panel j shows the scaling of the average total biomass $\langle M \rangle$ (blue crosses and dashed lines) and average total abundance $\langle N \rangle$ (black dots and dashed lines) with $A$. See Table \ref{tab:exploration} for estimates of exponents' values.}
\label{fig:z14alpha05}
\end{center}
\end{figure*}

\begin{figure*}
\begin{center}
\includegraphics[width=17.8cm,center]{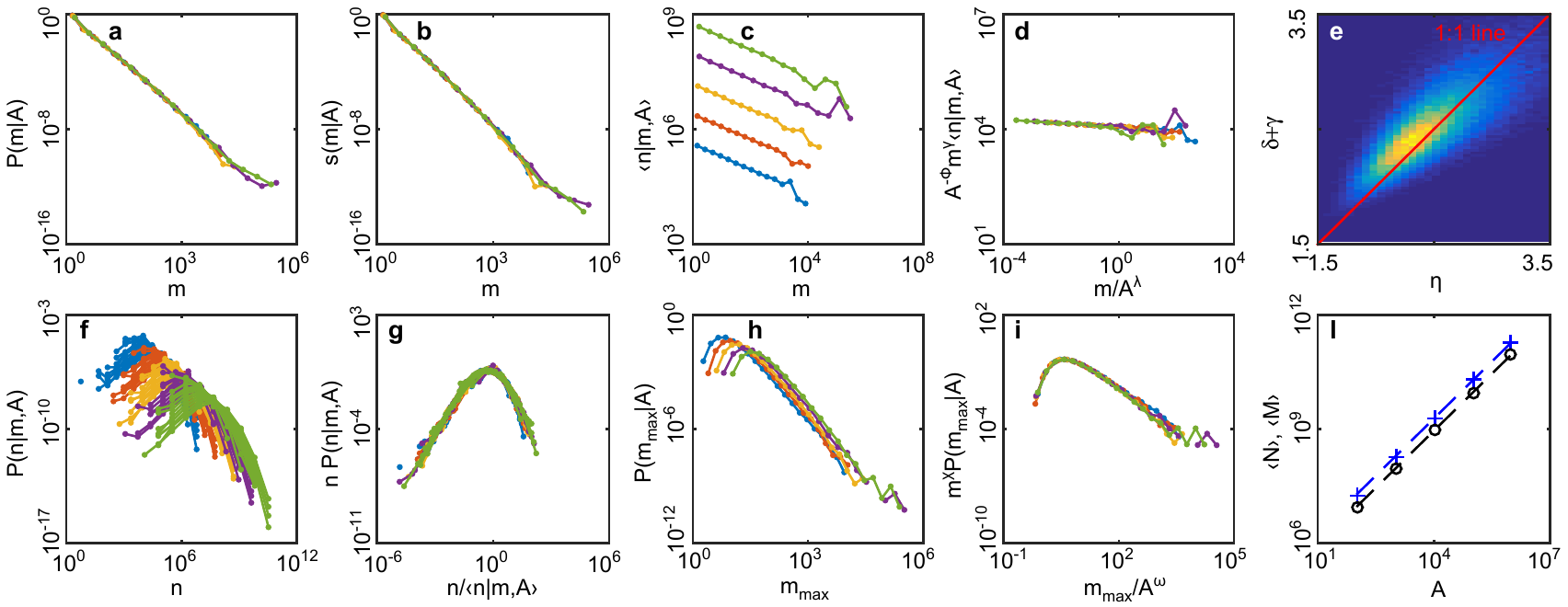}
\caption{Basic model statistics with parameters $z=1/4$, $w=10^{-3}$, $\alpha=1/4$, $\theta=1/4$. Different colors refer to different values of ecosystem area $A= 10^i$, with $i=2$ (cyan)$; 3$ (orange)$; 4$ (yellow)$; 5$ (purple)$; 6$ (green). Panels a-c, f and h show respectively $P(m|A)$, $s(m|A)$, $\langle n|m,A \rangle$,  $P(n|m,A)$ and $P(m_{\max}|A)$ estimated at stationarity. Panels d, g and i show collapses of simulation data for $\langle n|m,A \rangle$,  $P(n|m,A)$ and $P(m_{\max}|A)$, respectively. Panel e shows the density scatter-plot of $\delta+\gamma$ versus $\eta$. The density histogram is normalized to one, with blue representing the value zero and yellow the value one. Shown are simulation data for the largest area $A=10^6$. Panel j shows the scaling of the average total biomass $\langle M \rangle$ (blue crosses and dashed lines) and average total abundance $\langle N \rangle$ (black dots and dashed lines) with $A$. See Table \ref{tab:exploration} for estimates of exponents' values.}
\label{fig:z14alpha025}
\end{center}
\end{figure*}

\begin{figure*}
\begin{center}
\includegraphics[width=17.8cm,center]{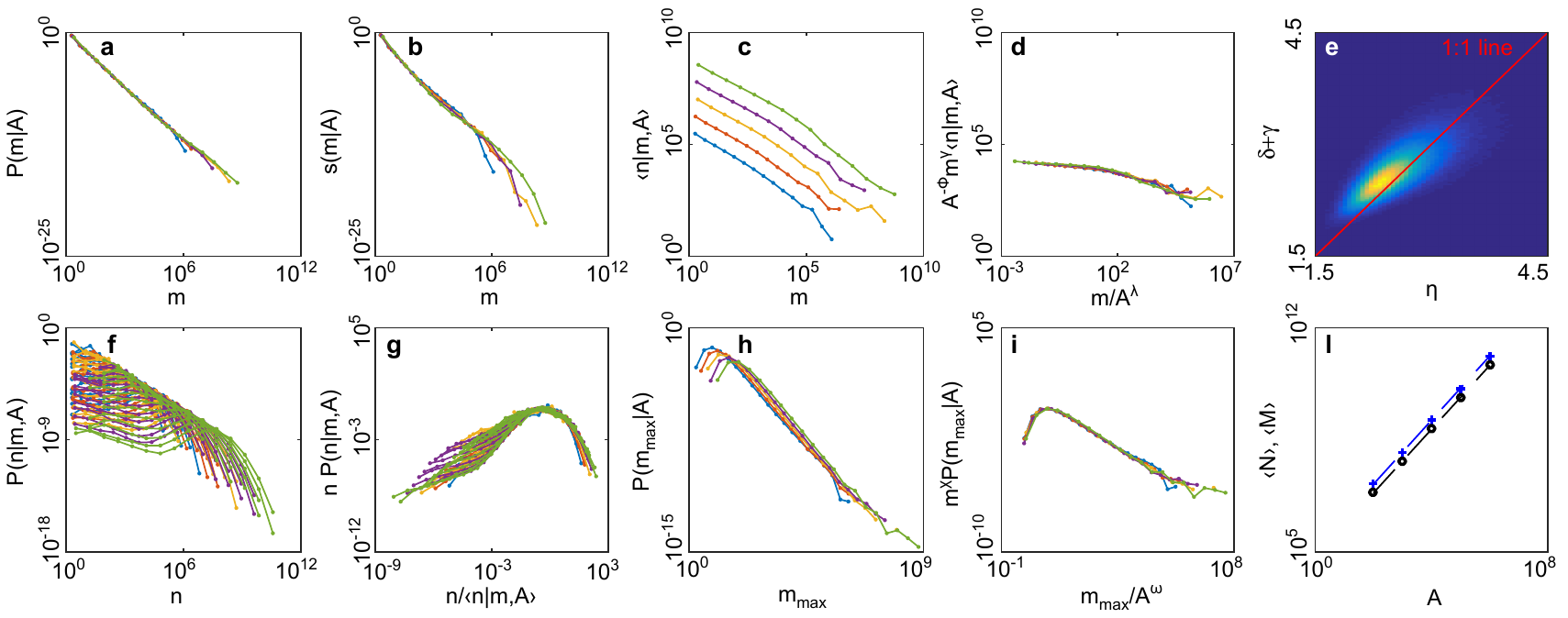}
\caption{Basic model statistics with parameters $z=1/4$, $w=10^{-3}$, $\alpha=3/4$, $\theta=1/2$. Different colors refer to different values of ecosystem area $A= 10^i$, with $i=2$ (cyan)$; 3$ (orange)$; 4$ (yellow)$; 5$ (purple)$; 6$ (green). Panels a-c, f and h show respectively $P(m|A)$, $s(m|A)$, $\langle n|m,A \rangle$,  $P(n|m,A)$ and $P(m_{\max}|A)$ estimated at stationarity. Panels d, g and i show collapses of simulation data for $\langle n|m,A \rangle$,  $P(n|m,A)$ and $P(m_{\max}|A)$, respectively. Panel e shows the density scatter-plot of $\delta+\gamma$ versus $\eta$. The density histogram is normalized to one, with blue representing the value zero and yellow the value one. Shown are simulation data for the largest area $A=10^6$. Panel j shows the scaling of the average total biomass $\langle M \rangle$ (blue crosses and dashed lines) and average total abundance $\langle N \rangle$ (black dots and dashed lines) with $A$. See Table \ref{tab:exploration} for estimates of exponents' values.}
\label{fig:z14theta05}
\end{center}
\end{figure*}

\begin{figure*}
\begin{center}
\includegraphics[width=17.8cm,center]{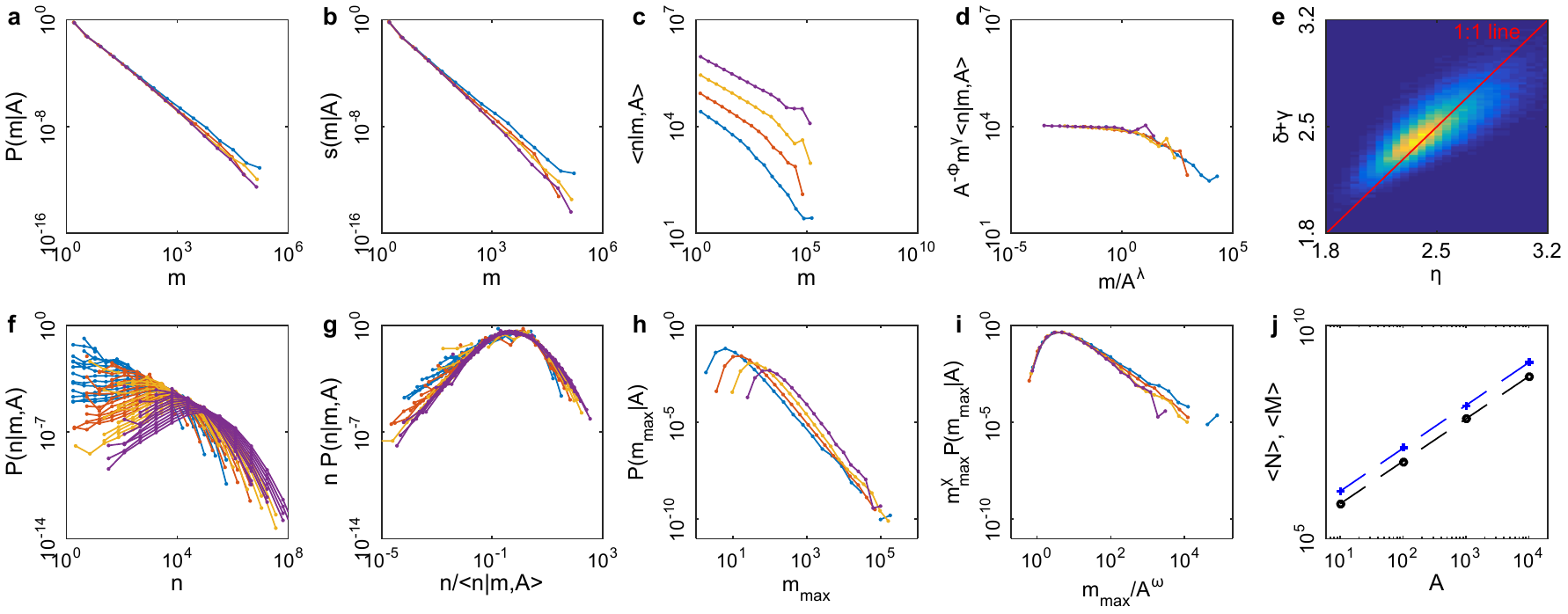}
\caption{Basic model statistics with parameters $z=1/2$, $w=10^{-3}$, $\alpha=3/4$, $\theta=1/4$. Different colors refer to different values of ecosystem area $A= 10^i$, where $i=1$ (cyan)$; 2$ (orange)$; 3$ (yellow)$; 4$ (purple). Panels a-c, f and h show respectively $P(m|A)$, $s(m|A)$, $\langle n|m,A \rangle$,  $P(n|m,A)$ and $P(m_{\max}|A)$ estimated at stationarity. Panels d, g and i show collapses of simulation data for $\langle n|m,A \rangle$,  $P(n|m,A)$ and $P(m_{\max}|A)$, respectively. Panel e shows the density scatter-plot of $\delta+\gamma$ versus $\eta$. The density histogram is normalized to one, with blue representing the value zero and yellow the value one. Shown are simulation data for the largest area $A=10^4$. Panel j shows the scaling of the average total biomass $\langle M \rangle$ (blue crosses and dashed lines) and average total abundance $\langle N \rangle$ (black crosses and dashed lines) with $A$. See Table \ref{tab:exploration} for estimates of exponents' values.}
\label{fig:z05}
\end{center}
\end{figure*}

\noindent \textbf{\\3.1.1 Basic model: exploration of parameters' space}
\label{basicmodelexp}
Our basic model for the community dynamics of an ecosystem depends on a number of parameters (see Methods). As explained in the Methods section, a thorough exploration of the parameters' space is computationally unfeasible. Nonetheless, we verified that varying the values of the parameters that are most meaningful for the dynamics (i.e. Kleiber's law exponent $\alpha$, the speciation rate $w$, the SAR exponent $z$ and the exponent $\theta$ that describes the scaling of vital rates with body mass), the scaling characterization of the stationary state always holds and the linking relationships in Eqs. 1, 6--9 of the main text are always satisfied. Starting from the set of parameters $w=10^{-3}$, $z=1/4$, $\alpha=3/4$ and $\theta=1/4$ (parameters used to generate Fig. 3 of the main text) we varied one or two parameters at a time, keeping the other ones fixed. The parameters $v_0$ and $c$ which appear in Eq. 20 of the main text were fixed to $v_0=1/2$ and $c=10^{-5}$.
Figs. (\ref{fig:z14standard}--\ref{fig:z05}) show the ecological patterns computed at stationarity for each set of parameters. Table \ref{tab:exploration} reports the estimates of the scaling exponents obtained for each set of parameters. All estimates were obtained as explained in the Methods section, unless otherwise stated. For each set of parameters, the relationships in Eqs. 1, 6--9 of the main text are satisfied within errors, the data collapses predicted by our scaling framework hold and the density scatter-plot of $\eta$ versus $\delta+\gamma$ estimated at each time-step (Figs. \ref{fig:z14standard}--\ref{fig:z05}, panel e) is peaked along the 1:1 line, implying that the linking relationship (1) is satisfied, on average, at all times.\\

\begin{figure*}
\begin{center}
\includegraphics[width=17.8cm,center]{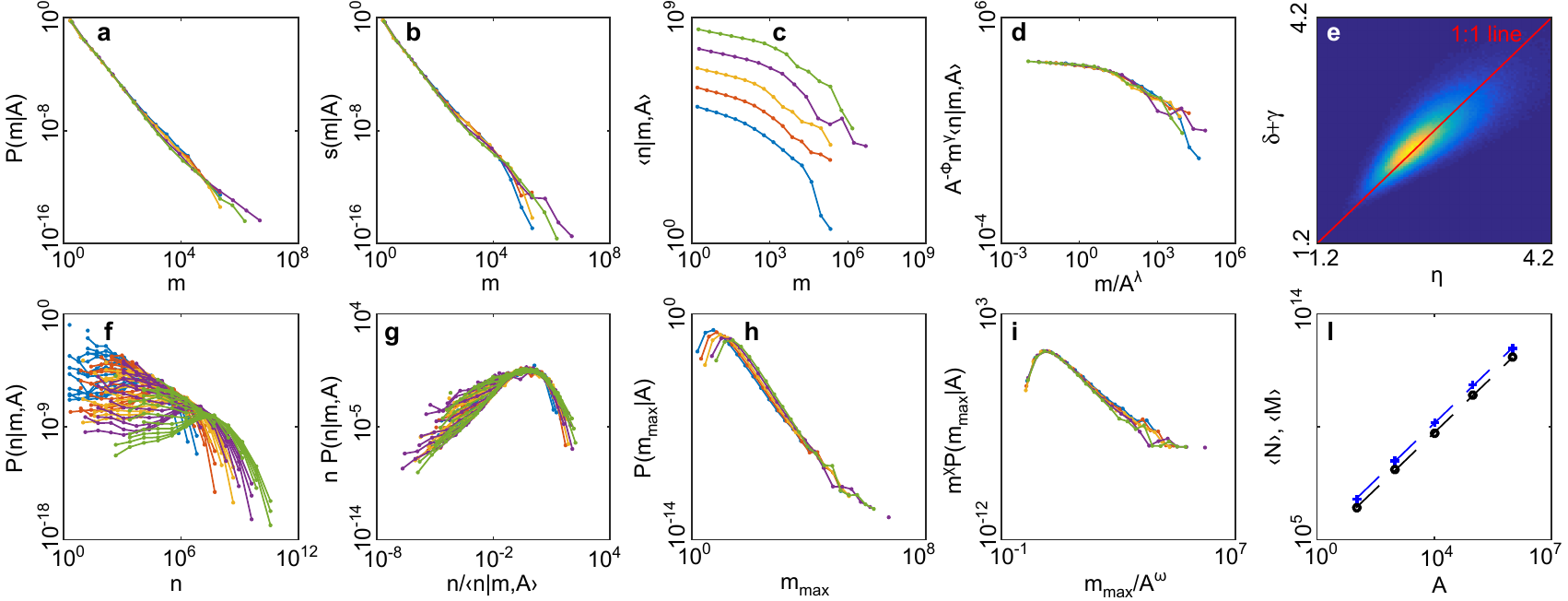}
\caption{Variation of the basic model, described in section \ref{variation}. Statistics computed with parameters $z=1/4$, $w=10^{-6}$, $\alpha=3/4$, $\theta=1/2$. Different colors refer to different values of ecosystem area $A= 10^i$, where $i=1$ (cyan)$; 3/2$ (orange)$; 2$ (yellow)$; 5/2$ (purple)$; 3$ (green). Panels a-c, f and h show respectively $P(m|A)$, $s(m|A)$, $\langle n|m,A \rangle$,  $P(n|m,A)$ and $P(m_{\max}|A)$ estimated at stationarity. Panels d, g and i show collapses of simulation data for $\langle n|m,A \rangle$,  $P(n|m,A)$ and $P(m_{\max}|A)$, respectively. Panel e shows the density scatter-plot of $\delta+\gamma$ versus $\eta$. The density histogram is normalized to one, with blue representing the value zero and yellow the value one. Shown are simulation data for the largest area $A=10^3$. Panel j shows the scaling of the average total biomass $\langle M \rangle$ (blue crosses and dashed lines) and average total abundance $\langle N \rangle$ (black crosses and dashed lines) with $A$. See Table \ref{tab:explorationvar} for estimates of exponents' values.}
\label{fig:z14variation}
\end{center}
\end{figure*}

\noindent \textbf{3.1.2 Variation on the speciation dynamics assumptions}
\label{variation}
In order to investigate the sensitivity of our results (i.e. the compatibility of the dynamic model with the scaling framework, Eqs. 6--9 of the main text) to changes in the dynamic model assumptions, we investigated a variation of our basic model in which the species that goes extinct at each speciation event (to maintain $S$ constant) is chosen randomly with a weight inversely proportional to its abundance, but independent of its mass. We ran this model with the same parameter values reported in the Methods section for the basic model and found that such modified model is compatible with our scaling framework, and thus with the predicted pattern covariations, which are verified within the errors. Table \ref{tab:exploration} reports the corresponding exponents values and Fig. \ref{fig:z14variation} displays the macroecological patterns in this model.

\subsection{\normalsize  Fluctuating number of species}
\label{fluctmodels}
In order to further investigate the sensitivity of our results to changes in the dynamic model assumptions, we relaxed the constraint of a fixed number of species $S$ (as in the basic model) and we let $\langle S \rangle = A^z$ be an emergent property of the stochastic model. We achieved this by maintaining the ecological dynamics of births and deaths as in the basic model and by modifying the speciation dynamics in two different ways:
\begin{itemize}
\item[a.] At each time step, the number of species that undergo speciation is drawn from a Poisson distribution with rate $w$. The species that undergo speciation are selected randomly. At speciation, a random number $n_i'<n_i$ of individuals of species $i$ maintains the original mass $m_i$, whereas the remaining individuals are assigned to a new species $j$ with mass $m_j=q m_i$, where $q$ is drawn from a lognormal distribution with constant mean and variance. To avoid instability (i.e. extreme fluctuations that lead to the extinction of the community), we impose that the sum of the consumption rates of species $i$ and $j$ after speciation is equal to the consumption rate of species $i$ before speciation. This is done by setting the abundance of species $j$ such that $n_i' m_i^\alpha+n_j m_j^\alpha=n_i m_i^\alpha$, i.e. $n_j=(n_i-n_i')m_i^\alpha/m_j^\alpha$.
\item[b.] Same as in model a, but the species that undergo speciation are selected randomly with a weight proportional to their abundance, so that more abundant species are more likely to speciate.
\end{itemize}

\begin{figure*}
\begin{center}
\includegraphics[width=17.8cm,center]{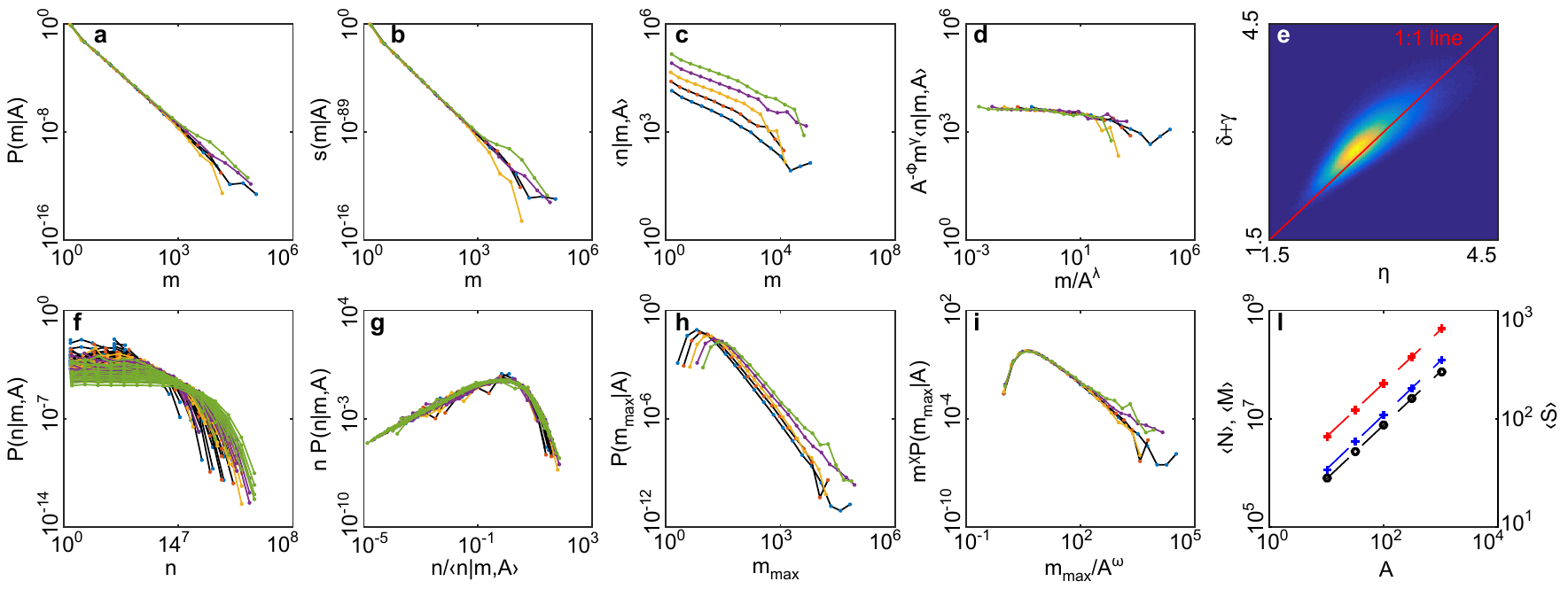}
\caption{Model \textit{a} (section 3.2) statistics computed with the parameter set: $\alpha=3/4$, $\theta=1/4$, $w=10^{-2}$, $v_0=1/2$ and $c=10^{-5}$. Different colors refer to different values of ecosystem area $A= 10^i$, where $i=1$ (cyan)$; 3/2$ (orange)$; 2$ (yellow)$; 5/2$ (purple)$; 3$ (green). Panels a-c, f and h show respectively $P(m|A)$, $s(m|A)$, $\langle n|m,A \rangle$,  $P(n|m,A)$ and $P(m_{\max}|A)$ estimated at stationarity. Panels d, g and i show collapses of simulation data for $\langle n|m,A \rangle$,  $P(n|m,A)$ and $P(m_{\max}|A)$, respectively. Panel e shows the density scatter-plot of $\delta+\gamma$ versus $\eta$. The density histogram is normalized to one, with blue representing the value zero and yellow the value one. Shown are simulation data for the largest area $A=10^3$. Panel j shows the scaling of the average total biomass $\langle M \rangle$ (blue crosses and dashed lines), average total abundance $\langle N \rangle$ (black crosses and dashed lines) and average number of species $\langle S \rangle$ (red crosses and dashed lines) with $A$. See Table \ref{tab:explorationfluct} for estimates of exponents' values.}
\label{fig:nonscaling_conserved_fission_no_n}
\end{center}
\end{figure*}

\begin{figure*}
\begin{center}
\includegraphics[width=17.8cm,center]{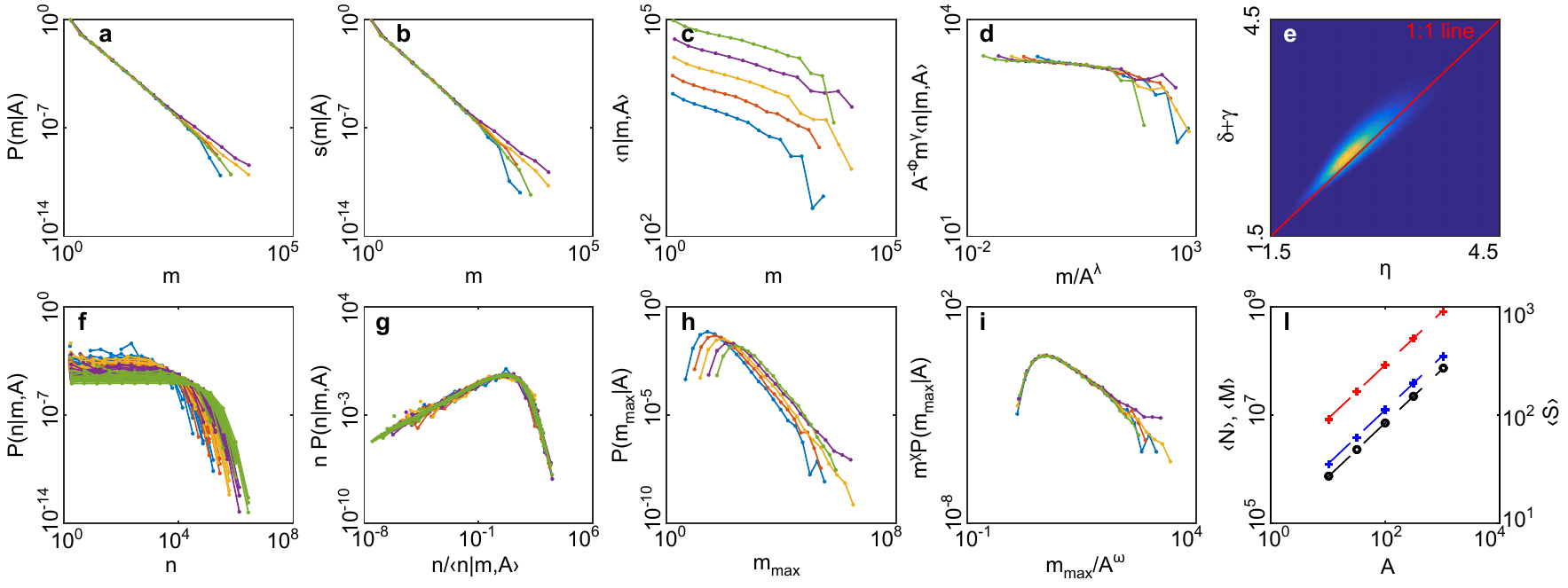}
\caption{Model \textit{b} (section 3.2) statistics computed with the parameter set: $\alpha=3/4$, $\theta=1/4$, $w=10^{-2}$, $v_0=1/2$ and $c=10^{-5}$. Different colors refer to different values of ecosystem area $A= 10^i$, where $i=1$ (cyan)$; 3/2$ (orange)$; 2$ (yellow)$; 5/2$ (purple)$; 3$ (green). Panels a-c, f and h show respectively $P(m|A)$, $s(m|A)$, $\langle n|m,A \rangle$,  $P(n|m,A)$ and $P(m_{\max}|A)$ estimated at stationarity. Panels d, g and i show collapses of simulation data for $\langle n|m,A \rangle$,  $P(n|m,A)$ and $P(m_{\max}|A)$, respectively. Panel e shows the density scatter-plot of $\delta+\gamma$ versus $\eta$. The density histogram is normalized to one, with blue representing the value zero and yellow the value one. Shown are simulation data for the largest area $A=10^3$. Panel j shows the scaling of the average total biomass $\langle M \rangle$ (blue crosses and dashed lines), average total abundance $\langle N \rangle$ (black crosses and dashed lines) and average number of species $\langle S \rangle$ (red crosses and dashed lines) with $A$. See Table \ref{tab:explorationfluct} for estimates of exponents' values.}
\label{fig:nonscaling_conserved_fission_n}
\end{center}
\end{figure*}

We found that these models give rise to the empirically observed set of macroecological laws described in the main text, and of course the exponents of such laws depend both on the model specifications and on the model parameters. Most importantly, despite the differences in the speciation dynamics, we found that these models are also compatible with our scaling framework (Eqs. 6--9 of the main text), which specifies the scaling properties of the joint distribution $P(n,m|A)$. Thereby, macroecological patterns in these models comply with our predicted pattern covariations. Tables \ref{tab:explorationfluct} and \ref{tab:explorationcope} report the exponents values measured in these models and Figs. \ref{fig:nonscaling_conserved_fission_no_n}--\ref{fig:nonscaling_conserved_fission_n_m12} display the corresponding macroecological patterns. Parameter values used to run the models are reported in the figures captions.

\subsection{\normalsize  Value of $\eta$ in the community dynamic models}
\label{valueeta}

\begin{figure*}
\begin{center}
\includegraphics[width=17.8cm,center]{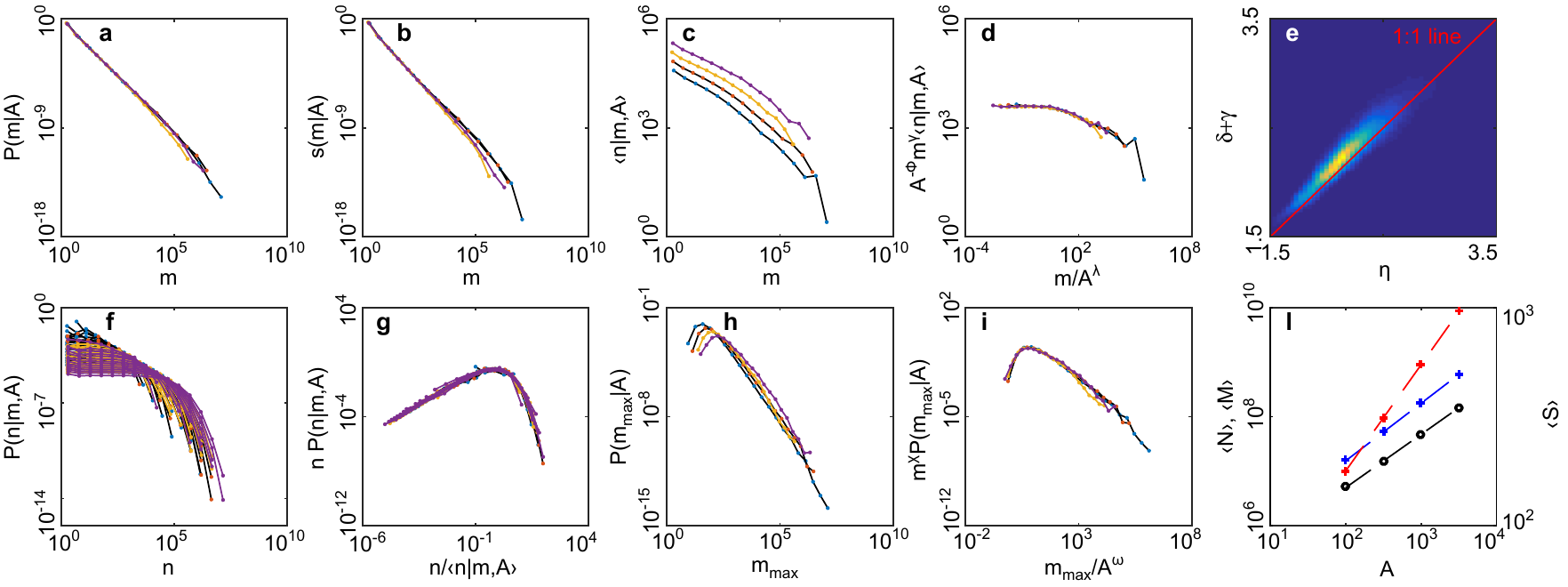}
\caption{Model \textit{a} (section 3.2) statistics computed with the parameter set: $\alpha=3/4$, $\theta=1/4$, $w=10^{-2}$, $v_0=1/2$, $c=10^{-5}$ and $\bar q=1.2$ ($\bar q$ is the mean of the multiplicative factor $q$ that defines the descendant species' mass at each speciation event, cfr. section 3.2). Different colors refer to different values of ecosystem area $A= 10^i$, where $i=1$ (cyan)$; 3/2$ (orange)$; 2$ (yellow)$; 5/2$ (purple)$; 3$ (green). Panels a-c, f and h show respectively $P(m|A)$, $s(m|A)$, $\langle n|m,A \rangle$,  $P(n|m,A)$ and $P(m_{\max}|A)$ estimated at stationarity. Panels d, g and i show collapses of simulation data for $\langle n|m,A \rangle$,  $P(n|m,A)$ and $P(m_{\max}|A)$, respectively. Panel e shows the density scatter-plot of $\delta+\gamma$ versus $\eta$. The density histogram is normalized to one, with blue representing the value zero and yellow the value one. Shown are simulation data for the largest area $A=10^3$. Panel j shows the scaling of the average total biomass $\langle M \rangle$ (blue crosses and dashed lines), average total abundance $\langle N \rangle$ (black crosses and dashed lines) and average number of species $\langle S \rangle$ (red crosses and dashed lines) with $A$. See Table \ref{tab:explorationcope} for estimates of exponents' values.}
\label{fig:nonscaling_conserved_fission_no_n_m12}
\end{center}
\end{figure*}

\begin{figure*}
\begin{center}
\includegraphics[width=17.8cm,center]{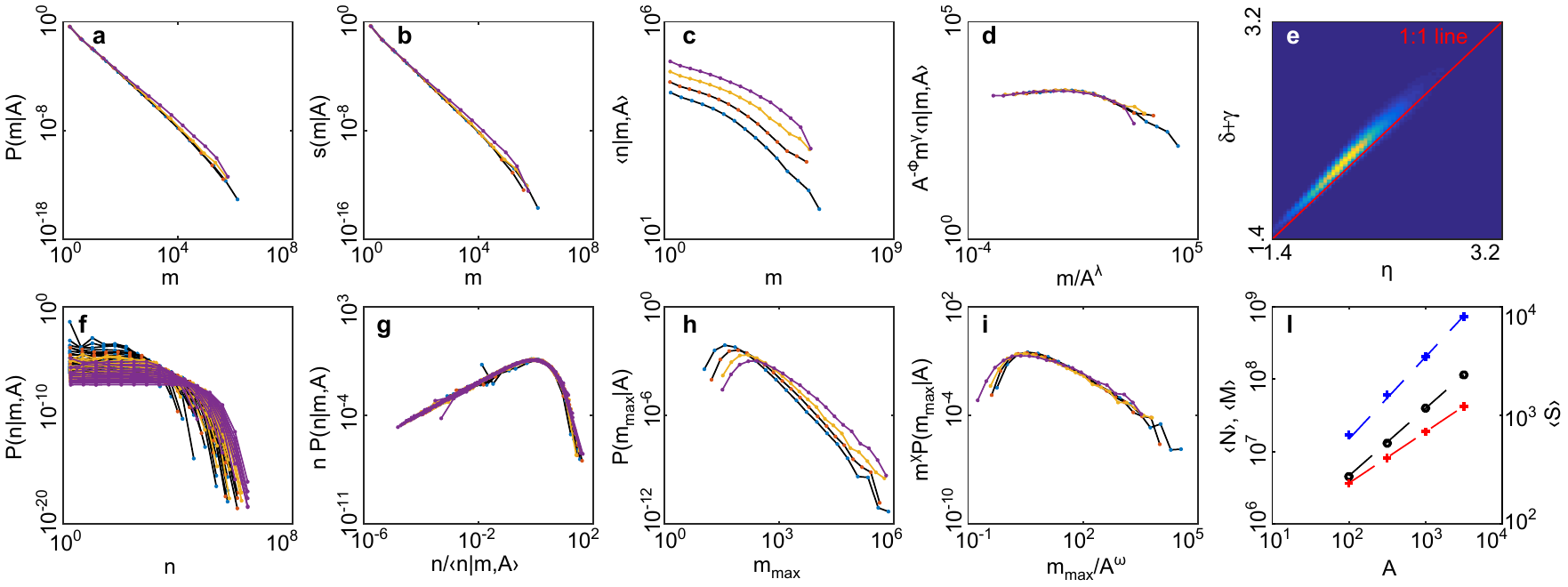}
\caption{Model \textit{b} (section 3.2) statistics computed with the parameter set: $\alpha=3/4$, $\theta=1/4$, $w=10^{-2}$, $v_0=1/2$, $c=10^{-5}$ and $\bar q=1.2$ ($\bar q$ is the mean of the multiplicative factor $q$ that defines the descendant species' mass at each speciation event, cfr. section 3.2). Different colors refer to different values of ecosystem area $A= 10^i$, where $i=1$ (cyan)$; 3/2$ (orange)$; 2$ (yellow)$; 5/2$ (purple)$; 3$ (green). Panels a-c, f and h show respectively $P(m|A)$, $s(m|A)$, $\langle n|m,A \rangle$,  $P(n|m,A)$ and $P(m_{\max}|A)$ estimated at stationarity. Panels d, g and i show collapses of simulation data for $\langle n|m,A \rangle$,  $P(n|m,A)$ and $P(m_{\max}|A)$, respectively. Panel e shows the density scatter-plot of $\delta+\gamma$ versus $\eta$. The density histogram is normalized to one, with blue representing the value zero and yellow the value one. Shown are simulation data for the largest area $A=10^3$. Panel j shows the scaling of the average total biomass $\langle M \rangle$ (blue crosses and dashed lines), average total abundance $\langle N \rangle$ (black crosses and dashed lines) and average number of species $\langle S \rangle$ (red crosses and dashed lines) with $A$. See Table \ref{tab:explorationcope} for estimates of exponents' values.}
\label{fig:nonscaling_conserved_fission_n_m12}
\end{center}
\end{figure*}

The size spectrum exponent $\eta$ in natural ecosystems typically assumes values $\eta \in (1,2]$ (see Tables \ref{exponents}, \ref{tab:bci} and \ref{tab:luquillo}), although values of $\eta>2$ can also be found in marine environments (Table \ref{exponents}). All our community dynamics models yield values of $\eta$ that are on average larger than $2$ (the average is performed over time, see Tables \ref{tab:exploration}, \ref{tab:explorationvar} and \ref{tab:explorationfluct}), although panels \textit{e} in Figs. \ref{fig:z14standard}--\ref{fig:nonscaling_conserved_fission_n_m12} show that $\eta$ can assume values smaller than $2$ at any fixed time point (i.e. in snapshots of the ecosystem). Unfortunately, a suitably broad exploration of the parameters space in our models is computationally unfeasible, as the estimation of scaling exponents requires several hours of computation in a high-performance computer in order to properly estimate the tails of the distribution $P(n,m|A)$. However, based on our exploration of parameters' space, $\langle \eta \rangle = 2$ does seem to be a lower limit in our community dynamics models. In the attempt to find parameter sets that may allow for $\langle \eta \rangle < 2$, we found that increasing the mean $\bar q$ of the multiplicative factor $q$ that specifies the mass of the descendant species at a speciation event (this may be seen as an implementation of Cope's rule \cite{Rensch1948}, which postulates that descendant lineages tend to increase in body size) causes a reduction of the mean size spectrum exponent $\langle \eta \rangle$ (Figs. \ref{fig:nonscaling_conserved_fission_no_n_m12} and \ref{fig:nonscaling_conserved_fission_n_m12}, and Table \ref{tab:explorationcope}). Nonetheless, parameter sets that yield $\langle \eta \rangle < 2$ lead to communities that are very unstable and that rapidly go towards extinction. It thus appears that our community dynamics models are missing processes that would allow multiple species to coexist at a stable equilibrium with $\langle \eta \rangle < 2$. We speculate that one reason for this behavior may be the fact that our models assume a well-mixed system, unlike terrestrial ecosystems such as forests. In this sense, it may not be coincidental that values of $\eta>2$ are typically found in aquatic ecosystems (Table \ref{exponents}) rather than terrestrial ones (Tables \ref{exponents}, \ref{tab:bci} and \ref{tab:luquillo}). Further research will be dedicated to the investigation of macroecological linkages in metacommunities \cite{Wilson1992,Leibold2004,Muneepeerakul2008,Bertuzzo2011}, with explicit focus on the implications of spatial structure and connectivity on scaling exponents values and linkages.

\subsection{\normalsize Specificity and universality}

Our investigation of dynamic birth, death and speciation models corroborates the generality of our scaling framework and the predicted pattern covariations. In fact, we found that all the models investigated that are compatible with the empirically observed macroecological patterns described in the main text are all characterized by the same scaling properties of $P(n,m|A)$, which are encapsulated in our scaling framework (Eqs. 2--5 of the main text) and univocally specify the pattern covariations in Eqs. 1, 6--9 of the main text. Therefore, Eqs. 2--5 of the main text do not rely on specific assumptions about the population and speciation dynamics of a community, but rather specify the universal scaling properties that possibly any dynamic model compatible with the empirically observed macroecological laws must satisfy. Furthermore, the pattern covariations predicted by our scaling framework agree with empirical evidence (section 2) and with heuristic arguments, i.e. the many back-of-the-envelope calculations reported in the main text. It must be understood, however, that the scaling framework does not predict the values of scaling exponents, but rather their covariations.
 
The various community dynamic models studied here, instead, do predict scaling exponents values, which emerge from the rates and assumptions concerning birth, death and speciation events. However, there may exist several dynamic models capable of reproducing quantitatively one specific set of exponents' values, as it is often the case that several processes lead to the same pattern \cite{Newman2007}. Furthermore, we do not claim that our dynamic models describe any real ecosystem in all its complexity, as such models are of course overly simplified to encompass the broad range of processes that may set the values of macroecological scaling exponents (e.g. species' interactions, landscape structure and ecological disturbances, to name a few). When modelling any natural process, the first step is that of abstraction: unnecessary details are removed, until one reaches the simplest model that is still compatible with the observed patterns, which for the purpose of this investigation are the functional forms of the various scaling relationships and distributions. In this sense, we believe that our models of birth, death and speciation capture the essential ingredients that produce the empirically-observed functional forms of macroecological laws and that set the scaling properties of the joint distribution of mass and abundance, and thus the pattern covariations. The exact values of the macroecological scaling exponents, instead, are most likely determined by several processes that are not included in our community dynamics models, but can be properly described by our scaling framework.

\clearpage


\begin{table*}
\caption{Estimates for the scaling exponents of the ecological patterns depicted in Fig. 1 (main text). Errors are SEM, CI stands for confidence interval. If no error is reported, none was given in the original paper. }
\label{exponents}
\centering
\small
\begin{tabular*}{\hsize}{@{\extracolsep{\fill}}cccc}
Law & Forests & Terrestrial & Aquatic\\
\hline
Kleiber's law (panel a) & $0.80 \pm 0.01$ & $0.67 \pm 0.2$ & $1.10,\  \mathrm{CI}\  95\%:\  [0.94,1.21]$\\
Damuth's law (panel b) &$0.26\pm 0.05$ & $0.57 \pm 0.08$ & $0.73,\  \mathrm{CI}\ 95\%:\ [0.73,0.92]$\\
SAR (panel c) & $0.27 \pm 0.01$ & $0.23 \pm 0.02 $ & $0.094$\\
$s(m)$ & $1.59,\  \mathrm{CI}\  95\%:\ [1.57,1.63]$ & $1.5 \pm 0.2$ & $2.11$
\end{tabular*}
\end{table*}

\begin{table*}
\caption{References to empirical measurements of the ecological patterns discussed in the main text. RSA stands for Relative Species Abundance and Max. body mass stands for the scaling of the maximum body mass with the area of the ecosystem.}
\label{citations}
\centering
\begin{tabular*}{\hsize}{@{\extracolsep{\fill}}l|ccc}
Law & Forests & Terrestrial & Aquatic\\
\hline
&&&\\
Kleiber's law & Mori \textit{et al.} (2010) \cite{Mori2010}  & Kleiber (1932) \cite{Kleiber1932}            & Nielsen and Sand-Jensen (1990) \cite{Nielsen1990}\\
 			  &                             &Dodds \textit{et al.} (2001) \cite{Dodds2001} & Finkel \textit{et al.} (2004) \cite{Finkel2004}\\
              &                             &                            & Mara{\~{n}}{\'{o}}n \textit{et al.} (2007) \cite{Maranon2007}\\ 
             &&&\\ \hline
              &&&\\
SAR           & Lomolino (1982) \cite{Lomolino1982}            & MacArthur and Wilson (1963) \cite{MacArthur1963} & Dodson (1992) \cite{Dodson1992}\\
             &                             & Newmark (1986) \cite{Newmark1986}              & Lonsdale (1999) \cite{Lonsdale1999}\\
              &                            & Okie \textit{et al.} (2009) \cite{Okie2009} & Smith \textit{et al.} (2005) \cite{Smith2005}\\
			  & 							& Preston (1962) \cite{Preston1962} 			   & \\ 
			 &&&\\ \hline
			  &&&\\
Damuth's law & Cohen \textit{et al.} (2012) \cite{Cohen2012} & Damuth (1981) \cite{Damuth1981}  			   & Cyr \textit{et al.} (1997) \cite{Cyr1997}\\
 			 & 								& Nee \textit{et al.} (1991) \cite{Nee1991}  & Cohen \textit{et al.} (2003) \cite{Cohen2003}\\ 
	&		 & Novosolov \textit{et al.} (2015) \cite{Novosolov2015}	&	\\ 
	&&&\\ \hline
	&&&\\
$s(m)$    & Muller-Landau \textit{et al.} (2006) \cite{Muller-Landau2006} & White \textit{et al.} (2007) \cite{White2007} & Sheldon (1972) \cite{Sheldon1972}\\
            & Stegen and White (2008) \cite{Stegen2008}             &             Halfpenny (2016) \cite{Niwot}                 & Cavender-Bares \textit{et al.} (2001) \cite{Cavender-Bares2001} \\
            &         Condit \textit{et al.} (2012) \cite{Condit2012}                            &                              & Rinaldo \textit{et al.} (2002) \cite{Rinaldo2002}\\
            &                                     &                              & Mara{\~{n}}{\'{o}}n \textit{et al.} (2015) \cite{Maranon2015}\\ 
           &&&\\ \hline
            &&&\\
$P(m)$    &                                     & Marquet and Taper (1998) \cite{Marquet1998} &\\
           & 										& Smith  \textit{et al.} (2003) \cite{Smith2003} &\\
			& 										& Marquet \textit{et al.} (2005) \cite{Marquet2005} &\\
			& 										&  Southwood \textit{et al.} (2006) \cite{Southwood2006}& \\ 
			&&&\\\hline
			&&&\\
Max. body mass &  									& Burness  \textit{et al.} (2001) \cite{Burness2001} &\\
				& 									& Okie  \textit{et al.} (2009) \cite{Okie2009} &\\ 
			&&&\\	\hline
				&&&\\
RSA 			& 									& Preston (1948) \cite{Preston1948} &\\ 
\hline
&&&\\
Taylor's law & Giometto  \textit{et al.} (2015) \cite{Giometto2015} & Taylor  \textit{et al.} (1961) \cite{Taylor1961} &\\
             &                                  & Taylor  \textit{et al.} (1980) \cite{Taylor1980} &\\ 
             &                                  & Anderson  \textit{et al.} (1982) \cite{Anderson1982} &
\end{tabular*}
\end{table*}
\clearpage

\begin{table*}
\caption{Estimates of scaling exponents $\eta$, $\delta$ and $\gamma$ in the BCI forest. Errors are computed as reported in the text.}
\label{tab:bci}
\centering
\small
\begin{tabular}{l|ccc}
BCI forest & Census $5$  & Census $6$ & Census $7$ \\ \hline
&&&\\
$s(m)$ & $\eta=1.43\pm0.04$ & $\eta=1.43\pm0.03$ & $\eta=1.44\pm0.04$\\
$P(m)$ & $\delta=1.03 \pm 0.03$ & $\delta=1.05 \pm 0.03$ & $\delta=1.07 \pm 0.05$\\
Damuth's law & $\gamma=0.41 \pm 0.07$ & $\gamma=0.40 \pm 0.06$& $\gamma=0.38 \pm 0.06$ \\ 
\end{tabular}
\end{table*}

\begin{table*}
\caption{Estimates of scaling exponents $\eta$, $\delta$ and $\gamma$ in the Luquillo forest \cite{Zimmerman2010}. Errors are computed as reported in the text.}
\label{tab:luquillo}
\centering
\small
\begin{tabular}{l|ccccc}
Luquillo forest & Census $1$  & Census $2$ & Census $3$ & Census $4$ & Census $5$ \\ \hline
&&&\\
$s(m)$ & $\eta=1.27\pm0.04$ & $\eta=1.18\pm0.03$ & $\eta=1.09\pm0.03$ & $\eta=1.09\pm0.04$ & $\eta=0.95\pm0.07$\\
$P(m)$ & $\delta=1.02 \pm 0.02$ & $\delta=1.01 \pm 0.03$ & $\delta=1.02 \pm 0.05$ & $\delta=1.02 \pm 0.03$ & $\delta=1.02 \pm 0.01$\\
Damuth's law & $\gamma=0.21 \pm 0.08$ & $\gamma=0.23 \pm 0.04$& $\gamma=0.16 \pm 0.04$ & $\gamma=0.09 \pm 0.06$& $\gamma=0.03 \pm 0.04$ \\ 
\end{tabular}

\end{table*}

\begin{table*}
\caption{Estimates of scaling exponents $z$, $\delta$, $\Phi$ and $\gamma$ for the LIZ dataset \cite{Novosolov2015}. Errors on $z$ and $\delta$ are SEM, the error on $\gamma$ is the SD, the error on $\Phi$ was obtained by bootstrapping. $P(\bar{m})$ is computed gathering together species from all the islands in the dataset.}
\label{tab:LIZ}
\centering
\small
\begin{tabular}{lcc}
$SAR$ & $z=0.17\pm 0.01 $ & R$^2$=0.46\\
Damuth's law &$\Phi=0.78 \pm 0.08$&\\
Damuth's law &$\gamma=0.53 \pm 0.03$ & R$^2$=0.89\\
$P(m)$ & $\delta=1.45 \pm 0.06$ & \\
\end{tabular}
\end{table*}

\begin{table*}
\caption{Estimates of scaling exponents $z$, $\xi$ and $\delta$ for the SSI dataset \cite{Okie2009}.  Errors on $z$, $\delta$ and $\xi$ are SEM. $R^2$ is the coefficient of determination. $P(\bar{m})$ is computed gathering together species from all the islands in the dataset.
}
\label{tab:SSI}
\centering
\small
\begin{tabular}{lcc}
$SAR$ & $z=0.23 \pm 0.02$ & $R^2=0.93$  \\
$M_{\max}$ & $\xi=0.49 \pm 0.09$ & $R^2=0.76$ \\
$P(m)$ & $\delta=1.6 \pm 0.2$  \\
\end{tabular}
\end{table*}
 \clearpage
 
 \begin{table*}
 \caption{Scaling exponents measured in the basic model for different sets of parameters' values. Each row (delimited by horizontal lines) refers to a set, indicated by the parameter which value has been changed with respect to the parameters set described in the main text (indicated as ``Main text set", specified in section 3.1). Under each value, the lower and upper ends of the confidence intervals are reported. The estimates and the confidence intervals were obtained as described in the Methods section.}
\label{tab:exploration}
\centering
\begin{tabular*}{\hsize}{@{\extracolsep{\fill}}l|ccccccccc}
& $\delta$&$\eta$&$\gamma$&$\Phi$&$\lambda$&$\chi$&$\omega$&$\mu$&$\nu$\\ &&&\\ \hline &&&\\
Main text set		&2.23 &2.54 &0.26 &0.74 &0.38 &0.99 &0.18&1.0016 &0.9982\\
 Fig. \ref{fig:z14standard}              			&2.05 &2.28 &0.20 &0.72 &0.32 &0.87 &0.17&1.0016 &0.9980\\
              			&2.41 &2.79 &0.34 &0.76 &0.42 &1.03 &0.21 & 1.0017 &0.9983\\ &&&\\ \hline &&&\\
$w=10^{-4}$ 	&2.27 &2.64 &0.33 &0.76 &0.52 & 0.98 & 0.19 &1.0030 &0.9968\\
Fig. \ref{fig:z14mu4}               		&2.01 &2.22 &0.28 &0.74 &0.50 & 0.92 & 0.17 &1.0027 &0.9967\\
              		&2.54 &2.94 &0.38 &0.78 &0.57 & 1.04 & 0.20 &1.0034 &0.9969\\ &&&\\ \hline &&&\\
$w=10^{-5}$ 	& 2.28 & 2.63 & 0.30 & 0.75 & 0.46 & 0.94 & 0.18 & 1.000 & 0.998\\
Fig. \ref{fig:z14mu5}               		& 2.00 & 2.21 & 0.25 & 0.74 & 0.41 & 0.73 & 0.16 & 0.996 & 0.995\\
              		& 2.54 & 2.94 & 0.36 & 0.79 & 0.52 & 1.05 & 0.22 & 1.005 & 1.001\\ &&&\\ \hline &&&\\
Fig. \ref{fig:z14alpha05}               		& 2.00 & 2.21 & 0.26 & 0.74 & 0.73 & 0.93 & 0.18 & 1.0029 & 0.9989 \\
              		& 2.54 & 2.93 & 0.31 & 0.78 & 0.87 & 1.03 & 0.21 & 1.0031 & 0.9989\\ &&&\\ \hline &&&\\
$\alpha=1/4$ 	& 2.38 & 2.64 & 0.29 & 0.76 & 0.63 & 0.98 & 0.19 & 1.000 & 1.000 \\
Fig. \ref{fig:z14alpha025}               		& 2.01 & 2.22 & 0.25 & 0.73 & 0.60 & 0.94 & 0.19 & 0.998 & 0.998 \\
              		& 2.54 & 2.94 & 0.35 & 0.80 & 0.69 & 1.03 & 0.20 & 1.002 & 1.000 \\ &&&\\ \hline &&&\\
$\theta=1/2$ 	& 2.13 & 2.62 & 0.49 & 0.78 & 0.47 & 0.95 & 0.22 & 1.002 & 0.997 \\
Fig. \ref{fig:z14theta05}               		& 1.88 & 2.26 & 0.26 & 0.75 & 0.44 & 0.87 & 0.21 & 1.002 & 0.996 \\
              		& 2.36 & 2.94 & 0.56 & 0.86 & 0.53 & 1.03 & 0.26 & 1.003 & 0.998 \\  &&&\\ \hline &&&\\
		$z=1/2$		& 2.23 & 2.41 & 0.32 & 0.50 & 0.93 & 0.98 &0.38&1.0 &0.99\\
Fig. \ref{fig:z05} 				& 2.05 & 2.28 & 0.31 & 0.49 & 0.92 & 0.96 &0.36 &1.00 &0.99\\
				& 2.54 & 2.79 & 0.34 & 0.51 & 0.96 & 1.05 &0.39 &1.01 &1.00\\
\end{tabular*}
\end{table*}

\clearpage

\begin{table*}
\caption{Scaling exponents measured in the variation of the basic model described in section 3.1.2, Fig. \ref{fig:z14variation}. Under each value, the lower and upper ends of the confidence intervals are reported. The estimates and the confidence intervals were obtained as described in the Methods section.}
\label{tab:explorationvar}
\centering
\begin{tabular*}{\hsize}{@{\extracolsep{\fill}}ccccccccc}
$\delta$&$\eta$&$\gamma$&$\Phi$&$\lambda$&$\chi$&$\omega$&$\mu$&$\nu$\\ \hline &&&\\
2.54 &2.78 &0.13 &0.76 &0.36 &0.94 &0.14&1.0021 &0.998\\
2.22 &2.29 &0.04 &0.74 &0.35 &0.7 &0.14&1.0021 &0.998\\
2.83 &3.11 &0.28 &0.78 &0.40 &1.1 &0.17 & 1.0022 &0.998\\ 
\end{tabular*}

\end{table*}

\begin{table*}
\caption{Scaling exponents measured in the community dynamics models with fluctuating numbers of species. Each row (delimited by horizontal lines) refers to a different model (see section 3.2). Model parameters are reported in Figs. \ref{fig:nonscaling_conserved_fission_no_n} and \ref{fig:nonscaling_conserved_fission_n}. Under each value, the lower and upper ends of the confidence intervals are reported. The estimates and the confidence intervals were obtained as described in the Methods section.}
\label{tab:explorationfluct}
\centering
\small
\begin{tabular*}{\hsize}{@{\extracolsep{\fill}}c|cccccccccc}
Model & $\delta$&$\eta$&$\gamma$&$\Phi$&$\lambda$&$\chi$&$\omega$&$\mu$&$\nu$ & $z$\\ \hline &&&\\
a	& 2.52 & 2.83 & 0.28 & 0.51 & 0.92 & 1.00 & 0.32 & 1.01 & 1.00 & 0.50 \\
Fig. \ref{fig:nonscaling_conserved_fission_no_n}      & 2.26 & 2.50 & 0.24 & 0.48 & 0.82 & 0.91 & 0.29 & 1.00 & 0.99 & 0.49 \\
      & 2.77 & 3.15 & 0.34 & 0.56 & 0.99 & 1.08 & 0.33 & 1.01 & 1.00 & 0.50 \\  &&&\\ \hline &&&\\
b	& 2.52 & 2.69 & 0.13 & 0.50 & 0.60 & 0.98 & 0.32 & 1.00 & 1.00 & 0.50 \\
Fig. \ref{fig:nonscaling_conserved_fission_n}      & 2.27 & 2.41 & 0.10 & 0.48 & 0.56 & 0.92 & 0.30 & 0.98 & 0.99 & 0.49 \\
      & 2.76 & 2.96 & 0.18 & 0.52 & 0.66 & 1.11 & 0.34 & 1.02 & 1.01 & 0.50 \\ 
\end{tabular*}
\end{table*}

\begin{table*}
\caption{Scaling exponents measured in the models with fluctuating numbers of species and $\bar q > 1$ (see section 3.3). Each row (delimited by horizontal lines) refers to a model and parameter set. Model parameters are reported in Figs. \ref{fig:nonscaling_conserved_fission_no_n_m12} and \ref{fig:nonscaling_conserved_fission_n_m12}. Under each value, the lower and upper ends of the confidence intervals are reported. The estimates and the confidence intervals were obtained as described in the Methods section. The parameter $\bar q$ is the mean of the multiplicative factor $q$ that defines the descendant species' mass at each speciation event (cfr. section 3.2).}
\label{tab:explorationcope}
\centering
\small
\begin{tabular*}{\hsize}{@{\extracolsep{\fill}}c|cccccccccc}
Model & $\delta$&$\eta$&$\gamma$&$\Phi$&$\lambda$&$\chi$&$\omega$&$\mu$&$\nu$ & $z$\\ \hline &&&\\
\textit a with				& 1.96 & 2.22 & 0.31 & 0.52 & 0.80 & 0.94 & 0.50 & 1.06 & 0.98 & 0.49 \\
 $\bar q=1.2$   			& 1.75 & 1.98 & 0.26 & 0.50 & 0.75 & 0.75 & 0.45 & 0.91 & 0.95 & 0.48 \\
Fig. \ref{fig:nonscaling_conserved_fission_no_n_m12}	& 2.16 & 2.44 & 0.34 & 0.55 & 0.90 & 1.18 & 0.58 & 1.20 & 1.00 & 0.50 \\  &&&\\ \hline  &&&\\
\textit b with				& 1.93 & 2.06 & 0.25 & 0.53 & 0.93 & 0.91 & 0.64 & 1.13 & 0.93 & 0.47 \\
$\bar q=1.2$			& 1.71 & 1.83 & 0.19 & 0.50 & 0.89 & 1.65 & 0.57 & 0.97 & 0.87 & 0.43 \\
Fig. \ref{fig:nonscaling_conserved_fission_n_m12}	& 2.14 & 2.27 & 0.28 & 0.55 & 1.10 & 0.46 & 0.70 & 1.30 & 0.99 & 0.52 \\ 
\end{tabular*}
\end{table*}

\begin{table*}
\caption{Summary of the empirical tests performed. References are to equations in the main text.}
\label{tab:summary_tests}
\centering
\small
\begin{tabular}{lll}
Dataset & Measured exponents & Relationship that was verified\\
\hline\\
BCI & $\eta$, $\gamma$, $\delta$ (Table \ref{tab:bci}) & Eq.(1): $\eta=\delta+\gamma$\\
Luquillo & $\eta$, $\gamma$, $\delta$ (Table \ref{tab:luquillo}) & Eq.(1): $\eta=\delta+\gamma$\\
LIZ & z, $\Phi$,$\gamma$, $\delta$ (Table \ref{tab:LIZ})& Eq.(6): $z=1-\Phi-\max\{0,\lambda(1+\alpha-\gamma-\delta)\}$\\
SSI & z, $\xi$, $\delta$ (Table \ref{tab:SSI}) & Eq.(9): $\xi=\frac{z}{\delta-1}$\\
\end{tabular}
\end{table*}

 \clearpage 

\clearpage

\clearpage

\clearpage

\clearpage

\clearpage

\clearpage

\clearpage

\clearpage

\clearpage

\clearpage

\clearpage

\clearpage
